\documentclass[rmp,aps,floatfix]{revtex4}
\usepackage[dvips]{graphicx}
\usepackage{epsfig}
\usepackage{pst-plot}
\usepackage{psfig}
\usepackage{bm}

\begin{document}
\title{Pairing in nuclear systems: from neutron stars to finite nuclei}
\author{D.~J.~Dean}
\affiliation{Physics Division, Oak Ridge National Laboratory \\
P.O. Box 2008, Oak Ridge, TN 37831-6373 USA} 
\author{M.~Hjorth-Jensen}
\affiliation{Department of Physics, University of Oslo, N-0316 Oslo, Norway}
\date{\today}
\begin{abstract}

% new abstract, september 2002
We discuss several pairing-related
phenomena in nuclear systems, ranging from superfluidity
in neutron stars to the gradual breaking of pairs in finite nuclei.
We focus on the links between
many-body pairing as it evolves from the underlying
nucleon-nucleon interaction
and the eventual experimental and theoretical manifestations
of superfluidity in infinite nuclear matter and
of pairing in finite nuclei.
We analyse the nature of pair correlations
in nuclei and their potential impact on nuclear structure
experiments.  We  also describe recent experimental evidence that
points to a relation between pairing and
phase transitions (or transformations) in finite nuclear
systems.  Finally, we discuss recent investigations of ground-state
properties of random two-body interactions where pairing plays little
role although
the interactions yield interesting nuclear properties such as 0$^+$
ground states in even-even nuclei.
   
% old abstract, august 12 2002

%In this work, we attempt to give an exposition of several pairing-related 
%phenomena in nuclear systems, ranging from superfluidity 
%in neutron stars to the gradual breaking of pairs in finite nuclei.
%We will focus on the links  between
%many-body pairing as it evolves from the underlying 
%nucleon-nucleon interaction 
%and the eventual experimental and theoretical manifestations
%of superfluidity in infinite nuclear matter and 
%of pairing in finite nuclei.  
%We analyse
%the nature of pair correlations
%in nuclei and their potential impact on nuclear structure
%experiments.
%We  also describe recent experimental evidence that 
%points to a relation between pairing and 
%phase transitions (or transformations) in finite nuclear 
%systems.  The origin of pairing
%in the presence of a random two-body interaction is also discussed. 
\end{abstract}

\maketitle

\tableofcontents

%\pagebreak

%\input{introduction}
% latest upgrade 3/25/02, 
% mhj added text to first part + neutron star sect. Figure addition as well
% latest 4-4-02 -- djd fixed typos. 
% 4/7/02 mhj minor typos
% june djd typo work
% july-aug final version djd and mhj plus more references
% october 2002 mhj included comments from Hubert
\section{INTRODUCTION}
\label{sec:introduction}

Pairing lies at the heart of nuclear physics and the quantum
many-body problem in general. In this review we address some
of the recent theoretical and experimental studies of pairing 
phenomena in finite nuclei and nuclear matter. 
In infinitely extended nuclear systems, such as neutron star matter 
and nuclear matter, 
the study of superfluidity and pairing has a 
long history, see e.g., \cite{migdal60,cms59,emerysessler}, 
even predating the 1967 discovery of pulsars \cite{hewish}, 
which were soon identified as rapidly rotating magnetic neutron 
stars \cite{gold}.  Interest in nucleonic pairing has intensified 
in recent years, owing primarily to experimental developments on two 
different fronts.  In the field of astrophysics, a series of $X$-ray 
satellites (including Einstein, EXOSAT, ROSAT, and ASCA) has brought a 
flow of data on thermal emission from neutron stars, comprising both upper 
limits and actual flux measurements.  The recent launching of the 
Chandra $X$-ray observatory provides further impetus for more incisive 
theoretical investigations.  On the terrestrial 
front, the expanding capabilities of radioactive-beam and heavy-ion 
facilities have stimulated a concerted exploration of 
nuclei far from stability, with a special focus on neutron-rich species 
\cite{riisager,mueller}.  Pairing plays a prominent role in modeling 
the structure and behavior of these newly discovered nuclei.  

Since the field
is quite vast, we limit our discussion to several recent 
advances that have taken place. We will
focus in particular on two overlapping questions: ({\it i}) how does
many-body pairing evolve from the bare nucleon-nucleon interaction, and
({\it ii}) what are the experimental (and perhaps theoretical) manifestations
of pairing in finite nuclei?  
Over fifty years ago, 
Mayer \cite{mayer_50} pointed out 
that a short-ranged, attractive, nucleon-nucleon interaction would 
yield $J=0$ ground states. The realistic bare nucleon-nucleon 
potential indeed contains short-range attractive 
parts (particularly in the singlet-$S$ and triplet-$P$ channels)
that give rise to pairing in infinite nuclear matter and nuclei. 
In this Review, we will
discuss various calculations that demonstrate this effect. We will
also demonstrate the link between superfluidity in nuclear matter
and its origin from realistic nucleon-nucleon interactions.
We then study the nature of pair correlations
in nuclei, their potential impact on nuclear structure
experiments, and the origin of pairing
in the presence of a random two-body interaction.
We conclude with recent experimental evidence that 
points to a relation between pairing and 
phase transitions (or transformations) in finite nuclear 
systems. 

Before we present the outline of this work, we feel that
some historical remarks about the particularity of the pairing 
problem in nuclear physics may be appropriate.

\subsection{Theory of pairing in nuclear physics}

In 1911 Kamerlingh Onnes discovered superconductivity in
condensed matter systems, and its microscopic explanation came 
about through the highly successful pairing theory proposed 
in 1957 by Bardeen, Cooper, and Schrieffer (BCS) 
\cite{bcs_theory}. A series of first applications to nuclear
structure followed \cite{bmp_58,belyaev_59,migdal_59}. 
The BCS theory also generalized
the seniority coupling scheme in which pair-wise coupling of 
equivalent nucleons to a state of zero angular momentum 
takes place. The scheme had been developed during
years previous to the discovery of BCS theory 
\cite{racah_42,mayer_50,racah_53}. 

BCS applications in nuclear structure calculations 
incorporate two inherent drawbacks. First, the BCS
wave function is not an eigenstate of the number operator, so that number
fluctuation is an issue. Second, there is a critical value
of the pairing-force strength for which no non-trivial solution exists. 
Several attempts were made to overcome these problems:
calculating the random phase approximation (RPA) in addition 
to BCS \cite{uw}; including particle number projection \cite{klm} 
after variation, valid 
for pairing strengths above the critical pairing strength, 
and a projection before variation that works well for all pairing strength 
values. A simplified prescription for the latter 
is a technique known as the Lipkin-Nogami method 
\cite{lipkin,nogami}. It has been quite successful in overcoming 
some of the shortfalls that
occur when BCS is applied to nuclei; see e.g., the recent works of Hagino 
{\em et al.}  \cite{hagino2000,hagino2002} and references therein. 
Of course, BCS is an approximate solution to the many-body problem
and assumes a particular form for the many-body wave function. Another,
more drastic, approximation to the many-body problem 
assumes that a single Slater determinant
suffices to describe the nuclear ground state. This mean-field 
solution to the many-body problem gives rise to 
Hartree-Fock (HF) theory. An effective nucleon-nucleon 
potential describes the nuclear interaction, and is
typically a 
parameterization of the Skyrme zero-range force \cite{sk_56,sk_59,vb_70,vb_72}.
Solutions of the HF equations 
describe various nuclear ground-state properties sufficiently \cite{qf_78}, 
but they do not include an explicit pairing interaction. 
Finite-range interactions, such as
the Gogny interaction \cite{dgg_75}, when used in Hartree-Fock calculations
has also no pairing by construction. 
A general way to include pairing into a mean-field 
description generated by e.g., a Skyrme interaction requires solving the 
Hartree-Fock-Bogoliubov (HFB) equations \cite{bog_58}. Recent
applications to both stable and weakly bound nuclei
may be found in, e.g., \cite{dob_96,heenen_2002b,heenen_2002a}. 
A renormalization scheme for the HFB equations 
was recently proposed by Bulgac and Yu for a zero range pairing interaction
\cite{aurel2001a,aurel2001b}.  
Rather than solving the full HFB equations, one may first
calculate the Hartree-Fock single-particle wave functions and
use these as a basis for solving the BCS equations \cite{tondeur_79,
nayak_95}.  
For stable nuclei with large one- or two-neutron 
separation energies, the HF+BCS approximation to HFB is valid, but
the technique is not able to adequately address weakly bound nuclei
due to the development of a particle (usually neutron) gas on or
near the nuclear surface. 

While nuclear mean-field calculations represent a well-founded 
method to describe nuclear properties, their results do not
represent complete solutions to the nuclear many-body problem. 
Short of a complete solution to the many-nucleon problem \cite{Vijay}, the
interacting shell model is widely regarded as the most broadly capable
description of low-energy nuclear structure and the one most 
directly traceable to the fundamental many-body problem. While
this is a widely accepted statement, applications of the shell model 
to finite nuclei encounter several difficulties. Chief among these
is the choice of the interaction. A second problem involves 
truncations of the Hilbert space, and a third problem involves 
the numerics of solving extremely large eigenvalue problems. 

Skyrme and Gogny forces are parameterized nuclear forces, but they
lack a clear link to the bare nucleon-nucleon interaction as
described by measured scattering phase shifts.  The same philosophy 
has been used for shell-model interactions, e.g., with the
USD $1s$-$0d$-shell interaction \cite{wilden}. While quite
successful, these types of interactions cannot be related 
directly to the nucleon-nucleon interaction either. The shell model
then becomes a true model with many parameters. Alternatively, 
many attempts have been made to derive an effective nucleon-nucleon
interaction in a given shell-model space from the bare nucleon-nucleon
interaction using many-body perturbation theory. (For a modern exposition
on this difficult problem, see \cite{mhj_95} and references therein.)
While this approach appears to work quite well for many nuclei, there 
are several indications \cite{Vijay,vijay2,vijay3} that an effective 
interaction based on a two-body force only fails to reproduce experimental
data. As shown in e.g., \cite{Vijay,vijay2,vijay3}, these
difficulties are essentially related to the absence of a real
three-body interaction. It should
be noted, however, that the deficiencies of the effective
interactions are minimal and affect the ground-state energies more
than they affect the nuclear spectroscopy. Thus, understanding 
various aspects of physics from realistic two-body interactions, or their 
slightly modified, yet more phenomenological, cousins,
is still a reasonable goal.

\subsection{Outline}

This work starts with an overview of pairing in infinite matter, with an 
emphasis on superfluidity and superconductivity in neutron stars.
As an initial theme, we focus on the link between superfluidity 
in nuclear matter 
and its origin from realistic nucleon-nucleon interactions. 
This is done 
in Sec.~\ref{sec:NN_to_pairing} where we discuss pairing in neutron star
matter and symmetric nuclear matter.
Thereafter, we focus on various aspects of pairing in finite nuclei, from
spectroscopic information in Sec.~\ref{sec:pairing_correlations}
to pairing from random interactions in Sec.~\ref{sec:randoms}
and thermodynamical properties in Sec.~\ref{sec:leveldensities_sec3}.
Concluding remarks are presented in Sec.~\ref{sec:conclusion}.
The paragraphs below serve as an introduction to the  exposed physics.

\subsubsection{Pairing in neutron stars}
\label{subsubsect:neutron_stars}

\begin{figure}
   \setlength{\unitlength}{1mm}
   \begin{picture}(100,60)
   \put(0,0){\epsfxsize=10cm \epsfbox{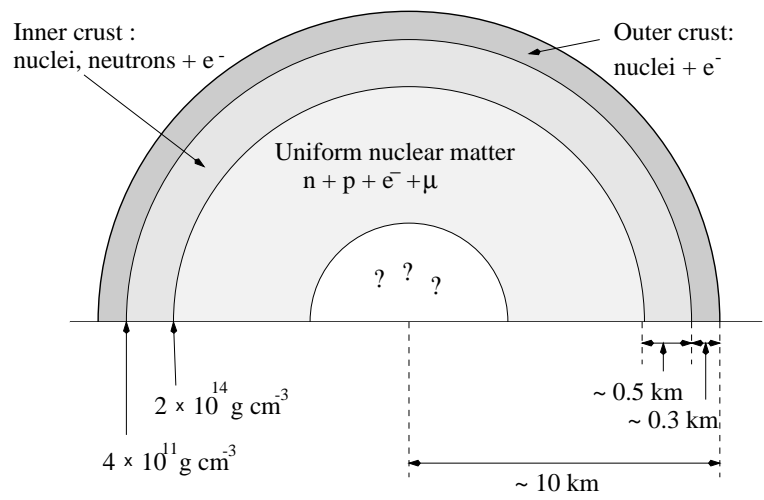}}
   \end{picture}
   \caption{Possible structure of a neutron star.}
   \label{fig:sec1fig_phases}
\end{figure}

The presence of neutron superfluidity in 
the crust and the inner part 
of neutron stars 
are considered well established 
in the physics of these compact stellar objects. 
To a first approximation, a neutron star 
is described as a neutral system of nucleons (and possibly heavier baryons) 
and electrons (and possibly muons) in beta equilibrium at zero temperature, 
with a central density several times the saturation density $\rho_0$ 
of symmetrical nuclear matter \cite{hh2000,pethick1992,shapiro,lamb,fiks,alparlives}. 
The gross structure of the star (mass, radius, pressure, and density 
profiles) is determined by solving the Tolman-Oppenheimer-Volkov general 
relativistic equation of hydrostatic equilibrium, consistently with 
the continuity equation and the equation of state (which embodies the 
microscopic physics of the system).  The star contains (i) an {\it outer 
crust} made up of bare nuclei arranged in a lattice interpenetrated 
by relativistic electrons, (ii) an {\it inner crust} where a similar 
Coulomb lattice of neutron-rich nuclei is embedded in Fermi seas of 
relativistic electrons and neutrons, (iii) a {\it quantum fluid interior} 
of coexisting neutron, proton, and electron fluids, and finally 
(iv) a {\it core region} of uncertain constitution and phase (but 
possibly containing hyperons, a pion or kaon condensate, and/or quark matter).
Fig.~\ref{fig:sec1fig_phases} gives a schematic portrait of a 
possible neutron star structure.

In the low-density outer part of a neutron star, 
the neutron superfluidity is expected 
mainly in the attractive singlet $^1S_0$ channel. 
Qualitatively, this phenomenon can be understood
as follows.  At the relatively large average particle spacing at the 
``low'' densities involved in this region, i.e., $\rho \sim \rho_0/10$ 
with  $\rho_0$ the saturation density 
of symmetrical nuclear matter, 
the neutrons experience mainly the attractive component 
of the $^1S_0$ interaction;  however, the pairing effect is quenched 
at higher densities, $\sim \rho_0$ and beyond, due to the strong 
repulsive short-range component of this interaction.  
At higher density, the nuclei in the crust dissolve, and one 
expects a region consisting of a quantum liquid of neutrons and 
protons in beta equilibrium. 
By similar reasoning, one 
thus expects $^1S_0$ proton pairing to occur in the quantum fluid interior, 
in a density regime where the proton contaminant (necessary for
charge balance and chemical equilibrium) reaches a partial 
density $\rho_p \sim \rho_0/10$. 
In this region, neutron superfluidity is expected to  
occur mainly in the coupled $^3P_2$-$^3F_2$ two-neutron channel. 
At such densities, one may also expect superfluidity from 
other baryons such as, e.g., 
hyperons to arise. The possibility for hyperon pairing is an entirely open
issue; see, for example, \cite{barnea1996}.
Neutron, proton, and eventual hyperon superfluidity in the $^1S_0$
channel, and neutron  superfluidity in the $^3P_2$ channel, have been shown
to occur with gaps of a few MeV or less \cite{pair2};
however, the density ranges in which gaps occur remain
uncertain. 
In the core of the star any superfluid 
phase should finally disappear, although the possibility of a
color superconducting phase may have interesting consequences.
At large baryon densities for which perturbative QCD
applies, pairing gaps for like quarks have been estimated to be a few
MeV \cite{bailin84}.  However, the pairing gaps of unlike quarks ($ud,~
us$, and $ds$) have been suggested to be several tens to
hundreds of MeV through non-perturbative studies \cite{qsf0}
kindling interest in quark superfluidity and superconductivity
\cite{qsf} and their effects on
neutron stars.

A realistic {\it ab initio} prediction of the 
microscopic physics of nucleonic superfluid components in the interiors 
of neutron stars is crucial to a quantitative understanding of neutrino 
cooling mechanisms \cite{cooling1,cooling2,cooling3,nstar} 
that operate immediately after their birth 
in supernova events, as well as the magnetic properties, vortex structure, 
rotational dynamics, and pulse timing irregularities of these 
superdense stellar objects.  
In particular, when nucleonic species enter 
a superfluid state in one or another region of the star, suppression 
factors of the form $\exp (-\Delta_F /k_B T)$ are introduced 
into the expression for the emissivity, $\Delta_F$ being an appropriate 
average measure of the energy gap at the Fermi surface.  
Pairing thus has a major effect on the
star's thermal evolution through suppressions of neutrino
emission processes and specific heats as well; see, for example,
\cite{page2000}.

\subsubsection{Pairing phenomena in nuclei}
\label{subsubsec:phenomena}

After the excursion to infinite matter, 
we return to the question concerning how to obtain 
information on pairing correlations in finite nuclei from 
abundantly available spectroscopic data.
We discuss this point in Sec.~\ref{sec:pairing_correlations}. 
Even in the presence of random interactions, 
signatures of pairing still remain in finite many-body 
systems.  In Sec.~\ref{sec:randoms} we present a discussion
of pairing derived from random interactions.

Apart from relatively weak electric forces, the interactions
between two protons are very similar to those between two neutrons. 
This yields the idea of charge symmetry of the nuclear forces. 
Furthermore, the proton-neutron interaction is also very similar. 
This led very early to the idea of isotopic invariance of the
nucleon-nucleon interaction. A nucleon with quantum number isospin $\tau=1/2$
may be in one of two states, $\tau_z=-1/2$ (proton) or $\tau_z=+1/2$ (neutron).
Of course, the symmetry is not exact, but is widely employed when 
discussing nuclei. It leads to a quantum number $T$ called isospin, and 
its projection $T_z=(N-Z)/2$, where the number of neutrons (protons) in 
the nucleus is $N$ ($Z$). 

We can define with this isospin symmetry two distinct states within the
two-nucleon system. A $T=1$ nucleon-nucleon system can have spin-projection
$T_z=1,0,-1$. $T_z=1$ corresponds to a neutron-neutron system, $T_z=0$ to 
a proton-neutron system, and $T_z=-1$ to a proton-proton system. The nucleons
in this case have total spin $J=0$ in order for the full wave function to 
maintain antisymmetry of the total nucleon-nucleon wave function. For the
same reason, $T=0$ proton-neutron systems can only have $T_z=0$ and $J=1$.
Thus, two different types of elementary particle pairs exist in the 
nucleus, and they
depend on both the spin and isospin quantum numbers of the two-particle system. 

This brief discussion of the general quantum numbers of a two-nucleon system
is a natural starting point for a discussion
of pairing found in nuclei. All even-even nuclei have a ground-state 
with total angular momentum quantum number and parity, $\pi$, $J^\pi=0^+$. 
One can postulate a pairing interaction that couples particles 
in time-reversed states.  Using this type of simple pairing 
interaction, one can also understand the fact that in even-even 
nuclei the ground state is rather well separated from excited 
states, although in the even-odd neighbor nucleus, several states 
exist near the ground state. 

The behavior of the even-even ground state is usually associated with
isovector ($T=1$) pairing of the elementary two-body system. Simplified
models of the nucleon-nucleon interaction, 
such as the seniority model \cite{talmi93}, 
predict a pair condensate in these systems. 
An open question concerns
evidence for isoscalar ($T=0$) pairing in nuclei. 
One unique aspect of 
nuclei with  $N$=$Z$ is that neutrons and protons occupy the same
shell-model orbitals. Consequently, 
the large spatial overlaps between neutron
and proton single-particle wave functions are expected to
enhance neutron-proton ($np$) correlations,  especially the $np$
pairing.  

At present, it is not clear what the specific experimental
fingerprints of the $np$ pairing are, whether the 
$np$ correlations
are strong enough to form
a static condensate, and what their 
main building blocks are. 
Most of our knowledge about nuclear pairing 
comes from nuclei with a  sizable neutron excess where
the isospin $T$=1 neutron-neutron ($nn$) and proton-proton 
($pp$) pairing  dominate. Now, for the first time,
there is an experimental opportunity to explore
nuclear  systems in the
vicinity of the $N$=$Z$ line which have
 {\em many} valence $np$ pairs;
that is,  to probe the interplay between the like-particle and
neutron-proton ($T$=0,1, $T_z$=0) pairing channels.
One evidence related to $T=0$ pairing
involves the Wigner energy, the extra binding that occurs in
$N=Z$ nuclei. We will discuss this in more detail in 
Sec.~\ref{sec:pairing_correlations}.

One possible way to experimentally access pair correlations
in nuclei is by neutron-pair transfer, see e.g., \cite{yoshida1962}. 
Simply stated, if the ground-state of a nucleus is made of  
BCS pairs of neutrons, then two-neutron transfer should be 
enhanced when compared to one-neutron transfer. 
Collective enhancement
of pair transfer is expected if nuclei with open shells are brought
into contact \cite{peter99}. Pairing fluctuations are also expected in rapidly rotating nuclei \cite{shimizu89}.
In lighter systems, such as $^6$He, two-neutron
transfer has been used for studying the wave function of the 
ground state \cite{ogan99}.  

Finally, we discuss phenomenological descriptions 
of nuclear collective motion where 
the nuclear ground state and its low-lying excitations are represented in
 terms of bosons.
In one such model, the Interacting Boson Model
(IBM), $L=0$ (S) and $L=2$ (D) bosons are identified with
nucleon pairs having the same quantum numbers \cite{Arima},
and the ground state can be
viewed as a condensate of such pairs.  Shell-model studies of the pair
structure of the ground state and its variation with the number of valence
nucleons can therefore shed light on the validity and microscopic foundations
of these boson approaches.

\subsubsection{Thermodynamic properties of nuclei and level densities}

The theory of pairing in nuclear physics is also strongly related 
to other fields of physics, such as distinct gaps in 
ultrasmall metallic grains in the 
solid state. These systems share in 
common the fact that the energy spectrum of a system of particles 
confined to a small region is quantized. 
It is only recently, through a series of experiments by Tinkham {\em et al.}
\cite{tinkham95,tinkham96,tinkham98}, that spectroscopic data on discrete energy levels from
ultrasmall metallic grains (with sizes of the order of a few nanometers 
and mean level spacings less than millielectron-volts) 
has been
obtained by way of single-electron-tunneling spectroscopy. 
Measurements in solid state have been much more elusive due to the size
of the system. The discrete spectrum could not be resolved due to the 
energy scale set by temperature.
Of interest here is the observation of so-called parity effects. 
Tinkham {\em et al.}
\cite{tinkham95,tinkham96,tinkham98} were able to observe the number parity (odd or even)
of a given grain by studying the evolution of the discrete spectrum 
in an applied magnetic field.
These effects were also observed in experiments on large Al grains.
It was noted that an even grain had a distinct spectroscopic gap 
whereas an odd grain did not. This is clear evidence of superconducting
pairing correlations in these grains.
The spectroscopic gap was driven to zero by an applied magnetic field;
hence the paramagnetic breakdown of pairing correlations could be studied in 
detail. For theoretical interpretations, 
see, for example, \cite{delft2000,balian1999,mastellone98,sierra99}.

In the smallest grains with sizes less than 3 nanometers, such 
distinct spectroscopic gaps could however not be observed. This vanishing gap
revived an old issue: what is the lower size of a system for the existence
of superconductivity is  such small grains? 

A  nucleus is also a small quantal system, with discrete spectra and strong
pairing correlations. However, whereas the statistical physics of the
above experiments on ultrasmall
grains can be well described through a canonical ensemble, i.e., 
a system in contact with a heat bath, the nucleus in the laboratory 
is an isolated system with no heat exchange with the environment.
The appropriate ensemble for its description is the 
microcanonical one \cite{balian1999}.
This poses significant interpretation problems. For example, 
is it possible to define 
a phase transition in an isolated quantal system such as a nucleus?  

In Sec.~\ref{sec:leveldensities_sec3} we attempt to link our discussion
to such topics via 
recent experimental evidence of pairing from studies
of level densities in rare-earth nuclei. 
The nuclear level density, the density of eigenstates of a nucleus
at a given excitation energy, is the important quantity that may be
used to describe thermodynamic properties of nuclei, such as the
nuclear entropy, specific heat, and temperature.
Bethe first described the level density using a
non-interacting fermi gas model
for the nucleons \cite{bethe1}. Modifications to this picture, such as the
back-shifted fermi gas which includes pair and
shell effects \cite{back_shift,newton56}
not present in Bethe's original formulation, are in
wide use.  These modifications incorporate 
long-range pair correlations that play an important role
in the low excitation region.
Experimentalists recently developed methods
\cite{oslo1,oslo2} to extract level densities at low spin from
measured $\gamma$-spectra. 

There is evidence for the existence of paired nucleons 
(Cooper pairs) at low temperature\footnote{The concept of temperature
in a microcanonical system such as the nucleus is highly non-trivial. 
Temperature itself is defined by a measurement process, involving
thereby the exchange of energy, a fact which is in conflict with the definition
of the  microcanonical ensemble. It is only in the thermodynamic 
limit that e.g., the caloric curves in the canonical and microcanonical
ensembles agree. The word temperature in nuclear physics should therefore
be used with great care.}. In high-spin nuclear physics, 
the backbending phenomenon is a beautiful manifestation of 
the breaking of pairs. The mechanism induced by Coriolis 
forces tends to align single particle angular momenta along 
the nuclear rotational axis 
\cite{stephenssimon,johnson1971,ried80,faessler76}. 
Theoretical models also 
predict a reduction in the pair correlations at 
higher temperatures \cite{mottelson60,muhlhans83,dossing95}.
There is also an interesting connection between quasiparticle spectra
in metallic grains and high-spin spectra in nuclei. In nuclei it is the 
Coriolis force that acts on pairs of nucleons and plays thus a
role similar to the magnetic field acting on Cooper pairs of electrons.

The breaking of pairs is difficult to observe as a 
function of intrinsic excitation energy. Recent 
theoretical \cite{dossing95} and experimental \cite{oslo2,oslo3} works 
indicate that the process of breaking pairs takes place 
over several MeV of excitation energy. Thus, the 
phenomenon of pair breaking in a finite-fermi system behaves 
somewhat differently than what would be expected in nuclear matter. 
The corresponding critical 
temperature in finite systems is measured to be 
$T_c \sim $ 0.5 MeV/$k_B$ \cite{schiller2001}, 
where $k_B$ is Boltzmann's constant.
Recent work extracted the entropy of the 
$^{161,162}$Dy and $^{171,172}$Yb isotopes and deduced the 
number of excited quasiparticles as a function of excitation energy. 
We describe this result in more detail in 
Sec.~\ref{sec:leveldensities_sec3}.

%********************   sect 2

%\input{NN_to_pairing}
% latest upgrade 3/26/2002 (MHJ). 
% Changes: polarization discussion added.
% latest upgrade 4-4-02 djd typos.
% 4/7/02 mhj typos and typos in figs 2 and 4
% 8/8    mhj referee upgrade and typos, final version
% october 2002 mhj included comments from Hubert
\section{PAIRING IN INFINITE MATTER AND THE NUCLEON-NUCLEON INTERACTION}
\label{sec:NN_to_pairing}

Pairing correlations and the phenomenon of superconductivity and superfluidity
are intimately related to the underlying interaction whether it 
is, for example,
the nucleon-nucleon
(NN) interaction or the interaction between $^{3}$He atoms.  
In  this section we discuss, through simple examples,
some of the connections between pairing correlations as they arise 
in nuclear systems and the bare NN
interaction itself, that is, the interaction of  a pair of nucleons
in free space. The latter is most conveniently expressed in terms
of partial waves (and their pertaining quantum numbers such as orbital angular 
momentum and total spin) and phase shifts resulting from nucleon-nucleon
scattering experiments. 
Actually, without specializing to some given fermionic systems 
and interactions, it is possible
to relate the pairing gap and the BCS theory of pairing 
to the experimental phase shifts. This means, in turn, that we can,
through an inspection of experimental scattering data, understand
which partial waves may yield a 
positive pairing gap and eventually lead to, e.g., 
a superfluid phase transition 
in an infinite
fermionic system.
We show this in subsec.~\ref{subsec:NN_to_pairing_sec1} (although we 
limit the attention to nuclear interactions),
after we have singled out those partial waves and 
interaction properties which are expected to be crucial for pairing
correlations in both nuclei and neutron stars. These selected features
of the NN interaction are discussed in the next subsection. A
brief overview of superfluidity in neutron stars and pairing
in symmetric nuclear matter
is presented in subsec.~\ref{subsec:NN_to_pairing_sec3}, with an emphasis
on those partial waves of the NN interaction which are expected  
to produce a finite pairing gap. Features of neutron-proton pairing
in infinite matter are reviewed in 
subsec.~\ref{subsubsec:NN_to_pairing_subsub2}.
Concluding remarks, open problems,
and perspectives are presented in the last subsection.

\subsection{Selected features of the nucleon-nucleon interaction}
\label{subsec:NN_to_pairing_sec2}

The interaction between nucleons is characterized 
by the existence of a strongly repulsive core at short distances,
with a characteristic radius $\sim 0.5-1$ fm. 
The interaction 
obeys several fundamental symmetries such as translational,
rotational, spatial-reflection, time-reversal invariance
and exchange symmetry. It also has 
a strong dependence on quantum numbers such as 
total spin $S$ and isospin $T$, and, 
through the nuclear tensor force
which arises from, e.g., one-pion exchange, 
it also depends on the angles between the nucleon spins and
separation vector. The tensor force thus mixes different angular 
momenta $L$ of the two-body system, that is, it couples two-body states
with 
total angular momentum
$J=L-1$ and $J=L+1$.
For example, for a proton-neutron two-body state, the tensor
force couples the states
$^3S_1$ and $^3D_1$, where we have used the standard spectroscopic
notation $^{2S+1}L_J$. 

Although there is no unique prescription for how to construct
an NN interaction, a description
of the interaction in terms of various meson exchanges is presently
the most quantitative representation, see for example  
\cite{mach89,v18,nijmegen94,cdbonn96,bonn-cd},
in the energy regime of nuclear structure physics.
We will assume that meson-exchange is an appropriate picture at low
and intermediate energies. Further, 
in our discussion of pairing, it suffices at the present stage
to limit our attention to the time-honored 
configuration-space version of the nucleon-nucleon (NN) interaction,
including only central, spin-spin, tensor and spin-orbit terms.
In our notation below, the mass of the nucleon $M_N$
is given by the average of the proton and neutron masses.
The interaction reads (omitting isospin)
\begin{equation}
V({\bf r})= \left\{ C^0_C + C^1_C + C_\sigma 
          \mbox{\boldmath $\sigma$}_1\cdot\mbox{\boldmath $\sigma$}_2
         + C_T \left( 1 + {3\over m_\alpha r} + {3\over
         \left(m_\alpha r\right)^2}\right) S_{12} (\hat r)+ 
         C_{SL} \left( {1\over m_\alpha r} + 
         {1\over \left( m_\alpha r\right)^2}\right) {\bf L}\cdot {\bf S}\right\} \frac{e^{-m_\alpha r}}{m_\alpha r},
       \label{eq:simpleNNv_sec2}
\end{equation}
where $m_{\alpha}$ is the mass of the relevant meson and
$S_{12}$ is the tensor term
\begin{equation}
   S_{12} (\hat r) = \mbox{\boldmath $\sigma$}_1
                      \cdot\mbox{\boldmath $\sigma$}_2
                      \mbox{\boldmath $\hat r$}^2
                      -\mbox{\boldmath $\sigma$}_1\cdot
                       \mbox{\boldmath $\hat r$}
                     \cdot\mbox{\boldmath $\sigma$}_2
                     \cdot \mbox{\boldmath $\hat r$},
\end{equation}
where $\mbox{\boldmath $\sigma$}$ is 
the standard operator for spin $1/2$
particles.
Within
meson-exchange models,  
we may have, e.g., the
exchange of $\pi , \eta , \rho , \omega , \sigma$, and $\delta$ mesons.
As an example, the coefficients for the
exchange of a $\pi$ meson are 
$C_\sigma =C_T = {g_{NN\pi}^2\over 4\pi}\frac{m^3_{\pi}}{12M_N^2}$, 
and
$C^0_C = C^1_C = C_{SL} = 0$
with the experimental value for $g_{NN\pi}^2\approx 13-14$; see, for example,
\cite{bonn-cd} for a recent discussion.

\begin{figure}
\setlength{\unitlength}{1mm}
   \begin{picture}(140,140)
   \put(-20,-50){\epsfxsize=16cm \epsfbox{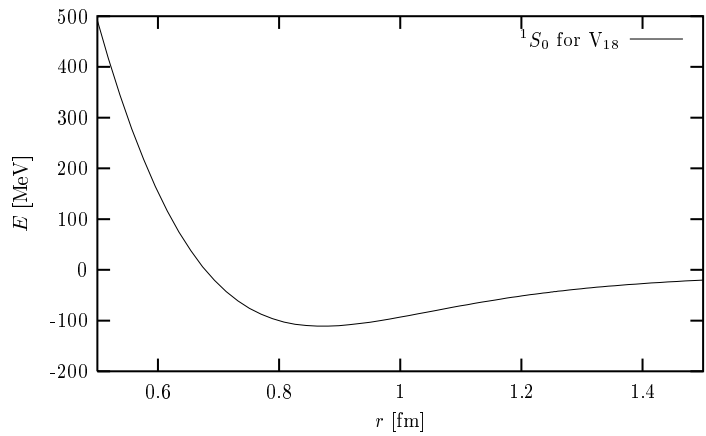}}
   \end{picture}
\caption{Plot of the NN interaction for the 
         $^1S_0$ channel employing the Argonne $V_{18}$ interaction
         \cite{v18}.
             \label{fig:singletspot_sec2}}
\end{figure}

The pairing gap is determined by the attractive part
of the NN interaction. In the $^1S_0$ channel 
the potential is attractive for momenta $k \leq 1.74$ fm$^{-1}$
(or for interparticle distances $r \geq 0.6$ fm), as can be
seen from Fig.~\ref{fig:singletspot_sec2}. 
In the weak coupling regime, where the 
interaction is weak and attractive,  a
gas of fermions may undergo a superconducting (or superfluid) 
instability at low temperatures, and a gas of Cooper pairs is formed.   
This gas of Cooper pairs will be surrounded by unpaired fermions and
the typical coherence length is large compared with the interparticle
spacing, and the bound pairs overlap. 
With weak coupling
we mean a regime where the coherence length is larger than the 
interparticle spacing.
In the strong-coupling limit, the formed bound pairs 
have only a small
overlap, the coherence length is small, and the bound pairs
can be treated as a gas of point bosons. One expects then the system
to undergo a Bose-Einstein condensation into a single quantum
state with total momentum $k=0$ \cite{nsr85}.  
For the $^1S_0$ channel in nuclear physics, we may actually expect to have 
two weak-coupling limits,
namely when the potential is weak and attractive for large interparticle
spacings and when the potential becomes repulsive at $r\approx 0.6$ fm.
In these regimes, the potential has values of typically some few
MeV. One may also loosely speak of a strong-coupling limit 
where the NN potential is large 
and attractive. This takes place where the NN potential
reaches its maximum, with an absolute value of typically $\sim 100$ MeV, 
at roughly $\sim 1$ fm, see again Fig.~\ref{fig:singletspot_sec2}.
We note that
fermion pairs in the $^1S_0$ wave in neutron and nuclear matter will not
undergo the above-mentioned Bose-Einstein condensation, since, even though
the NN potential is large and attractive for certain Fermi momenta, the 
coherence length will always be larger than the interparticle spacing, 
as demonstrated by De Blasio {\em et al.} \cite{deblasio97}.
The inclusion of in-medium effects, such as screening
terms, are expected to further reduce the pairing gap 
and thereby enhance further the coherence
length. This does not imply that such a transition
is not possible in nuclear matter. A recent analysis by Lombardo
{\em et al.} \cite{lombardo2001,schuck2001} of
triplet $^3S_1$ pairing in low-density symmetric and asymmetric nuclear
matter indicates that such a transition
is indeed possible. 

\begin{figure}
\setlength{\unitlength}{1mm}
   \begin{picture}(140,140)
   \put(-20,-50){\epsfxsize=16cm \epsfbox{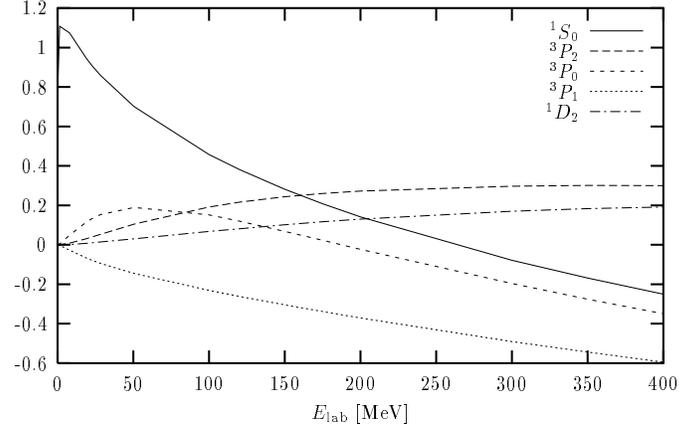}}
   \end{picture}
\caption{Phase shifts for $J\leq 2$  partial waves as function of
         the incoming energy of two $T=1$ nucleons in the lab system.
         \label{fig:t1partialwaves}}
\end{figure}

\begin{figure}
\includegraphics[totalheight=11.cm,angle=0,bb=-50 170 350 720]{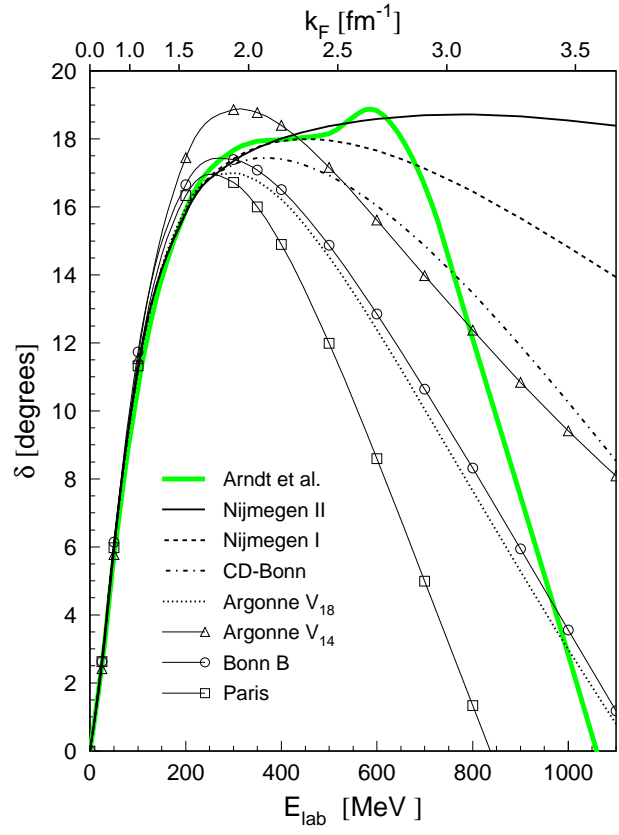}
\caption{$^3P_2$ phase-shift predictions of different potentials
         up to $E_{\rm lab}=1.1\;{\rm GeV}$, compared with the
         phase shift analysis of Arndt {\em et al.}~\protect\cite{arndt97}.}
\label{fig:3p2phaseshift}
\end{figure}

Hitherto we have limited our attention to one single partial wave,
the $^1S_0$ channel. 
Our discussion about the relation among the NN interaction, its
pertinent phase shifts, and
the pairing gap, can be extended to higher partial waves as well. 
An inspection
of the experimental phase shifts for waves with $J \leq 2$ and total 
isospin $T=1$, see
Fig.~\ref{fig:t1partialwaves},
reveals that there are several partial waves which exhibit
attractive (positive phase shifts) contributions
to the NN interaction. Such attractive terms are in turn expected to
yield a possible positive pairing gap. This means that
the energy dependence of the nucleon-nucleon ($NN$) phase shifts in different 
partial waves offers some guidance in judging what nucleonic pair-condensed 
states are possible or likely in different regions of a neutron star.
A rough correspondence between baryon density and $NN$ bombardment energies
can be established through the Fermi momenta assigned to the nucleonic
components of neutron-star matter.
The lab energy relates to the Fermi energy through 
$E_{\mathrm{lab}}=4\epsilon_F=4\hbar^2k_F^2/2M_N$. 
This is demonstrated in Fig.~\ref{fig:3p2phaseshift} for various NN interaction
models that fit scattering data up to $E_{\mathrm{lab}}\approx 350$ MeV.
For comparison, we include results for older 
potential models such as the Paris 
\cite{paris}, $V_{14}$ \cite{v14} and Bonn B \cite{mach89} 
interactions. Note, as well, that 
beyond the point where these potential models 
have been fit, there is a considerable variation. This has important 
consequences for reliable predictions of the $^3P_2$ pairing gap.

In pure neutron matter, only $T=1$ partial waves are allowed.  
Moreover, one need only consider partial 
waves with $L\leq 4$ in the range of baryon density -- optimistically,
$\rho < (3-4) \rho_0$ -- where a nucleonic model of neutron-star material 
is tenable, where $\rho_0=0.16$ fm$^{-3}$ is
the saturation density of nuclear matter.  
We have already seen that the $^1S_0$ phase shift is positive at low energy 
(indicating an attractive in-medium force) but turns negative (repulsive) 
at around 250 MeV lab energy.  Thus, unless the in-medium pairing force 
is dramatically different from its vacuum counterpart, the situation 
already suggested above should prevail:  S-wave pairs should form at 
low densities but should be inhibited from forming when the density 
approaches that of ordinary nuclear matter.  

\begin{figure}
\setlength{\unitlength}{1mm}
   \begin{picture}(140,140)
   \put(-20,-50){\epsfxsize=16cm \epsfbox{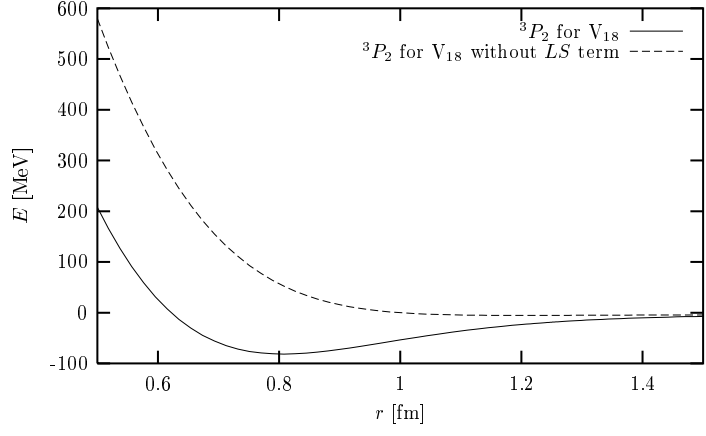}}
   \end{picture}
\caption{Plot of the Argonne $V_{18}$ \cite{v18} 
$^3P_2$ partial wave contribution  
with and without the
spin-orbit contribution.
         \label{fig:tripletwaves}}
\end{figure}

The next lowest $T=1$ partial waves are the three triplet 
P waves $^3P_J$, with $J=0,1,2$.
For the the $^3P_0$ state, the
phase shift is positive  at low energy, turning
negative  at a lab energy of 200 MeV. The attraction is, however,
not sufficient to produce a finite pairing gap
in neutron star matter. The $^3P_1$ phase
shift is negative  at all energies, indicating a repulsive 
interaction.  The
$^3P_2$ phase shift is positive for energies up to 1 GeV and 
is the most attractive $T=1$ phase shift at energies above
about 160 MeV.  
Whereas the $^1S_0$ partial wave is dominated by the central force 
contribution of the NN interaction, see Eq.~(\ref{eq:simpleNNv_sec2}),
the main contribution to the attraction seen in the  
$^3P_2$ partial wave stems from the two-body spin-orbit force
for intermediate ranges
in Eq.~(\ref{eq:simpleNNv_sec2}), i.e., the term proportional with
${\bf L}\cdot {\bf S}$. This is demonstrated in 
Fig.~\ref{fig:tripletwaves} where we plot the coordinate space
version of the Argonne $V_{18}$ interaction \cite{v18} 
with and without the
spin-orbit contribution. Moreover, there is an additional enhancement
due to the $^3P_2$--$^3F_2$ tensor force.  
A substantial pairing effect in the
$^3P_2$--$^3F_2$ channel may hence be expected at densities somewhat 
in excess of $\rho_0$, again assuming that the relevant in-vacuum 
interaction is not greatly altered within the medium.

The remaining $T=1$ partial waves with $L\leq 4$ are both
singlets: $^1D_2$ and $^1G_4$.  However, the phase shifts of
these partial waves, albeit being positive over the energy domain of interest,
do not provide any substantial contribution to the pairing gap.
Thus, 
only the $^1S_0$ and $^3P_2$ partial waves yield enough attraction
to produce a finite pairing gap in pure neutron matter.
Singlet and triplet pairing are hence synonomous with 
$^1S_0$ and $^3P_2$--$^3F_2$ pairing, respectively.

\subsection{Pairing gap equations}

The gap equation for pairing in non-isotropic partial waves is, in general,
more complex than in the simplest singlet $S$-wave case, 
in particular in neutron and 
nuclear matter, where the tensor interaction can couple two 
different partial waves \cite{tam70,taka93,bls95}. 
This is indeed the situation for the $^3P_2$-$^3F_2$ neutron channel
or the $^3S_1$-$^3D_1$ channel for symmetric nuclear matter. 
For the sake of simplicity, we disregard for the moment
spin degrees of freedom and the tensor interaction.
Starting  with the Gorkov equations \cite{gorkov}, 
which involve the propagator
$G(\bm{k},\omega)$, the anomalous propagator $F(\bm{k},\omega)$, and the
gap function $\Delta(\bm{k})$, we have 
\begin{eqnarray}
 \left( \begin{array}{rr} \omega-\epsilon(\bm{k}) & -\Delta(\bm{k}) \\ 
  -\Delta^\dagger(\bm{k}) & \omega+\epsilon(\bm{k})
 \end{array} \right)
 \left(\begin{array}{c} G \\ F^\dagger \end{array}\right)(\bm{k},\omega) =
 \left( \begin{array}{c} 1 \\ 0 \end{array} \right) \:,
  \label{eq:gorkoveq}
\end{eqnarray}
where $\epsilon(\bm{k}) = e(\bm{k}) - \mu$, $\mu$ being the 
chemical potential and $e(\bm{k})$ the single-particle spectrum. 
The quasi-particle energy $E(\bm{k})$ is the solution of the corresponding
secular equation and is given by
\begin{equation}
  E(\bm{k})^2 = \epsilon(\bm{k})^2 + |\Delta(\bm{k})|^2 \:.
 \label{eq:qua1}
\end{equation}
The anisotropic gap function $\Delta(\bm{k})$ 
is to be determined from the gap
equation
\begin{equation}
  \Delta(\bm{k}) = - \sum_{\bm{k}'} \langle \bm{k} | V | \bm{k}' 
\rangle
  {\Delta(\bm{k}')\over 2E(\bm{k}') } \:.
\end{equation}
The angle-dependent energy denominator in this equation prevents 
a straightforward separation into the different partial wave components
by expanding the potential,
\begin{equation}
  \langle \bm{k} | V | \bm{k}' \rangle = 
  4\pi \sum_L (2L+1) P_L(\bm{\hat{k}\cdot\hat{k}'}) V_L(k,k') \:,
\end{equation}
and the gap function, 
\begin{equation}
  \Delta(\bm{k}) = 
  \sum_{L,M} \sqrt{4\pi\over 2l+1} Y_{LM}(\bm{\hat{k}}) \Delta_{LM}(k),
\end{equation}
with $L$ and $M$ being the total orbital momentum and its projection, respectively.
The functions $ Y_{L,M}$ are the spherical harmonics. 
However, after performing an angle average approximation for the gap in the
quasi-particle energy,
\begin{equation}
  |\Delta(\bm{k})|^2 \rightarrow D(k)^2 \equiv 
  {1\over 4\pi} \int d\bm{\hat{k}}\, |\Delta(\bm{k})|^2 =
  \sum_{L,M} {1\over 2L+1} |\Delta_{LM}(k)|^2 \:,
\end{equation}
the kernels of the coupled integral equations become isotropic, and one can 
see that the different $m$-components become uncoupled and all equal. 
One obtains the following equations for the 
partial wave components of the gap function:
\begin{equation}
  \Delta_L(k) = - {1\over\pi} \int_0^\infty 
  k' dk'{V_L(k,k') \over \sqrt{ \epsilon(k')^2 + 
  \left[\sum_{L'} \Delta_{L'}(k')^2 \right] }}
  \Delta_L(k') \:.
\end{equation}
Note that there is no dependence on the quantum number $M$ in these 
equations; however, they still couple the components of the
gap function with different orbital momenta $L$ 
($^1S_0$, $^3P_0$, $^3P_1$, $^3P_2$, $^1D_2$, $^3F_2$, etc.~in neutron matter) 
via the energy denominator.
Fortunately, in practice the different components $V_L$ of the potential
act mainly in non-overlapping intervals in density, 
and therefore also this coupling can usually be disregarded.

The addition of spin degrees of freedom and of the tensor force does not
change the picture qualitatively and is explained in detail 
in \cite{taka93,bls95}.
The only modification is the introduction 
of an additional $2\times2$ matrix structure due to the tensor coupling
of the $^3P_2$ and $^3F_2$ channels. Such coupled channel equations can
be written as 
\begin{eqnarray}
 \left( \begin{array}{c} \Delta_L \\ \Delta_{L'} \end{array} \right)(k) &=&
 - {1\over\pi} \int_0^\infty dk' k'^2 {1\over E(k')}
 \left( \begin{array}{rr}
  V_{LL} & -V_{LL'} \\ -V_{L'L} & V_{L'L'}
 \end{array} \right)(k,k')
 \left(\begin{array}{c} \Delta_L \\ \Delta_{L'} \end{array}\right)(k') \:,
\\
  E(k)^2 &=& [\epsilon(k)-\epsilon(k_F)]^2 + D(k)^2 \:,
\\
  D(k)^2 &=& \Delta_{L}(k)^2 + \Delta_{L'}(k)^2 \:.
\label{eq:gap2c}
\end{eqnarray}
Here $\epsilon(k)=k^2/2m + U(k)$ are the single-particle energies of a neutron with 
momentum $k$, and $k_F$ is the Fermi momentum. The orbital
momenta  $L$ and $L'$ could, e.g., represent the $^3P_2$ and $^3F_2$ channel,
respectively. Restricting the attention to only one partial wave,
it is easy to get the equation for an  uncoupled channel like the 
$^1S_0$ wave, i.e., we obtain
\begin{equation}
      \Delta(k)_L=-\frac{1}{\pi}\int_{0}^{\infty}dk'k'^2 
                 V_{LL}(k,k'){{\Delta(k')}\over{E(k')}},
       \label{eq:pairinggap2_sec2}
\end{equation}
where $V_{LL}(k,k')$ is now the bare momentum-space NN interaction in the 
$^1S_0$ channel, and $E(k)$ is the quasiparticle energy given by 
$E(k)=\sqrt{(\epsilon(k)-\epsilon(k_F))^2+\Delta(k)_L^2}$. 

The quantities 
\begin{equation}
  V_{LL'}(k,k') =  \int_0^\infty dr r^2 j_{L'}(k'r) V_{LL'}(r) j_L(kr)
\label{e:v}
\end{equation}
are the matrix elements of the bare interaction in 
the different coupled channels, e.g.,  
$(T=1;\;S=1;\,J=2;\,L,L'=1,3)$. 
It has been shown that the angle average approximation is an 
excellent approximation to the true solution that involves a gap function 
with ten components \cite{taka93,kkc96}, as long as one is 
only interested in the average value of the gap at the Fermi surface, 
$\Delta_F\equiv D(k_F)$, and not the angular dependence of the gap functions 
$\Delta_L(\bm{k})$ and $\Delta_{L'}(\bm{k})$.  

Recently  Khodel, Khodel, and Clark \cite{kkc2001,kkc1998} proposed a
separation method for the triplet pairing gap, based on \cite{kkc96},
which allows a generalized solution of 
the BCS equation that is numerically reliable, 
without employing an angle-average approach.
We refer the reader to \cite{kkc96,kkc2001,kkc1998} for more details.
In this approach, the pairing matrix elements are written as a separable
part plus a remainder that vanishes when either momentum variable is on the
Fermi surface.  This decomposition effects a separation of the problem
of determining the dependence of the gap components in a spin-angle
representation on the magnitude of the momentum (described by a set of
functions independent of magnetic quantum number) from the problem of
determining the dependence of the gap on angle or magnetic projection.
The former problem is solved through a set of nonsingular, quasilinear
equations \cite{kkc2001,kkc1998}. 
There is, in general, a good agreement between their approach and the angle 
average scheme. However, the general scheme of Khodel, Khodel, and Clark
offers a much more stable algorithm 
for solving the pairing gap equations for 
any channel and starting with the bare interaction itself. In nuclear physics
the interaction typically has a strongly repulsive core, a fact  
that can complicate
significantly the iterative solution of the BCS equations.

An important ingredient in the calculation of the pairing gap is
the single-particle potential $U(k)$. The gap equation 
is extremely sensitive to both many-body renormalizations of the 
interaction and the similar corrections to the single-particle 
energies. Many-body renormalizations of the interaction will be 
discussed in Sec.~\ref{subsec:polarizationterms}. 
In our discussion below, we will present results for various
many-body approaches to $U(k)$, from $U(k)=0$ to results with  
different Brueckner-Hartree-Fock (BHF) calculations, with both a
discontinuous choice, a model-space BHF approach and  
within the ``continuous-choice'' scheme \cite{jjm76}.

The single-particle energies appearing in the quasiparticle energies  
(\ref{eq:qua1}) and (\ref{eq:gap2c}) are typically 
obtained through a self-consistent 
BHF calculation, 
using a $G$-matrix defined through the Bethe-Brueckner-Goldstone
equation as  
\begin{equation}
   G=V+V\frac{Q}{\omega -H_0}G,
\end{equation}
where $V$ is the nucleon-nucleon potential, $Q$ is the Pauli operator
which prevents scattering into intermediate 
states prohibited by the Pauli
principle, $H_0$ is the unperturbed 
Hamiltonian acting on the intermediate
states, and $\omega$ is the starting energy, 
the unperturbed energy
of the interacting particles. Methods to solve this equation are reviewed in
\cite{mhj_95}.
The single-particle energy for state $k_i$ ($i$ 
encompasses all relevant
quantum numbers like momentum, 
isospin projection, spin, etc.)
in nuclear matter is assumed to have
the simple quadratic form
\begin{equation}
   \epsilon_{k_i}=
   {\displaystyle\frac{k_{i}^2\hbar^2}
   {2M^{*}_N}}+\delta_i ,
   \label{eq:spen}
\end{equation}
where $M_N^{*}$ is the effective mass.
The terms $M_N^{*}$ and $\delta$, the latter being 
an effective single-particle
potential related to the $G$-matrix, are obtained through the
self-consistent BHF  procedure. The 
model-space BHF (MBHF) method 
for the single-particle spectrum has also been used, see,
for example, \cite{mhj_95}, with a cutoff momentum  
$k_M=3.0$ fm$^{-1}>k_{F}$.
In this approach the single-particle spectrum is defined by 
\begin{equation}
\epsilon_{k_{i}}=\frac{k_{i}^{2}\hbar^2}{2M_N}+u_{i}, 
\label{eq:modsp}
\end{equation}
with the single-particle potential $u_{i}$ given by 
\begin{equation}
    u_{i}=
    \left\{\begin{array}{cc}\sum_{k_{h}\leq k_{F}}\left\langle 
 k_{i}k_{h}|G(\omega=\epsilon_{k_{i}}+\epsilon_{k_{h}})|k_{i}k_{h}
\right\rangle_{AS},&k_{i}\leq k_{M},\\
    0,k_{i}>k_{M},\end{array}\right.,
\label{eq:modpot}
\end{equation}
where the subscript $AS$ denotes antisymmetrized matrix elements.  
This prescription reduces the discontinuity in the single-particle spectrum 
as compared with the standard BHF choice $k_{M}=k_{F}$.  
The self-consistency scheme 
consists of choosing adequate initial values of the
effective mass and $\delta$. The obtained $G$-matrix is then used to 
calculate the single-particle potential $u_{i}$, from which we obtain 
new values for $m^{*}$ and $\delta$.  
This procedure continues until these parameters vary little. 

Recently,  Lombardo {\em et al.}  \cite{lsz2001,ls2000} have reanalyzed the importance 
of the various approaches to the single-particle energies. Especially, 
they demonstrate that 
the energy dependence
of the self-energy can deeply affect the magnitude of the energy gap in a 
strongly correlated Fermi system; see also the recent works of Bozek in
\cite{bozek1999,bozek2000,bozek2002}. 
We will discuss these effects in Subsec.~\ref{subsec:polarizationterms}.

\subsection{Simple relations between the interaction and the pairing gap for identical particles}
\label{subsec:NN_to_pairing_sec1}

\subsubsection{The low density limit}

A general two-body Hamiltonian 
can be written in the form $\hat{H}=\hat{H}_1+\hat{H}_2$ where
\begin{eqnarray}
\hat{H}_1&=&\sum_\alpha \varepsilon_\alpha a^\dagger_\alpha a_{\alpha} \;, \\
\hat{H}_2&=&\sum_{\alpha\beta\gamma\delta}V_{\alpha\beta\gamma\delta}
a^\dagger_\alpha a^\dagger_\beta a_\delta a_\gamma \;,
\label{eq:hamiltonian_sec1}
\end{eqnarray}
where $a^\dagger$ and $a$ are fermion creation and annihilation operators,
and $V$ are the uncoupled matrix elements of the two-body interaction. 
The sums run over all possible single-particle quantum numbers.

We limit the discussion in this section to 
a Fermi gas model with two-fold degeneracy and 
a pairing-type interaction as an example; i.e., 
the degeneracy of the 
single-particle levels is set to $2s+1=2$, with $s=1/2$ being the 
spin of the particle. 
We specialize to 
a singlet two-body interaction with quantum numbers $l=0$ and
$S=0$, that is a $^1S_0$ state, with $l$ the relative orbital
momentum and $S$ the total spin. 
For this partial wave, the NN interaction is dominated by the
central component in Eq.~(\ref{eq:simpleNNv_sec2}), 
which, within a meson-exchange picture, can be portrayed
through $2\pi$ (leading to an effective $\sigma$ meson) 
and higher $\pi$ correlations in order to yield enough
attraction at intermediate distances.

At low densities, the interaction can be characterized by its 
scattering length only in order to get expansions for the energy
density or the excitation spectrum. For the nucleon-nucleon
interaction, the 
scattering length is 
$a_0=-18.8\pm 0.3$ fm for  
neutron-neutron scattering in the $^1S_0$ channel.
If we first assume discrete single-particle
energies, 
the scattering length approximation leads to the following
approximation of the two-body
Hamiltonian of Eq.~(\ref{eq:hamiltonian_sec1}) 
\begin{equation}
   H=\sum_i \varepsilon_i a^{\dagger}_i a_i +\frac{1}{2} G\sum_{ij>0}
           a^{\dagger}_{i}
     a^{\dagger}_{\bar{\imath}}a_{\bar{\jmath}}a_{j}.
     \label{eq:pairHamiltonian1_sec2}
\end{equation}
The indices $i$ and $j$ run 
over the number of levels $L$, and the label $\bar{\imath}$ stands 
for a time-reversed state. The parameter $G$ is now the 
strength of the pairing force, while $\varepsilon_i$ is the single-particle 
energy of level $i$. 
Introducing the pair-creation operator 
$S^+_i=a^{\dagger}_{im}a^{\dagger}_{i-m}$,
one can rewrite the Hamiltonian in 
Eq.\ (\ref{eq:pairHamiltonian1_sec2}) as
\begin{equation}
   H=d\sum_iiN_i+
     \frac{1}{2} G\sum_{ij>0}S^+_iS^-_j,
     \label{eq:pairH2_sec}
\end{equation}
where  $N_i=a^{\dagger}_i a_i$
is the number operator, and 
$\varepsilon_i = id$ so that the single-particle orbitals 
are equally spaced at intervals $d$. The latter commutes with the 
Hamiltonian $H$. In this model, quantum numbers like seniority 
$\cal{S}$ are good quantum numbers, and the eigenvalue problem 
can be rewritten in terms of blocks with good seniority. Loosely 
speaking, the seniority quantum number $\cal{S}$ is equal to 
the number of unpaired particles; see  \cite{talmi93} for 
further details. As it stands 
Eq.~(\ref{eq:pairHamiltonian1_sec2}), lends itself for shell-model
studies. 
Furthermore, in  a series of papers, Richardson 
\cite{richardson1,richardson2,richardson3,richardson4,richardson5,richardson6,richardson7} obtained the exact solution of the pairing Hamiltonian, with 
semi-analytic (since there is still the need for a numerical solution) 
expressions for the eigenvalues and eigenvectors. The exact solutions
have had important consequences for several fields, from Bose condensates to
nuclear superconductivity.
 
We will come back to this model 
in our discussion of level densities
and thermodynamical features of the pairing Hamiltonian in
finite systems in Sec.~\ref{sec:leveldensities_sec3}. 

Here we are interested in features of infinite matter with
identical particles, and using
$\sum_k\rightarrow V/(2\pi)^3\int_0^{\infty}d^3k$, we rewrite 
Eq.~(\ref{eq:pairHamiltonian1_sec2}) as
\begin{equation}
   H=V\sum_{\sigma=\pm}\int\frac{d^3k}{(2\pi)^3} 
     \epsilon_{k\sigma}a_{k\sigma}^{\dagger}a_{k\sigma}
     +GV^2\int\frac{d^3k}{(2\pi)^3}\int\frac{d^3k'}{(2\pi)^3}
     a_{k+}^{\dagger}a_{-k-}^{\dagger}a_{-k'-}a_{k'+}.
\end{equation}
The first term represents the kinetic energy, with 
$\epsilon_{k\sigma}=k^2/2m$. The label 
$\sigma=\pm 1/2$ stands for the spin, while $V$ is the volume. 
The second term is the 
expectation value of the two-body
interaction with a constant interaction strength $G$.
The energy gap in infinite matter is obtained by solving the BCS equation 
for the gap function $\Delta(k)$. For our simple model  
we see that Eq.~(\ref{eq:pairinggap2_sec2}) reduces to
\begin{equation}
      1=-\frac{GV}{2(2\pi)^3}\int_{0}^{\infty}dk'k'^3 
                 \frac{1}{E(k')},
       \label{eq:pairinggap1_sec2}
\end{equation}
with $E(k)$ the quasiparticle energy given by 
$E(k)=\sqrt{(\epsilon(k)-\epsilon(k_F))^2+\Delta(k)^2}$, where 
$\epsilon(k)$ is the single-particle energy of a neutron with 
momentum $k$, and $k_F$ is the Fermi momentum.  
Medium effects should 
be included in $\epsilon(k)$, but we will use free single-particle 
energies $\epsilon(k)=k^{2}/2M_N$.

Papenbrock and Bertsch \cite{pb98} obtained an analytic
expression for the pairing gap in the low-density limit
by combining Eq.~(\ref{eq:pairinggap1_sec2}) with the 
equation for the scattering length $a_0$ and its relation
to the interaction
\begin{equation}
    -\frac{M_NGV}{4\pi a_0}+1 = -\frac{GV}{2(2\pi)^3} 
     \int d^3k\frac{1}{\sqrt{(\epsilon(k)-\epsilon(k_F))^2}},
     \label{eq:a0_sec2}
\end{equation}
which is divergent. However, the authors of \cite{pb98}
showed that by subtracting Eq.~(\ref{eq:pairinggap1_sec2})
and Eq.~(\ref{eq:a0_sec2}), one obtains 
\begin{equation}
\frac{M_NG}{4\pi a_0} = - \frac{G}{2 (2\pi)^3} 
\int d^3k\left[\frac{1}{E(k)}-\frac{1}
{\sqrt{(\epsilon(k)-\epsilon(k_F))^2}}\right],
\end{equation}
which is no longer divergent. Moreover, we can divide out the 
interaction strength and obtain 
\begin{equation}
\frac{M_N}{4\pi a_0} = - \frac{1}{2 (2\pi)^3} 
\int d^3k\left[\frac{1}{E(k)}-\frac{1}
{\sqrt{(\epsilon(k)-\epsilon(k_F))^2}}\right].
\end{equation}
Using dimensional regularization techniques, Papenbrock and Bertsch
\cite{pb98} obtained the analytic expression
\begin{equation}
      \frac{1}{k_F a_0} = (1+x^2)^{1/4} \,P_{1/2}\left(-1/\sqrt{1+x^2}\right),
\end{equation}
where $x=\Delta(k_F)/\epsilon(k_F)$ and $P_{1/2}$ 
denotes a Legendre function. 
With a given fermi momentum, we can thus obtain the 
pairing gap. For small values of $k_F a_0$,
one obtains the well-known result \cite{gorkovmelik,kkc96}
\begin{equation}
    \label{eq:lowdensgap_sec2}
    \Delta(k_F) = {8\over e^2}\lambda\exp{\left(-\pi\over 2 k_F |a_0|\right)}.
\end{equation}
This comes about by the behavior of $P_{1/2}(z)$,
which has a logarithmic singularity at $z=-1$ (see \cite{pb98x}).
For large values of $k_Fa_0$, the gap is proportional 
to $\epsilon(k_F)$, approaching
$\Delta\approx 1.16\epsilon(k_F)$.
The large value of the scattering length ($a_0=-18.8 \pm 0.3$fm)
clearly limits the domain of validity of the Hamiltonian 
in Eq.~(\ref{eq:pairHamiltonian1_sec2}).
However, Eq.~(\ref{eq:lowdensgap_sec2}) provides us with a useful
low-density result to compare with results arising from numerical
solutions of the pairing gap equation. 
The usefulness of Eq.~(\ref{eq:lowdensgap_sec2}) cannot be underestimated:
one experimental parameter, the scattering length, allows us to make
quantitative statements about pairing at low densities. Polarization 
effects arising from renormalizations of the in-medium effective interaction
can however change this behavior, as demonstrated recently 
in \cite{henning2000,spr2001} (see the discussion in 
Subsec.~\ref{subsec:polarizationterms}).

\subsubsection{Relation to phase shifts}

With the results from Eq.~(\ref{eq:lowdensgap_sec2}) in mind, we ask the
question whether we can obtain information about the 
pairing gap at higher densities, without resorting to a detailed model
for the NN interaction. 

Here we show that this is indeed the case.
Through the experimental phase shifts, we show that one can determine
fairly accurately the $^1S_0$ pairing gap in pure neutron matter
without needing an explicit model for the NN interaction. 
It ought to be mentioned that this was demonstrated long ago 
by e.g., Emery and Sessler, see \cite{emerysessler}. 
Their approach is however slightly different from ours.

As we saw in the previous subsection, 
a characteristic feature of $^1S_0$ NN scattering is the large, negative 
scattering length, indicating the presence
of a nearly bound state at zero scattering energy.  Near a bound state, 
where the NN $T$-matrix has a pole, it can be written in separable form, 
and this implies that the NN interaction itself to a good approximation is 
rank-one separable near this pole \cite{kkc96,kohler96}.   
Thus, at low energies, we approximate
\begin{equation}
       V(k,k')=\lambda v(k)v(k'),
       \label{eq:separableV1_sec2} 
\end{equation}
where $\lambda$ is a constant.  Then it is easily seen from 
Eq.~(\ref{eq:pairinggap2_sec2}) 
that the gap function can be rewritten as 
\begin{equation}
      1=-\frac{1}{\pi}
      \int_{0}^{\infty}dk'k'^2\frac{\lambda v^2(k')}{E(k')}.
      \label{eq:pairinggap3_sec2}
\end{equation}
Numerically, the integral on the right-hand side of this equation depends 
very weakly on the momentum structure of $\Delta(k)$, so in our 
calculations we could take $\Delta(k)\approx \Delta_F$ in $E(k)$.  
Then Eq.~(\ref{eq:pairinggap3_sec2})  
shows that the energy gap $\Delta_F$ is 
determined by the diagonal elements $\lambda v^2(k)$ of the NN interaction.  
The crucial point is that in scattering theory it can be shown that 
the inverse scattering problem, that is, the determination of a 
two-particle potential from the knowledge of the phase shifts at all 
energies, is exactly, and uniquely, solvable for rank-one 
separable potentials \cite{cs92}.  Following the notation 
of \cite{bj76}, we have 
\begin{equation}
       \lambda v^2(k)=-{{k^2+\kappa_B^2}\over{k^2}}
                       {{\sin \delta(k)}\over{k}}e^{-\alpha(k)},
       \label{eq:separableV2_sec2} 
\end{equation}
for an attractive potential with a bound state at energy $E=-\kappa_B^2$. 
In our case $\kappa_B\approx 0$.    
Here $\delta(k)$ is the $^1S_0$ phase shift as a function of momentum $k$, 
while $\alpha(k)$ is given by a principal value integral: 
\begin{equation}
       \alpha(k)={{1}\over{\pi}}{\rm P}\int_{-\infty}^{+\infty}dk'
                 {{\delta(k')}\over{k'-k}},   
       \label{eq:phaseshift1_sec2} 
\end{equation}
where the phase shifts are extended to negative momenta through 
$\delta(-k)=-\delta(k)$ \cite{kohler96}.

 From this discussion we see that $\lambda v^2(k)$, and therefore also 
the energy gap $\Delta_F$, is completely determined by the $^1S_0$ 
phase shifts.  However, there are two obvious limitations on the 
practical validity of this statement.  First of all, the separable 
approximation can only be expected to be good at low energies, near the 
pole in the $T$-matrix.  Secondly, we see from 
Eq.~(\ref{eq:phaseshift1_sec2}) that 
knowledge of the phase shifts $\delta(k)$ at all energies is required.  
This is, of course, impossible, and most phase shift 
analyses stop at a laboratory energy $E_{\rm lab}=350$ MeV.  The 
$^1S_0$ phase shift changes sign from positive to negative at 
$E_{\rm lab}\approx 248.5$ MeV;
however, at low values of $k_F$, knowledge 
of $v(k)$ up to this value of $k$ may actually be enough to determine 
the value of $\Delta_F$, as the integrand in 
Eq.~(\ref{eq:pairinggap3_sec2}) is 
strongly peaked around $k_F$.  

The input in our calculation is the $^1S_0$ phase shifts taken from  
the recent Nijmegen nucleon-nucleon phase shift analysis \cite{nijpwa}. 
We then evaluated $\lambda v^2(k)$ from Eqs.~(\ref{eq:separableV2_sec2})  
and (\ref{eq:phaseshift1_sec2}), using methods described in 
\cite{bj76} to 
evaluate the principal value integral in Eq.~(\ref{eq:phaseshift1_sec2}). 
Finally, we evaluated the energy gap $\Delta_F$ for various values 
of $k_F$ by solving Eq.~(\ref{eq:pairinggap3_sec2}), which is an algebraic 
equation due to the approximation $\Delta(k)\approx \Delta_F$ in the 
energy denominator.

\begin{figure}
\setlength{\unitlength}{1mm}
   \begin{picture}(140,140)
   \put(-20,-50){\epsfxsize=16cm \epsfbox{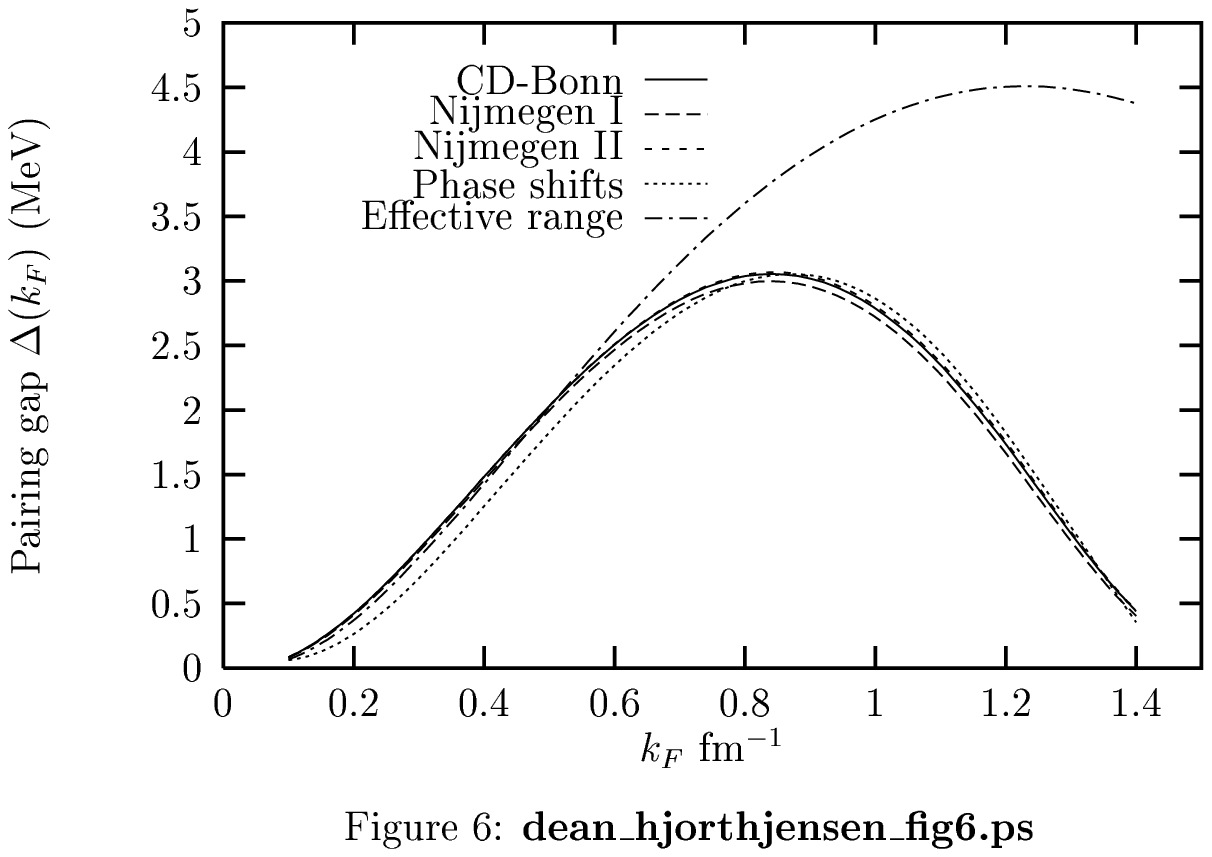}}
   \end{picture}
\caption{$^1S_0$ energy gap in neutron matter with the CD-Bonn, 
             Nijmegen I and Nijmegen II potentials. In addition, we 
             show the results obtained from phase shifts only, 
             Eqs.~(\ref{eq:pairinggap3_sec2})-(\ref{eq:phaseshift1_sec2}), 
             and the effective range approximation of 
             Eq.~(\ref{eq:effrange2_sec2}). Taken from \cite{eh98}.
             \label{fig:energygap_sec2}}
\end{figure}

The resulting energy gap obtained from the experimental phase
shifts only is plotted in Fig.~\ref{fig:energygap_sec2}.
In the same figure we also report the results (dot-dashed line) 
obtained using the effective range approximation to the phase shifts: 
\begin{equation}
       k\cot \delta(k)=-{{1}\over{a_0}}+{{1}\over{2}}r_0 k^2, 
\end{equation}
where $a_0=-18.8\pm 0.3$ fm and $r_0=2.75\pm 0.11$ fm are the singlet 
neutron-neutron scattering length and effective range, respectively.  
In this case an analytic expression can be obtained for $\lambda v^2(k)$, as 
shown in \cite{cs92}:
\begin{equation}
  \lambda v^2(k)=-{{1}\over{\sqrt{k^2+{{r_0^2}\over{4}}(k^2+\alpha^2)^2}}}
                \sqrt{{k^2+\beta_2^2}\over{k^2-\beta_1^2}},
   \label{eq:effrange2_sec2}
\end{equation}
with $\alpha^2=-2/ar_0$, $\beta_1\approx-0.0498\;{\rm fm}^{-1}$, 
and $\beta_2\approx 0.777\;{\rm fm}^{-1}$.  
The phase shifts using this approximation are positive at all energies, 
and this is reflected in Eq.~(\ref{eq:effrange2_sec2})  
where $\lambda v^2(k)$ 
is attractive for all $k$.  From Fig.~\ref{fig:energygap_sec2} we see that 
below $k_F=0.5\;{\rm fm}^{-1}$ the energy gap can, with reasonable 
accuracy, be calculated with the interaction obtained directly from 
the effective range approximation.  
One can therefore say that 
at densities below $k_F=0.5\;{\rm fm}^{-1}$, and at the crudest level 
of sophistication in many-body theory,  the superfluid properties 
of neutron matter are determined by just two parameters, namely 
the free-space scattering length and effective range. At such densities,
more complicated many-body terms are also less important.
Also interesting is the fact that the phase shifts predict the position 
of the first zero of $\Delta(k)$ in momentum space, since we see from 
Eq.~(\ref{eq:effrange2_sec2}) 
that $\Delta(k)=\Delta_F v(k)=0$ first for $\delta(k)=0$, 
which occurs at $E_{\rm lab}\approx 248.5$ MeV (pp scattering) 
corresponding to $k\approx 1.74\;{\rm fm}^{-1}$.  
This is in good agreement with the results of 
Khodel {\em et al.}~\cite{kkc96}.  In \cite{kkc96}, it is 
also shown that this first zero of the gap function determines the 
Fermi momentum at which $\Delta_F=0$.  Our results therefore indicate 
that this Fermi momentum is in fact given by the energy at which 
the $^1S_0$ phase shifts become negative. 

In Fig.~\ref{fig:energygap_sec2} we show also results obtained
with recent NN interaction models parametrized to reproduce the
Nijmegen phase shift data. We have here employed  
the CD-Bonn potential \cite{cdbonn96},  
the Nijmegen I and Nimegen II potentials \cite{nijmegen94}. 
The results are virtually identical, with the maximum value 
of the gap varying from 2.98 MeV for the Nijmegen I potential to 3.05 MeV 
for the Nijmegen II potential.  
As the reader can see, the agreement 
between the direct calculation from the phase shifts and the CD-Bonn and
Nijmegen 
calculation of $\Delta_F$ is satisfying, even 
at densities as high as $k_F=1.4\;{\rm fm}^{-1}$.  
The energy gap 
is to a remarkable extent determined by the available $^1S_0$ phase shifts. 
Thus, the quantitative features 
of $^1S_0$ pairing in neutron matter can be obtained directly from 
the $^1S_0$ phase shifts. This happens because the NN interaction 
is very nearly rank-one separable in this channel due to the presence 
of a bound state at zero energy, even for densities as high as 
as $k_F=1.4\;{\rm fm}^{-1}$ \footnote{This is essentially due to the 
fact that the integrand in the gap equation is strongly peaked around  
the diagonal matrix elements.}. This explains why all bare NN interactions 
give nearly identical results for the $^1S_0$ energy gap in lowest-order 
BCS calculations.  {\em Combined with Eq.~(\ref{eq:lowdensgap_sec2}), we have a
first approximation to the pairing gap 
with experimental inputs only, phase shifts, and scattering length.}

However, 
it should be mentioned that this agreement 
is not likely to survive in a more refined calculation, for instance, 
if one includes the density and spin-density fluctuations in the 
effective pairing interaction or renormalized single-particle energies.
Other partial waves will then be involved, and the simple arguments 
employed here will, of course, no longer apply.

\subsection{Superfluidity in neutron star matter and nuclear matter}
\label{subsec:NN_to_pairing_sec3}

\subsubsection{Superfluidity in neutron star matter}

As we have seen, the  
presence of two different superfluid regimes 
is suggested by the known trend of the 
nucleon-nucleon (NN) phase shifts 
in each scattering channel. 
In both the $^1S_0$ and $^3P_2$-$^3F_2$ channels the
phase shifts indicate that the NN interaction is attractive. 
In particular for the $^1S_0$ channel, the occurrence of 
the well-known virtual state in the neutron-neutron channel
strongly suggests the possibility of a 
pairing condensate at low density, 
while for the $^3P_2$-$^3F_2$ channel the 
interaction becomes strongly attractive only
at higher energy, which therefore suggests a possible 
pairing condensate
in this channel at higher densities. 
In recent years, the BCS gap equation
has been solved with realistic interactions, 
and the results confirm
these expectations. 

The $^1S_0$ neutron superfluid is relevant for phenomena
that can occur in the inner crust of neutron stars, like the 
formation of glitches, which may be related to vortex pinning  
of the superfluid phase in the solid crust \cite{glitch}. 
The results of different groups are in close agreement
on the $^1S_0$ pairing gap values and on 
its density dependence, which
shows a peak value of about 3 MeV at a Fermi momentum close to
$k_F \approx 0.8\; {\rm fm}^{-1}$ \cite{bcll90,kkc96,eh98,sclbl96}. 
All these calculations adopt the bare
NN interaction as the pairing force, and it has been pointed out
that the screening by the medium of the interaction 
could strongly reduce
the pairing strength in this channel \cite{sclbl96,chen86,ains89a,ains89b}. 
The issue of the 
many-body calculation of the pairing 
effective interaction is a complex
one and still far from a satisfactory solution (see
also the discussion in Sec.~\ref{subsec:polarizationterms}).

The precise knowledge of the $^3P_2$-$^3F_2$ pairing gap is of 
paramount relevance for, e.g., the cooling of neutron stars, 
and different values correspond to drastically
different scenarios for the cooling process.
Generally, the gap suppresses the cooling by a factor
$\sim\exp(-\Delta/T)$ (where $\Delta$ is the energy gap),
which is severe for
temperatures well below the gap energy.
Unfortunately, only few and partly
contradictory calculations of the $^3P_2$-$^3F_2$ 
pairing gap exist in the literature, 
even at the level of the bare NN interaction 
\cite{amu85,bcll92,taka93,elga96,kkc96}. 
However, when comparing the results, one should note that the  
NN interactions used in these calculations are not phase-shift 
equivalent, i.e.,  they do not 
predict exactly the same NN phase shifts.  
Furthermore, for the interactions used in 
\cite{amu85,bcll92,taka93,elga96} the predicted 
phase shifts do not agree accurately with modern phase shift 
analyses, and the fit of the NN data has typically 
$\chi^2/{\rm datum}\approx 3$.  

\begin{figure}
\includegraphics[totalheight=17.5cm,angle=0,bb=0 80 350 730]{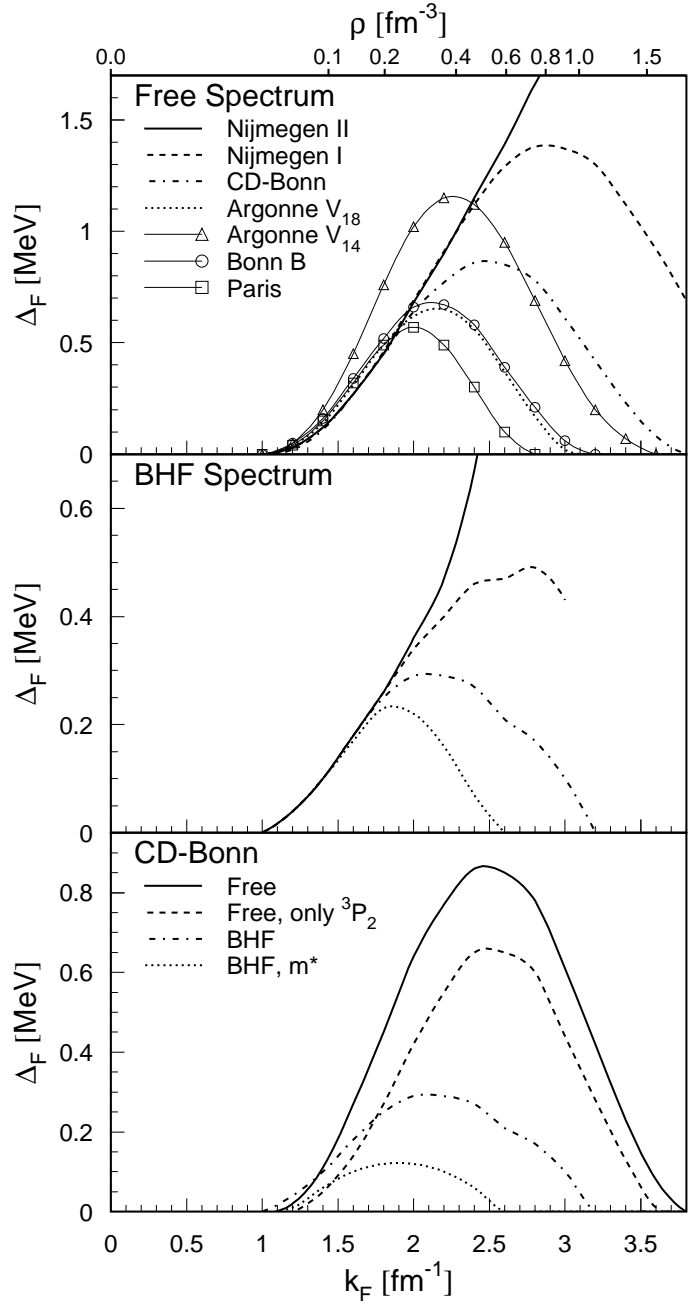}
\caption{Top panel: The angle-averaged $^3P_2$-$^3F_2$ gap in neutron matter
         depending on the Fermi momentum, evaluated with free 
         single-particle spectrum and different nucleon-nucleon potentials. 
         Middle panel: The gap evaluated with BHF spectra.
         Bottom panel: The gap with the CD-Bonn potential in different 
         approximation schemes. Taken from \cite{pair2}.\label{fig:gaps}}

\end{figure}

Fig.~\ref{fig:gaps} contains a comprehensive collection of our results for
the $^3P_2$-$^3F_2$ pairing gaps with different potential models. 
We start with the top part of the figure that displays the results
calculated with free single-particle energies.  
Differences between the results are therefore solely due to differences 
in the $^3P_2$-$^3F_2$ matrix elements of the potentials.
The plot shows results obtained with the old as well as with the modern
potentials.
The results (with the notable exception of 
the Argonne $V_{14}$ interaction model)
are in good agreement at densities below $k_F\approx 2.0\;{\rm fm}^{-1}$, 
but differ significantly at higher densities.  
This is in accordance with the fact that 
the diagonal matrix elements of the potentials are very similar 
below $k_F\approx 2.0\;{\rm fm}^{-1}$, corresponding  
to a laboratory energy for free NN scattering 
of $E_{\rm lab} \approx 350\;{\rm MeV}$.  
This indicates that
within this range the good fit of the potentials 
to scattering data below 350 MeV makes the ambiguities in the 
results for the energy gap quite small,
although there is, in general, no unique relation between phase shifts and gaps.

We would also like to calculate the gap at densities above 
$k_F=2.0\;{\rm fm}^{-1}$.  
Then we need the various potentials at higher energies, 
outside of the range where they are fitted to scattering data.  
Thus there is no guarantee that 
the results will be independent of the model chosen, and in fact 
the figure shows that there are considerable differences 
between their predictions at high densities,
following precisely the trend observed in the phase-shift predictions:
the Argonne $V_{18}$ is the most repulsive of the modern potentials,
followed by the CD-Bonn \cite{cdbonn96} and Nijmegen I and II 
\cite{nijmegen94}.
Most remarkable are the results obtained with Nijm-II: 
we find that the predicted gap 
continues to rise unrealistically even at $k_F \approx 3.5\;{\rm fm}^{-1}$, 
where the purely nucleonic description of matter surely breaks down.

Since the potentials fail to reproduce the measured phase shifts 
beyond $E_{\rm lab}=350\;{\rm MeV}$, the predictions for the $^3P_2$-$^3F_2$ 
energy gap in neutron matter cannot be trusted above 
$k_F\approx 2.0\;{\rm fm}^{-1}$.  
Therefore, the behavior of the $^3P_2$-$^3F_2$ energy gap at high densities 
should be considered as unknown, and cannot be obtained until potential models 
which fit the phase shifts in the inelastic region 
above $E_{\rm lab}=350\;{\rm MeV}$ are constructed.  
These potential models need the flexibility to 
include both the flat structure in the phase shifts above 600 MeV, 
due to the ${\rm NN}\rightarrow{\rm N}\Delta$ channel, as well as the 
rapid decrease to zero at $E_{\rm lab}\approx 1100\;{\rm MeV}$.  
 
We proceed now to the middle part of Fig.~\ref{fig:gaps}, where
the results for the energy gap using Brueckner-Hartree-Fock (BHF) 
single-particle energies are shown.  
For details on the BHF calculations, see, e.g., \cite{jjm76}.  
From this figure, two trends are apparent.
First, the reduction of the in-medium nucleon mass leads to a sizeable 
reduction of the $^3P_2$-$^3F_2$ energy gap, as observed in 
earlier calculations \cite{amu85,bcll92,taka93,elga96}.  
Secondly, the new NN interactions give again similar results 
at low densities, while beyond $k_F\approx 2.0\;{\rm fm}^{-1}$ 
the gaps differ, as in the case with free single-particle energies.   

\begin{figure}
\setlength{\unitlength}{1mm}
   \begin{picture}(140,140)
   \put(-20,-50){\epsfxsize=16cm \epsfbox{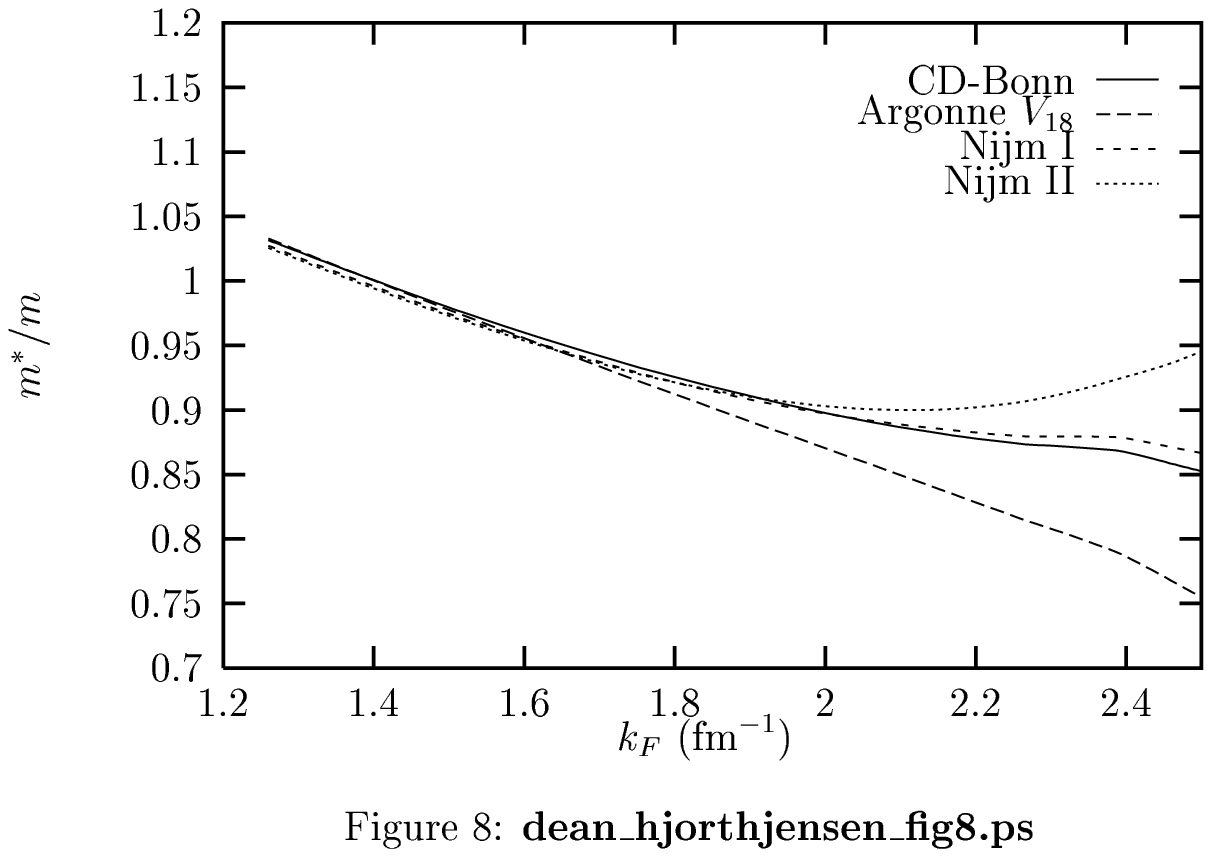}}
   \end{picture}
\caption{Effective masses derived from various interactions in 
         the BHF approach. Taken from \cite{pair2}.\label{fig:mstar}}
\end{figure}

The single-particle energies at moderate densities obtained from the 
new potentials are rather similar, particularly in the 
important region near $k_F$.  
This is illustrated by a plot, Fig.~\ref{fig:mstar},
of the neutron effective mass,
\begin{equation}
 {m^*\over m} = \left( 1 + {m\over k_F} 
 \left.{dU\over dk}\right|_{k_F} \right)^{-1} \:,
\label{eq:effmass}
\end{equation}
as a function of density.
Up to $k_F \approx 2.0\;{\rm fm}^{-1}$ all results agree satisfactorially,
but beyond that point the predictions diverge in the same manner as observed
for the phase shift predictions.
The differences in the BHF gaps at densities slightly above  
$k_F\approx 2.0\;{\rm fm}^{-1}$ are therefore mostly 
due to the differences in the $^3P_2$-$^3F_2$ waves of the potentials,
but at higher densities the differences between the gap are enhanced 
by differences in the single-particle potentials.
An extreme case is again the gap obtained with Nijm-II.  
It is caused by the very attractive $^3P_2$ matrix elements, 
amplified by the fact that the effective mass 
starts to increase at densities above $k_F\approx 2.5\;{\rm fm}^{-1}$ 
with this potential.

Finally, in the lower panel of Fig.~\ref{fig:gaps}, we illustrate the
effect of different approximation schemes with an individual NN potential
(CD-Bonn), namely we compare the energy gaps obtained 
with the free single-particle spectrum, the BHF spectrum,
and an effective mass approximation,
\begin{equation}
  e(k) = U_0 + \frac{k^2}{2m^*} \:,
\label{eq:mstarapp}
\end{equation}
where $m^*$ is given in Eq.~(\ref{eq:effmass}).
In addition, also the gap in the uncoupled $^3P_2$ channel, 
i.e., neglecting the tensor coupling, is shown.

It becomes clear from the figure that the BHF spectrum forces a 
reduction of the gap by about a factor of 2--3.
However, an effective mass aproximation should not be used when 
calculating the gap,  
because details of the single-particle spectrum around the Fermi 
momentum are important in order to obtain a correct value.   
The single-particle energies in the effective mass 
approximation are too steep near $k_F$.
We also emphasize that it is important to solve the coupled 
$^3P_2$-$^3F_2$ gap equations.  
By eliminating the $^3P_2$-$^3F_2$ and $^3F_2$ channels, 
one obtains a $^3P_2$ gap that is considerably lower than the 
$^3P_2$-$^3F_2$ one.  
The reduction varies with the potential,
due to different strengths of the tensor force.
For more detailed discussions of the importance of the tensor force, 
the reader is referred to \cite{amu85,taka93,elga96,kkc2001,kkc1998}.

We end this subsection
with a discussion of pairing for $\beta$-stable
matter of relevance for neutron star cooling, see for example 
 \cite{nstar,pethick1992}.
We will also omit a discussion on neutron pairing gaps in the
$^1S_0$ channel, since these appear at densities corresponding 
to the crust of the neutron star, see for example
\cite{barranco1997}. The gap in the crustal material 
is unlikely
to have any significant effect on cooling processes \cite{pr95}, 
though
it is expected to be important in the explanation 
of glitch phenomena.
Therefore, the relevant pairing gaps for neutron star cooling
should stem from 
the proton contaminant 
in the $^1S_0$ channel, and superfluid neutrons yielding energy gaps 
in the coupled $^3P_2$-$^3F_2$ two-neutron channel. 

To obtain an effective interaction and pertinent single-particle
energies at the BHF level, we can easily solve
the BHF equations 
for different proton fractions.
The conditions for $\beta$ equilibrium require
that 
\begin{equation}
     \mu_{n}=\mu_{p}+\mu_{e},
\end{equation}
where $\mu_i$ is the chemical potential of particle type $i$, 
and that charge is conserved
\begin{equation}
     n_{p}=n_{e},
\end{equation}
where $n_{i}$ is the particle number density for particle $i$.  If 
muons are present, the condition for charge conservation becomes 
\begin{equation}
n_{p}=n_{e}+n_{\mu},
\label{eq:chcon2}
\end{equation}
and conservation of energy requires that 
\begin{equation}
\mu_{e}=\mu_{\mu}.
\end{equation}
We assume that neutrinos escape freely from the neutron star.  
The proton and neutron chemical potentials are determined from the 
energy per baryon, calculated self-consistently in the MBHF approach.  
The electron chemical potential, and thereby the muon chemical potential, 
is then given by $\mu_{e}=\mu_{n}-\mu_{p}$.  The Fermi momentum of lepton 
type $l=e,\mu$ is found from 
\begin{equation}
k_{F_{l}}=\mu_{l}^{2}-m_{l}^{2}
\end{equation}
where $m_{l}$ is the mass of lepton $l$, and we get the particle density 
using $n_{l}=k_{l}^{3}/3\pi^{2}$.  The proton fraction is then determined 
by the charge neutrality condition (\ref{eq:chcon2}).

\begin{figure}
\begin{center}
      {\epsfxsize=20pc \epsfbox{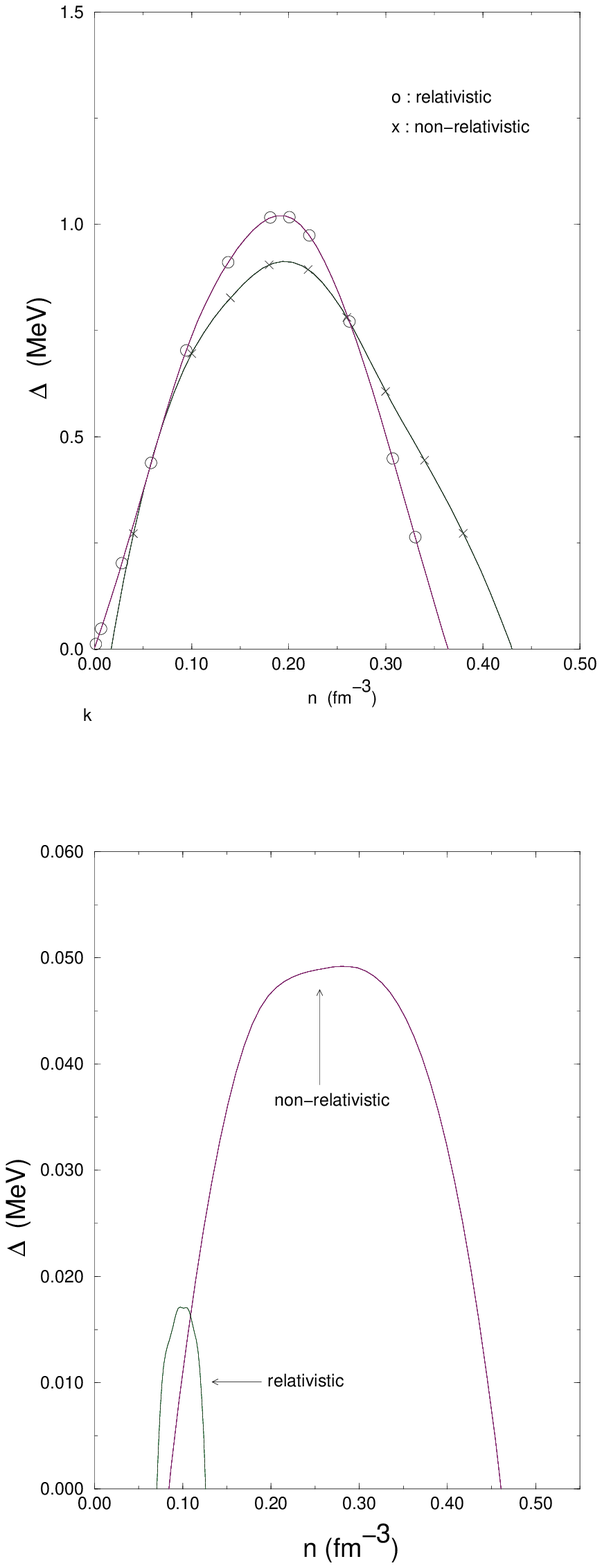}}
%\hspace{1cm}{\epsfxsize=10pc \epsfbox{dean_hjorthjensen_fig10.eps}}
      \caption{Upper box: Proton pairing in $\beta$-stable matter for 
          the $^1S_0$ partial wave. Lower box: Neutron pairing in 
          $\beta$-stable matter for the $^3P_2$
          partial wave. Taken from \cite{pair1}.}
     \label{fig:figgap}
\end{center}
\end{figure}

Since the relevant total baryonic densities for these types of
pairing will be higher than the saturation
density of nuclear matter, we will account for relativistic
effects as well in the calculation of the pairing gaps.
As an example, consider the evaluation of the proton
$^1S_0$ pairing gap using a Dirac-Brueckner-Hartree-Fock  approach,
see \cite{elga96,pair1} for details.
In Fig.\ \ref{fig:figgap} we plot as a function of the total baryonic 
density the pairing gap for protons in the $^1S_0$
state, together with the results from a standard non-relativistic 
BCS approach.
These results are all 
for matter in $\beta$-equilibrium. In Fig.\ \ref{fig:figgap} 
we also plot the 
corresponding relativistic 
results for the neutron energy gap in the $^3P_2$ channel. 
For the 
$^3P_0$ and the $^1D_2$ channels,
the non-relativistic and the relativistic
energy gaps vanish. 

As can be seen from Fig.\ \ref{fig:figgap}, there are only small
differences (except for higher densities) between the non-relativistic
and relativistic proton gaps in the $^1S_0$ wave.
This is expected since the proton fractions (and their respective Fermi
momenta) are rather small; however,
for neutrons, 
the Fermi momenta are larger, and we would 
expect relativistic effects to be important. At Fermi momenta
which correspond to the
saturation point of nuclear matter, $k_F=1.36$ fm$^{-1}$,
the lowest relativistic correction to the kinetic energy per 
particle is of the order of 2 MeV. 
At densities higher than the saturation
point, relativistic effects should be even 
more important.
Since we are dealing with
very small proton fractions a Fermi momentum
of $k_F=1.36$ fm$^{-1}$ would correspond to a total baryonic 
density $\sim 0.09$  fm$^{-3}$. Thus, at larger densities,
relativistic effects for neutrons should
be important.
This is also reflected in Fig.\ \ref{fig:figgap} for the pairing
gap in the $^3P_2$ channel.
The maximum of the relativistic $^3P_2$ gap is less  than half
the corresponding non-relativistic one and the 
density region over which it does not vanish is also much smaller; 
see  \cite{pair1} for further details.

This discussion  can be summarized as follows.
\begin{itemize}
      \item The $^1S_0$ proton gap in $\beta$-stable matter
            is $ \le 1$ MeV, and if polarization
            effects were taken into account \cite{sclbl96},
            it could be further reduced by a factor of 2--3.
      \item The $^3P_2$ gap is also small, of the order
            of $\sim 0.1$ MeV in $\beta$-stable matter.
            If relativistic effects are taken into account,
            it is almost vanishing. However, there is
            quite some uncertainty with the value for this
            pairing gap for densities above $\sim 0.3$
            fm$^{-3}$ due to the fact that the NN interactions
            are not fitted for the corresponding lab energies. 
      \item Higher partial waves give essentially vanishing
            pairing gaps in $\beta$-stable matter.
\end{itemize}
Thus, the $^1S_0$ and $^3P_2$ partial waves are crucial for our
understanding of superfluidity in neutron star matter. 

As an exotic aside, at densities greater than two-three times nuclear
matter saturation density, model calculations based on baryon-baryon
interactions \cite{stoks99,stoks2000,isaac2000,schulze1998,schulze2000b} or relativistic mean field calculations \cite{glendenning2000} indicate that hyperons like $\Sigma^-$ 
and $\Lambda$ are likely to appear in neutron star matter.
The size of the pairing gaps arising from these baryons is, however,
still an open problem, as it  depends entirely on the parametrization 
of the interaction models, see \cite{barnea1996,balberg1999,taka2002} 
for a critical 
discussion. 
Preliminary calculations of the pairing gap for $\Lambda$-hyperons
using recent  meson-exchange models for the hyperon-hyperon interaction  
\cite{stoks99} indicate a vanishing gap, while $\Sigma^-$-hyperon
has a gap of the size of several MeVs \cite{shulzeelgaroy_priv}.
 At large baryon densities for which perturbative QCD
applies, pairing gaps for like quarks have been estimated to be a few
MeV \cite{bailin84}.  However, the pairing gaps of unlike quarks ($ud,~
us$, and $ds$) have been suggested to be several tens to
hundreds of MeV through non-perturbative studies \cite{qsf0}.

The cooling of a young (age $<10^5$ yr) neutron star is mainly
governed by neutrino emission processes and the specific heat
\cite{page2000,schaab1996a,schaab1996b}.  
Due to the extremely high thermal conductivity of
electrons, a neutron star becomes nearly isothermal within a time
$t_w\approx1-100$ years after its birth, depending upon the thickness
of the crust~\cite{pr95}.  After this time, its thermal evolution is
controlled by energy balance: 
\begin{equation}
 \frac{dE_{th}}{dt} = C_V \frac{dT}{dt} = -L_{\gamma} -L_{\nu} + \Phi,
\label{equ:balance}
\end{equation} 
where $E_{th}$ is the total thermal energy and $C_V$ is the
specific heat.  $L_{\gamma}$ and $L_{\nu}$ are the total luminosities
of photons from the hot surface and neutrinos from the interior,
respectively.  Possible internal heating sources, due, for example, to
the decay of the magnetic field or friction from differential
rotation, are included in $\Phi$.  
Cooling simulations are typically performed
by solving the heat transport and hydrostatic equations including
general relativistic effects, see for example the work of Page 
{\em et al.} \cite{page2000}.  

The most powerful energy losses are expected to be given by the 
direct URCA mechanism
\begin{equation}
    n\rightarrow p +e +\overline{\nu}_e, \hspace{1cm} p+e \rightarrow
    n+\nu_e .
    \label{eq:directU}
\end{equation}
However, in the outer cores of massive neutron stars and in the
cores of not too massive neutron stars ($M < 1.3-1.4 M_{\odot}$), the direct
URCA process is allowed at densities
where the momentum conservation $k_F^n < k_F^p + k_F^e$ is
fulfilled. This happens
only at densities $\rho$ several times
the nuclear matter saturation density $\rho_0 =0.16$ fm$^{-3}$.

Thus, for a long time the dominant processes for neutrino emission
have been the modified URCA processes.
See, for example, \cite{pethick1992,nstar} for a discussion, 
in which the two reactions
\begin{equation}
    n+n\rightarrow p+n +e +\overline{\nu}_e,
    \hspace{0.5cm} p+n+e \rightarrow
    n+n+\nu_e ,
    \label{eq:ind_neutr}
\end{equation}
occur in equal numbers.
These reactions are just the usual processes of neutron
$\beta$-decay and electron capture on protons of Eq.\ (\ref{eq:directU}),
with the addition of an extra bystander neutron. They produce
neutrino-antineutrino pairs, but leave the composition of matter constant
on average. Eq.\ (\ref{eq:ind_neutr}) is referred to as the
neutron branch of the modified URCA process. Another branch is the
proton branch
\begin{equation}
    n+p\rightarrow p+p +e +\overline{\nu}_e, \hspace{0.5cm} p+p+e
    \rightarrow
    n+p+\nu_e .
    \label{eq:ind_prot}
\end{equation}
Similarly, at higher densities, if muons are present, we may also
have  processes where the muon and the muon neutrinos 
($\overline{\nu}_{\mu}$ and $\nu_{\mu}$) 
replace the electron and the electron neutrinos
($\overline{\nu}_e$ and $\nu_e$) in the above equations.
In addition, one also has the possibility of neutrino-pair
bremsstrahlung,
processes with baryons more massive than the nucleon
participating, such as isobars or hyperons
or neutrino emission from more exotic states like pion and kaon
condensates or quark matter.

There are 
several cooling calculations including both superfluidity and many of 
the above processes, see for example \cite{page2000,schaab1996a,schaab1996b}. Both
normal neutron star matter and exotic states such as hyperons are included. 
The recent simulation of Page {\em et al.} \cite{page2000} seems to indicate
that available observations of thermal emissions from pulsars can aid
in constraining hyperon gaps. However, all these calculations suffer from the
fact that the microscopic inputs, pairing gaps, composition of matter, emissivity rates, etc.
are not computed at the same many-body theoretical level. This leaves a 
considerable uncertainty. 
  
These calculations deal however with the interior of a neutron star.
The thickness of the crust and an eventual superfluid state in the crust
may have important consequences for the surface temperature.
The time needed for a 
temperature drop in the core to affect the surface temperature should 
depend on the thickness of the crust and on its thermal properties, such 
as the total specific heat, which is strongly influenced 
by the superfluid state of matter inside the crust.

It has recently been proposed that the 
Coulomb-lattice structure of a neutron 
star  crust may influence significantly  the thermodynamical 
properties  of the superfluid neutron gas \cite{broglia1994}. 
The authors of \cite{pr95}  have  proposed that in the crust 
of a neutron star non-spherical nuclear shapes could be present at
densities ranging from $\rho=1.0\times 10^{14}$ gcm$^{-3}$ to
$\rho=1.5\times 10^{14}$ gcm$^{-3}$, a density 
region which represents about $20\%$ of the whole crust. The 
saturation density of nuclear matter is $\rho_0 =2.8\times 10^{14}$
gcm$^{-3}$.
These unusual 
shapes  are supposed \cite{pr95}  
to be disposed in a Coulomb lattice embedded in an
almost uniform background of relativistic electrons. According to the 
fact that the neutron drip point is supposed to occur at lower density
($\rho\sim 4.3 \times 10^{11}$ gcm$^{-3}$), 
and considering the characteristics of the nuclear force in 
this density range, we expect  these
unusual nuclear shapes to be  
surrounded by a gas of superfluid neutrons. 

To model the influence on the heat conduction due to pairing in the crust,
Broglia {\em et al} \cite{broglia1994} studied various 
nuclear shapes for nuclei immersed in a neutron fluid using phenomenological
interactions and employing a local-density approach. They found
an enhacement of the fermionic specific heat due to these shapes
compared to uniform neutron matter. 
These results seem to indicate that the inner part 
of the crust may play a more relevant role on the heat diffusion time 
through the crust.  Calculations with realistic nucleon-nucleon 
interaction were later repeated by Elgar{\o}y {\em et al} \cite{elgaprd},
with qualitatively similar results.

\subsubsection{Proton-neutron pairing in symmetric nuclear matter}
\label{subsubsec:NN_to_pairing_subsub2}

The calculation of the $^1S_0$ gap in symmetric nuclear matter is  
closely related to the one for neutron matter.  Even with modern 
charge-dependent interactions, the resulting pairing gaps for 
this partial wave are fairly similar,
see for example \cite{eh98}.

The size of the neutron-proton (np) $^3S_1$-$^3D_1$ energy gap in 
symmetric or asymmetric 
nuclear matter has, however, been a much debated issue since the 
first calculations of this quantity appeared.  While 
solutions of the BCS equations with bare nucleon-nucleon (NN) forces 
give a large energy gap of several MeVs at the saturation density 
$k_F=1.36\;{\rm fm}^{-1}$ ($\rho=0.17\;{\rm fm}^{-3}$) 
\cite{alm90,vonder91,taka93,bls95,sedrakian1997,sedrakian2000,garrido2001},  
there is little empirical evidence from 
finite nuclei for such strong np pairing correlations, except
possibly for isospin $T=0$ and $N=Z$, see also
the discussion in Sec.~\ref{sec:pairing_correlations} and the recent
work of Jenkins {\em et al} \cite{jenkins}.  
One possible resolution of this problem lies in the fact 
that all these calculations have neglected contributions from the 
induced interaction.  Fluctuations in the isospin and the spin-isospin 
channel will probably make the pairing interaction more repulsive, 
leading to a substantially lower energy gap.
One often-neglected aspect is that all non-relativistic calculations 
of the nuclear matter equation of state (EOS) with two-body NN forces 
fitted to scattering data fail to reproduce the empirical saturation 
point, seemingly regardless of the sophistication of the 
many-body scheme employed. 
For example, a BHF calculation of the 
EOS with recent parametrizations of the NN interaction 
would typically give saturation at $k_F=1.6$-$1.8\;{\rm fm}
^{-1}$.  In a non-relativistic approach, it seems necessary to invoke 
three-body forces to obtain saturation at the empirical equilibrium 
density, see for example \cite{akmal98}.  
This leads one to be cautious when talking about pairing at 
the empirical nuclear matter saturation density when the energy gap 
is calculated within a pure two-body force model, as this density 
will be below the calculated saturation density for this two-body force, 
and thus one is calculating the gap at a density where the 
system is theoretically unstable.   
One even runs the risk, as pointed out in 
 \cite{jack83}, that 
the compressibility is negative at the empirical saturation density, 
which means that the system is unstable against collapse into a 
non-homogeneous phase.  
A three-body force need not have dramatic consequences 
for pairing, which, after all, is a two-body phenomenon, but still it 
would be of interest to know what the $^3S_1$-$^3D_1$ gap is in 
a model in which the saturation properties of nuclear matter are reproduced.  
If one abandons a non-relativistic description, the empirical saturation 
point can be obtained within the Dirac-Brueckner-Hartree-Fock (DBHF) approach, 
as first pointed out by Brockmann and Machleidt \cite{brock90}.    
This might be fortuitous, since, among other things,  
important many-body effects are neglected 
in the DBHF approach.  Nevertheless, it is interesting to investigate 
$^3S_1$-$^3D_1$ pairing in this model and compare our results with a  
corresponding non-relativistic calculation.
Furthermore, several groups have recently developed 
relativistic formulations of 
pairing in nuclear matter \cite{ring91,guim96,matera97,milena2001} and have   
applied them to $^1S_0$ pairing.  
The models are of the Walecka-type \cite{sw86} in the sense that 
meson masses and coupling constants are fitted 
so that the mean-field EOS of nuclear 
matter meets the empirical data.  In this way, however, the relation of 
the models to free-space NN scattering becomes somewhat unclear.  
An interesting result found in  \cite{ring91,guim96,matera97} 
is that the $^1S_0$ energy gap vanishes at densities slightly below the 
empirical saturation density.  This is in contrast with 
non-relativistic calculations which generally give a relatively small, 
but non-vanishing $^1S_0$ gap at this density, see for instance  
\cite{kuch89,bcll90,chen93,elg961}.   

\begin{figure}
\setlength{\unitlength}{1mm}
   \begin{picture}(140,140)
   \put(-20,-50){\epsfxsize=16cm \epsfbox{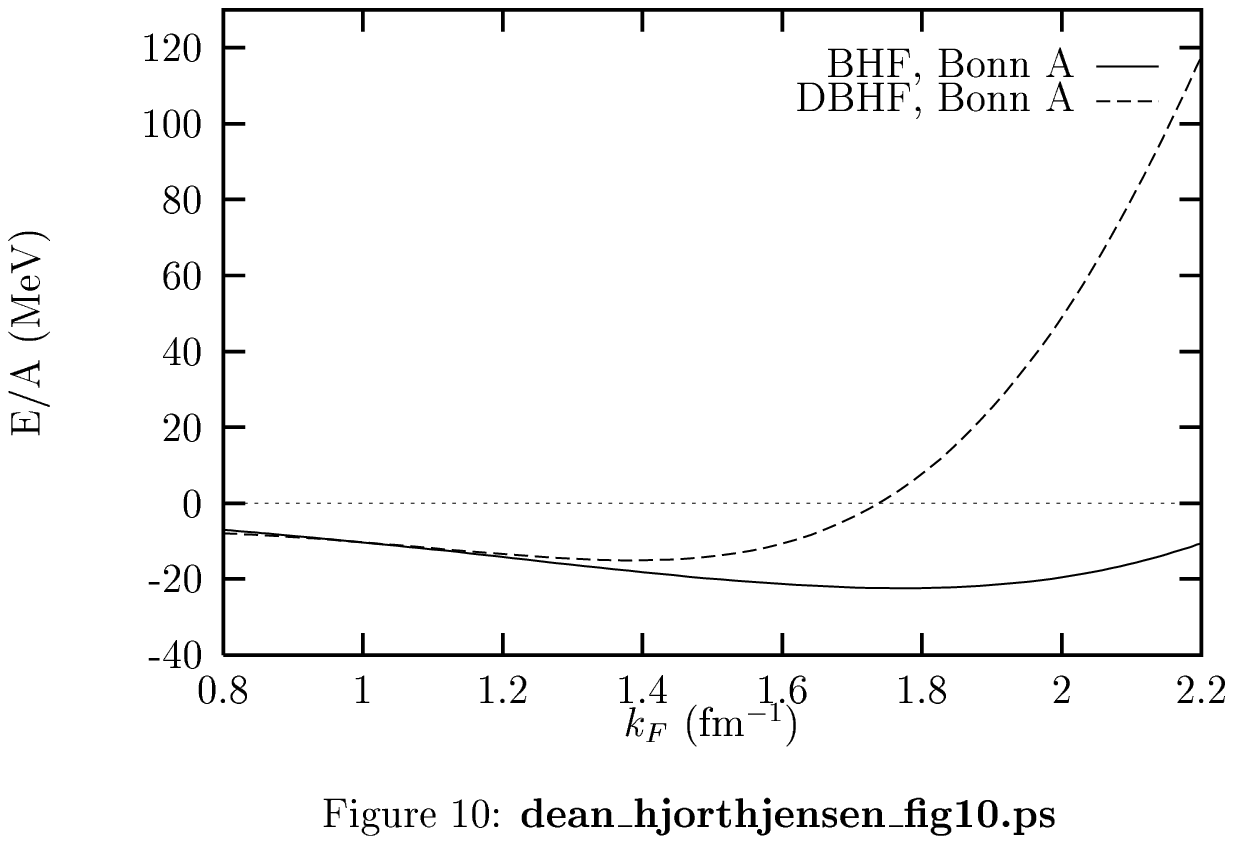}}
   \end{picture}
    \caption{EOS for symmetric nuclear matter with the NN potentials 
 and many-body methods described in the text. Taken from \cite{eeho98}.}
    \label{fig:figprc58_1}
\end{figure}

\begin{figure}
\setlength{\unitlength}{1mm}
   \begin{picture}(140,140)
   \put(-20,-50){\epsfxsize=16cm \epsfbox{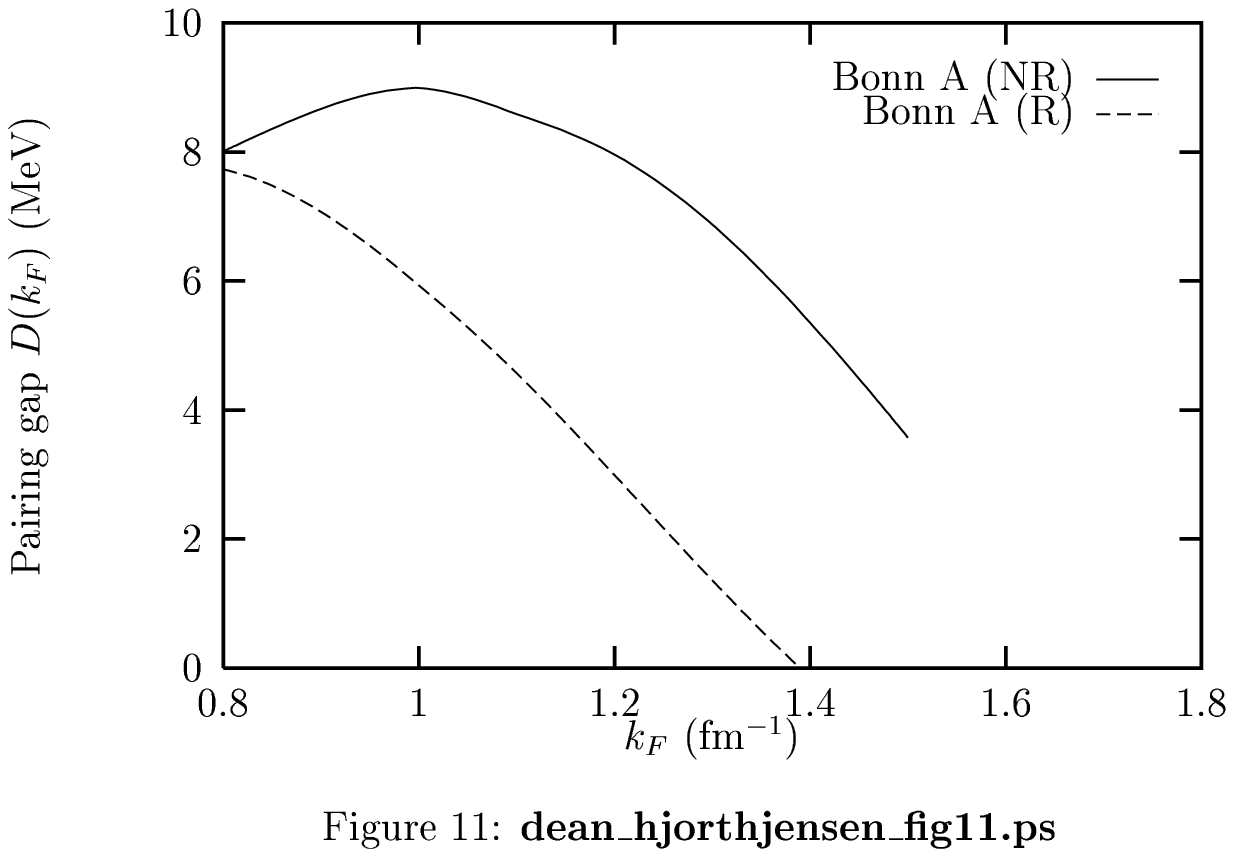}}
   \end{picture}
    \caption{$^3S_1$-$^3D_1$ energy gap in nuclear matter, calculated in 
  non-relativistic (full lines) and relativistic 
 (dashed line) approaches. Taken from \cite{eeho98}. }
    \label{fig:figprc58_2}
\end{figure}

In Fig. \ref{fig:figprc58_1} we show the EOS obtained in our non-relativistic 
and relativistic calculations.   The non-relativistic one fails to 
meet the empirical data, while the relativistic calculation very nearly 
succeeds. 
In these calculations, we employed the non-relativistic and relativistic
one-boson exchange models from the 
Bonn A interaction  defined in \cite{mach89}.  
A standard BHF calculation was done in the non-relativistic case, whereas
in the relativistic case we incorporate 
minimal relativity in the gap equation, thus using DBHF single-particle 
energies in the energy denominators and modifying the free NN interaction 
by a factor $\tilde{m}^2/\tilde{E}_k\tilde{E}_{k'}$ \cite{pair1}.  
The resulting pairing gaps are shown in 
Fig.\ \ref{fig:figprc58_2}.   For the non-relativistic 
calculation, we see a 
large energy gap at the empirical saturation density around $6$ MeV 
at $k_F=1.36\;{\rm fm}^{-1}$, in agreement with  
earlier non-relativistic calculations 
\cite{alm90,vonder91,taka93,bls95}.  
In the relativistic calculation,
we find that the gap is vanishingly small  at this density.  

Since non-relativistic calculations with two-body 
interactions will, in general, give a saturation density that is too high
(an example is shown in Fig. \ref{fig:figprc58_1}), 
this implies that in a  
non-relativistic approach we are actually calculating the 
gap at a density below the theoretical saturation density, and one 
may question the physical relevance of a large gap at a density where 
the system is theoretically unstable.  If one considers the gap 
at the {\em calculated} saturation density for a non-relativistic
approach with a two-body force only, it is in fact close 
to zero.  
In the DBHF calculation, we come 
very close to reproducing the empirical saturation density and binding 
energy, and when this is used as a starting point for a BCS calculation, 
we find that the gap vanishes, both at the empirical and the calculated 
saturation density.  That the DBHF calculation meets the empirical 
points is perhaps fortuitous, as important many-body diagrams are 
neglected and only medium modifications of the nucleon mass are 
accounted for.  An increased repulsion in the non-relativistic 
may thus reduce the gap dramatically.

We end this section with a comment on the interesting possibility of
a transition from BCS pairing to a Bose-Einstein
condensantion in asymmetric nuclear matter at low densities. 
For the singlet
$^1S_0$ partial wave we do not expect
to see a transition, essentially because the coherence length is much 
bigger than the interparticle spacing.
The inclusion of medium effects such as screening
terms are expected to further reduce the pairing gap,
see \cite{deblasio97} and thereby enhance the coherence
length. However, this does not imply that such a transition
is not possible in nuclear matter or asymmetric nuclear matter as present 
in  a neutron star. A recent analysis by Lombardo
{\em et al.} \cite{lombardo2001}, see also the work of Baldo {\em et al.} 
\cite{bls95}, of
triplet $^3S_1$ pairing in low-density symmetric and asymmetric nuclear
matter, indicates that such a transition
is indeed possible. As the system is diluted, the BCS state with large
overlapping Cooper pairs evolves smoothly into a Bose-Einstein condensation
of tightly bound deuterons, or neutron-proton pairs. A neutron excess in this
low-density regime does not affect these deuterons due to the large spatial
separation of the deuterons and neutrons. Even at large asymmetries, these
deuterons are only weakly affected. This effect can have interesting 
consequences for the understanding of e.g., exotic nuclei and asymmetric and 
expanding nuclear matter in heavy-ion collisions.

\subsection{Conclusions and open problems beyond BCS}
\label{subsec:polarizationterms}

We have seen that pairing in neutron star matter is essentially
determined by singlet pairing in the $^1S_0$ channel and triplet pairing
in the $^3P_2$ channel. These two partial waves exhibit a contribution
to the NN interaction which is attractive for a large range of densities.
These partial waves are also crucial for our understanding
of pairing correlations in finite nuclei.
Whether it is possible to have a strong neutron-proton 
pairing gap for symmetric
matter in the $^3S_1$ channel is still an open question. Relativistic
calculations indicate a vanishing gap at nuclear matter saturation density.
The results we have discussed have all been within the frame of a simple
many-body approach; however, the analyses that have 
been performed are not contingent upon these simplifications. Combined with,
e.g., the separation analysis of \cite{kkc96,kkc2001,kkc1998}, we believe the 
calculation procedures will retain their validity when more 
complicated many-body terms are inserted.

\begin{figure}
   \includegraphics[totalheight=10cm,angle=90,bb=0 80 500 700]{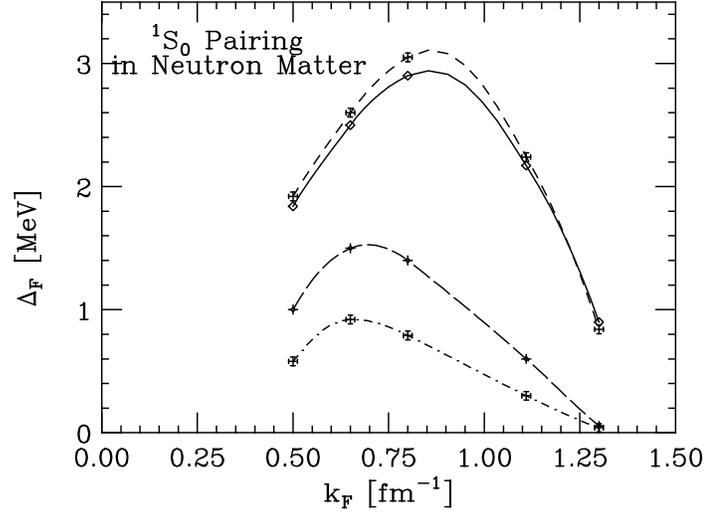}
\caption{Energy gap in different approximations for the self-energy. 
The upper line (dashes with crosses)  
stands for the free single-particle spectrum with a standard BCS approach
while the upper solid line
arises from the BHF approach of Eq.~(\ref{eq:mstarapp})
and the standard BCS approach. The lower lines arise from solving 
Eq.~(\ref{eq:gorkoveq}) for the pairing gap with different 
approaches to the self-energy, for further details see \cite{lsz2001,ls2000}. Taken from \cite{lsz2001,ls2000}.
\label{fig:lombardo2001}}
\end{figure}

A complete and realistic treatment of pairing in a given 
strongly coupled Fermi system such as neutron matter 
demands {\it ab initio} calculation of both the single-particle 
energies and the interaction in the medium. 
The dependence of, e.g., $^3P_2$ pairing upon various 
approaches to the single-particle energies is a clear signal of the need
for a consistent many-body scheme, see for example Fig.~\ref{fig:gaps}.
Whether we employ a density-dependent effective mass approach as
in Eq.~(\ref{eq:effmass}) or a standard effective mass approach as in
Eq.~(\ref{eq:mstarapp}), the results are different contributions
to the pairing gap.
Recently, Lombardo {\em et al.}  \cite{lsz2001,ls2000}
reexamined the role played by ground-state correlations in the self-energy.
Solving the Gorkov equations, see Eq.~(\ref{eq:gorkoveq}), they found a 
substantial suppression of the $^1S_0$ pairing due to changes in the 
quasiparticle strength around the Fermi surface.  
Their results are shown in Fig.~\ref{fig:lombardo2001} for a set of different 
$k_F$-values. 

This figure shows that self-energy 
effects are an important ingredient in our 
understanding of the pairing gap in infinite matter.

\begin{figure}
   \setlength{\unitlength}{1mm}
   \begin{picture}(100,60)
   \put(-10,-50){\epsfxsize=14cm \epsfbox{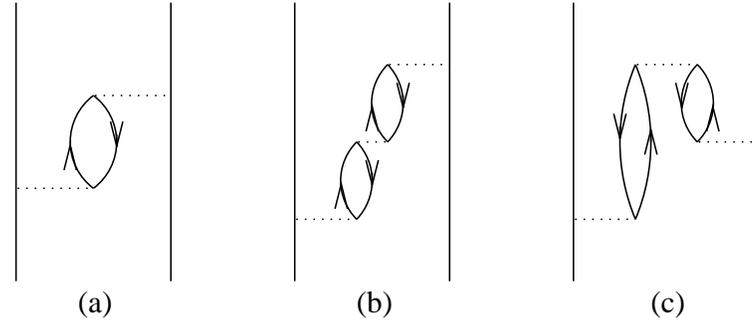}}
   \end{picture}
\caption{Diagram (a) is the 
second-order diagram with particle-hole intermediate states. 
The external legs can be particles or holes. Diagrams (b) and (c) are examples
of third-order TDA or RPA diagrams. 
The dotted lines represent the interaction vertex.\label{fig:secondordercorelpol}}
\end{figure}

A correct treatment of the self-energy entails a self-consistent
scheme where the renormalization of the interaction is done at an
equal footing. Of special interest for the pairing interaction are
polarization corrections. 
At low densities we may expect that the dominant polarization term stems from 
a second-order perturbative correction with particle-hole intermediate
states, as depicted in Fig.~\ref{fig:secondordercorelpol}.

For contributions around the Fermi surface, one can evaluate diagram (a) 
analytically and obtain a result in terms of 
the Fermi momentum and the scattering length.
As shown by Heiselberg {\em et al.} and Schulze {\em et al.}
\cite{henning2000,spr2001}, even the low-density expression of 
Eq.~(\ref{eq:lowdensgap_sec2}) is reduced by a factor of $\approx 2.2 $
when polarization terms are included.

To go beyond diagram (a) and simple low-density 
approximations requires considerable efforts 
and has not been accomplished yet. This means that there is still a large
uncertainty regarding the value of the pairing gap in infinite matter.
There are few calculations of the pairing gap from the point of view of an 
{\it ab initio} approach.

One such scheme is the one favored by Clark and co-workers, based on 
correlated-basis (or CBF) theory \cite{bishop,chen86,chen93}. Within the CBF scheme, 
the following approach to the quantitative physics of pairing in
extended nucleonic systems has been undertaken:
\begin{enumerate}
\item[(a)]
Dressing of the pairing interaction by Jastrow correlations within 
CBF theory \cite{kroclark,ksj}
\item[(b)]
Dressing of the pairing interaction by dynamical collective effects
within CBF theory \cite{ksj,chen86,chen93} (including polarization effects 
arising from exchange of density and spin-density fluctuations, etc.) 
\item[(c)]
Consistent renormalization of single-particle energies by short-
and long-range correlations within CBF theory (cf.\ \cite{kcj})
\end{enumerate}
This approach has already been explored in the $^1S_0$ neutron pairing
problem \cite{chen86,chen93}, although the assumed Jastrow correlations
have not been optimized and only a second-order CBF perturbation
treatment is available for step (b).  Application of this scheme 
to $^3P_2$--$^3F_2$ pairing in neutron-star matter is 
still an unexplored topic.
Alternatively, coupled-cluster (CC) \cite{bishop} or Fermi-hypernetted chain 
inspired approaches could be used \cite{adelchi97}.
Another approach, followed by Wambach, Ainsworth and Pines \cite{ains89a,ains89b} and
Schulze {\em et al.} \cite{sclbl96} departs from the  Landau theory inspired 
many-body approach to screening of  Babu and Brown 
\cite{bb72,back85,jack82,dick81,dick83,dickoffherbert87}. 
This microscopic derivation 
of the effective interaction starts from the following physical idea: 
the particle-hole (p-h) interaction can be considered as made of a 
{\em direct} component containing the short-range correlations and an 
{\em induced} component due to the exchange of the collective excitations 
of the medium. 

Finally, another alternative is to solve the full set of the 
Parquet equations, as discussed in \cite{jls82,mhj98}. 
This  self-consistent scheme entails the summation to all orders 
of all two-body diagrams
with particle-particle and hole-hole (ladder diagrams) 
and particle-hole (polarization and screening diagrams) intermediate states,
accompanied with the solution of Dyson's equation for the single-particle
propagator.
Recently, Bozek \cite{bozek2002} has studied the generalized ladder diagram
resummation in the superfluid phase of nuclear matter. This is the first step
towards the solution of the Parquet diagrams.  

\begin{figure}
\includegraphics[totalheight=12cm,angle=0,bb=0 350 500 700]{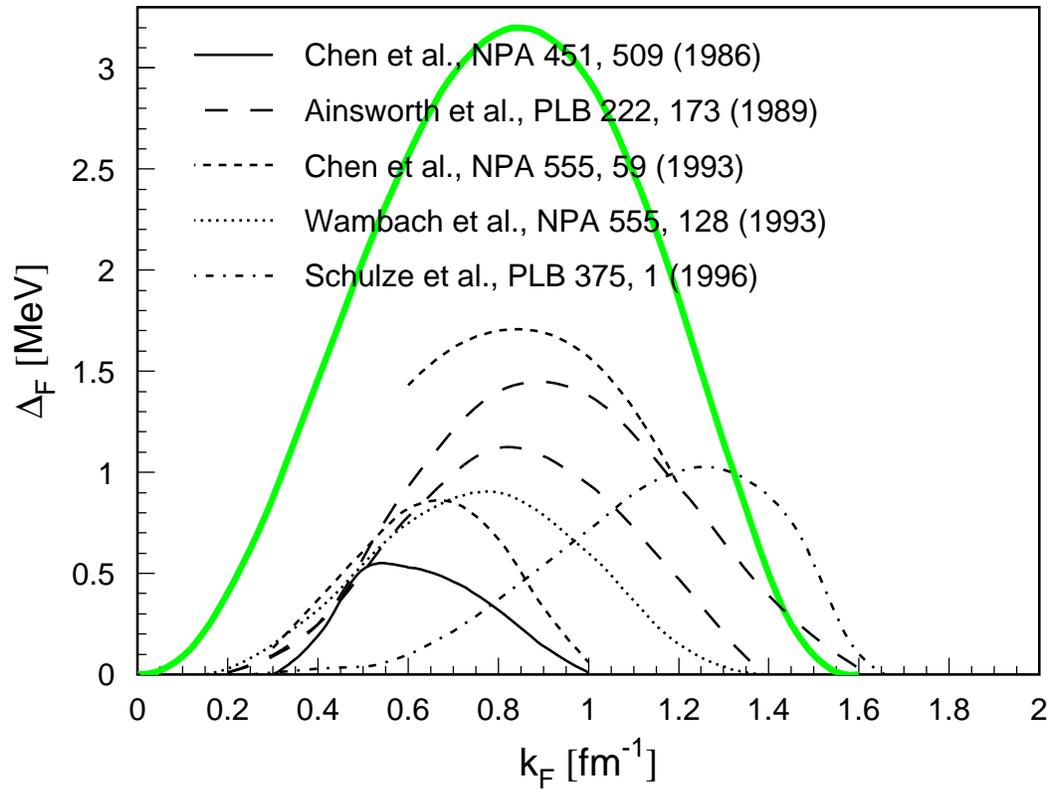}
\caption{The $^1S_0$ gap in pure neutron matter 
predicted in several publications
taking account of polarization effects.
Taken from \cite{ls2000}\label{fig:screening}}.
\end{figure}

We conclude by summarizing this section through 
Fig.~\ref{fig:screening}. 
This figure exhibits the influence of various
approaches which include 
screening corrections  to the pairing gap. 
The curve in the background is given by the  
calculation with free single-particle energies and the bare nucleon-nucleon
interaction. 
These calculations are similar,
except for the potential model employed, to those discussed in,
e.g., Fig.~\ref{fig:energygap_sec2}.  
This means that the calculations of Subsec.~\ref{subsec:NN_to_pairing_sec1}, 
with only experimental inputs, phase shifts and scattering length, 
yield an upper limit for the $^1S_0$ pairing gap. 
How such renomalizations will affect the $^3P_2$ gap is an entirely open issue.
This gap is crucial since it extends to large densities 
and can reasonably be expected to occur at the centers of neutron
stars.  Unfortunately, one cannot constrain at present 
the size of the pairing gap
from  data on thermal emission from neutron stars, see also the discussion
in Sec.~\ref{subsec:NN_to_pairing_sec3}.

%%%%   section 3

%\input{pairing_correlations}
% new section on pairing and sn isotopes, mhj 26/march/02
% latest typo upgrade 7/4/02 mhj
% june djd typos
% final version aug 9, djd and mhj
% october 2002 mhj included comments from Hubert
\section{PAIRING CORRELATIONS IN FINITE NUCLEI}

\label{sec:pairing_correlations}

\subsection{Introduction to the nuclear shell model}

Our tool for analyzing pairing correlations in finite nuclei
is the nuclear shell model, with appropriately defined model spaces
and effective interactions. In this section we extract information
on pairing correlations through large-scale  
shell-model calculations of
several nuclear systems, from nuclei in the $sd$-shell to 
heavy tin isotopes. 

We define the nuclear shell model by a set of spin-orbit coupled
single-particle states
with quantum numbers $ljm$ denoting the orbital angular momentum ($l$) and
the total angular momenta ($j$) and its $z$-component, $m$.  In a
rotationally invariant basis, the one-body states have energy
$\varepsilon_{lj}$ that are independent of $m$.  
The single-particle states and energies may be different
for neutrons and protons, in which case it is 
convenient to include also the isospin component
$t_z=\pm 1/2$ in the state description.  
We will use the label $\alpha$ for the set of quantum
numbers $ljm$ or $ljmt_z$, as appropriate.
These orbits define the 
valence $P$-space, or model space for the shell model, 
while remaining single-particle
orbits define the so-called  excluded space, or $Q$-space. We can express
these spaces through the operators
\begin{equation}
P = \sum_{i = 1}^n \left | \psi_i\right\rangle \left \langle \psi_i\right |  ,
\;\;\;
Q = \sum_{i = n+1}^{\infty} \left |\psi_i\right\rangle \left \langle \psi_i\right |,
\end{equation}
where $n$ defines the dimension of the model space while the wave functions
$\psi_i$ could represent a many-body Slater determinant built on 
the chosen single-particle basis. 
As an example, if we consider 
the chain of tin isotopes from $^{100}$Sn to $^{132}$Sn, the
neutron single-particle orbits 
$2s_{1/2}$, $1d_{5/2}$, $1d_{3/2}$, $0g_{7/2}$, and $0h_{11/2}$ could define
an eventual model space. 
We could then choose 
$^{132}$Sn as a closed-shell core. Neutron holes from 
$^{131}$Sn to $^{100}$Sn define then the valence-space or model-space 
degrees of freedom. We could, however,
have chosen $^{100}$Sn as a closed-shell core.
In this case, neutron particles from $^{101}$Sn to $^{132}$Sn
define the model space.

The shell-model Hamiltonian $\hat{H}$ is thus built upon such a 
single-particle basis. 
The shell-model problem requires normally the solution of a real and symmetric
$n \times n$ matrix eigenvalue equation
\begin{equation}
\hat{H}\left | \Psi_k\right\rangle  = 
       E_k \left | \Psi_k\right\rangle,
\end{equation}
with $k = 1,\ldots, n$, where the size of this matrix is defined by the 
actual shell-model space. 
The dimensionality
$n$ of the eigenvalue matrix $H$ is increasing
with an increasing number of valence particles or holes. As an example,
for $^{116}$Sn with the above mentioned single-particle basis,
the dimensionality of the Hamiltonian matrix is of  
the order of  $n \sim 10^{8}$. For nuclei in the rare-earth 
region, this dimensionality can be of the order of
$n \sim 10^{12}-10^{14}$.

The
shell-model Hamiltonian can be written in the form $\hat{H}=\hat{H}_1+
\hat{H}_2+\hat{H}_3+\dots$ where
$\hat{H}_1$
is a one-body term typically represented by experimental single-particle
energies, see Eq.~(\ref{eq:hamiltonian_sec1}).
The two-body term, see Eq.~(\ref{eq:hamiltonian_sec1}), 
is given in 
terms of the uncoupled matrix elements $V$ of the two-body interaction. 
These matrix elements must obey rotational invariance, parity conservation,
and (when implemented) isospin invarience. To make explicit the 
rotational and isospin invariance, we rewrite the two-body Hamiltonian 
as 
\begin{equation}
\hat{H}_2 = \frac{1}{4}\sum_{\alpha\beta\gamma\delta}\sum_{JT}
\left[\left(1+\delta_{\alpha\beta}\right)\left(1+\delta_{\gamma\delta}\right)\right]^{1/2}
V_{JT}(\alpha\beta,\gamma\delta)\sum_{MT_z}\hat{A}^\dagger_{JT;MT_z}(\alpha\beta)
\hat{A}_{JT;MT_z}(\gamma\delta) \;, 
\end{equation}
where the pair operator is 
\begin{equation}
\hat{A}^\dagger_{JT;MT_z}(\alpha\beta)=\sum_{m_{\alpha},m_{\beta},t_{\alpha},t_{\beta}}
\left(j_{\alpha} m_{\alpha} j_{\beta} m_{\beta} \mid JM\right)
\left(\frac{1}{2}t_{\alpha}\frac{1}{2} t_{\beta}\mid TT_z\right)
a^\dagger_{j_{\beta}m_{\beta}t_{\beta}}a^\dagger_{j_{\alpha}m_{\alpha}t_{\alpha}}\;.
\end{equation}
In these expressions $(JM)$ are the coupled angular momentum quantum
numbers and $(TT_z)$ are the coupled isospin quantum numbers. The coupled
two-body matrix elements $V_{JT}$ define the valence particle interactions
within the given shell-model space. They are matrix elements of a 
scalar potential $V(\vec{r}_1,\vec{r}_2)$ and are defined as 
\begin{equation}
\left\langle 
\left[\psi_{j_{\alpha},t_{\alpha}}
(\vec{r}_1)\times\psi_{j_{\beta},t_{\beta}}(\vec{r}_2)\right]^{JM;TT_z} 
\mid V(\vec{r}_1,\vec{r}_2) \mid  
\left[\psi_{j_{\delta},t_{\delta}}(\vec{r}_1)
\times\psi_{j_{\gamma},t_{\gamma}}(\vec{r}_2)\right]^{JM;TT_z}
\right\rangle \;,
\end{equation}
and are independent of $M$ and $T_z$. The antisymmetrized matrix elements
are $V^A_{JT}(\alpha\beta,\gamma\delta)$ and are then given by
\begin{equation}
V^{A}_{JT}(\alpha\beta,\gamma\delta)=
\left[\left(1+\delta_{\alpha\beta}\right)\left(1+\delta_{\gamma\delta}\right)\right]^{-1/2}
\left[V_{JT}(\alpha\beta,\gamma\delta)-(-1)^{J+j_a+j_b+T-1}V_{JT}
(\beta\alpha,\gamma\delta)\right]\;.
\end{equation}
We remark here that
three-body or higher-body terms such as $\hat{H}_3$ are normally
not included in a shell-model effective interaction, although shell-model
analyses with three-body interactions 
have been made in \cite{herbert89,eh2002}.

In the following subsections, we discuss how to extract information
about pairing correlations 
within the framework
of large-scale shell-model and shell-model Monte Carlo (SMMC) 
calculations. 
In Subsec.~\ref{subsec:seniority}, we discuss selected features
of the tin isotopes such as the near constancy of the energy difference 
between the first excited state with $J=2$ and the ground state
with $J=0$ for the whole chain of even isotopes from
$^{102}$Sn to $^{130}$Sn. 
These are nuclei whose excited states are well reproduced by the neutron
model space mentioned above.
We relate this near constancy to strong pairing
correlations and the same partial waves which contribute
to superfluiditity in neutron stars, namely the $^1S_0$ and 
$^3P_2$ components of the nucleon-nucleon interaction.
The $^1S_0$ component is generally the dominating partial wave, 
a well-known fact in nuclear
physics.
We show also that a truncation scheme like generalized
seniority \cite{talmi93} is a viable first approximation to large-scale
shell-model calculations.  

In Subsec.~\ref{subsec:isoscalarvector}, we discuss isoscalar and isovector
pairing correlations, whereas proton-neutron pairing and Wigner energy are
discussed in Subsec.~\ref{subsec:wigner}. Various thermal properties
are discussed in the remaining subsections. These results are obtained 
through large-scale SMMC calculations, see for example \cite{kdl97}.

\subsection{Tin isotopes, seniority, and the nucleon-nucleon interaction}
\label{subsec:seniority}

Nuclei far from the line of $\beta$-stability are at present in
focus of the nuclear structure physics community. 
Considerable attention is being devoted to the
experimental and theoretical 
study of nuclei near 
$^{100}$Sn from studies of the chain of Sn isotopes
up to $^{132}$Sn to, e.g., nuclei 
near the proton drip line like $^{105,106}$Sb. 

Our scheme to obtain an effective two-body interaction for 
shell-model studies
starts with a free nucleon-nucleon  interaction $V$, which is
appropriate for nuclear physics at low and intermediate energies. 
Here we employ the charge-dependent
version of the Bonn potential models, 
see \cite{bonn-cd} and the discussion in Sec.~\ref{sec:NN_to_pairing}.
The next step 
in our many-body scheme is to handle 
the fact that the repulsive core of the nucleon-nucleon potential $V$
is unsuitable for perturbative approaches. This problem is overcome
by introducing the reaction matrix $G$, which in 
a diagrammatic language represents  the sum over all
ladder types of diagrams. This sum is meant to renormalize
the repulsive short-range part of the interaction. The physical interpretation
is that the particles must interact with each other an infinite number
of times in order to produce a finite interaction. 
We calculate $G$ using the double-partitioning scheme discussed
in \cite{mhj_95}.
Since the $G$-matrix represents just
the summation to all orders of particle-particle
ladder diagrams, there are obviously other terms which need to be included
in an effective interaction. Long-range effects represented by 
core-polarization terms are also needed.
In order to achieve this,  the $G$-matrix elements
are renormalized by the so-called $\hat{Q}$-box method.
The $\hat{Q}$-box is made up of non-folded diagrams which are irreducible
and valence-linked. Here we include all non-folded diagrams to third
order in $G$ \cite{mhj_95}.
Based on the $\hat{Q}$-box, we compute 
an effective interaction
$\tilde{H}$ in terms of the $\hat{Q}$-box
 using the folded-diagram expansion method
(see for example   \cite{mhj_95} for further details).

The effective two-particle interaction is then used in 
large-scale shell-model
calculations.
For the shell-model calculation, we employ the Oslo $m$-scheme 
shell-model code
\cite{eh2002}, which is based
on the Lanczos algorithm, an iterative method which gives the solution of
the lowest eigenstates. 
The technique is described in detail in  \cite{whit77}. 
The shell-model space consists of the orbits 
$2s_{1/2}$, $1d_{5/2}$, $1d_{3/2}$, $0g_{7/2}$ and $0h_{11/2}$.

Of interest in this study is the fact that 
the chain of even tin isotopes from $^{102}$Sn to $^{130}$Sn 
exhibits a near constancy of the 
$2^+_1-0^+_1$ excitation energy, a constancy which can be related
to strong pairing correlations and the near degeneracy in energy 
of the relevant single-particle orbits. As an example, we show the 
experimental\footnote{We limit the discussion to even isotopes
from  $^{116}$Sn to $^{130}$Sn, since a qualitatively similar picture
is obtained from $^{102}$Sn to $^{116}$Sn.}
$2^+_1-0^+_1$ excitation energy 
from  $^{116}$Sn to $^{130}$Sn in Table \ref{tab:table_tincalc1}. 
Our aim is to see whether partial waves which play a crucial
role in superfluidity of neutron star matter, 
viz. $^1S_0$ and $^3P_2$, are equally
important in reproducing the near-constant spacing in the chain
of even tin isotopes shown in  Table \ref{tab:table_tincalc1}.

In order to test whether the $^1S_0$ and $^3P_2$ partial waves are equally
important in reproducing the near constant spacing in the chain
of even tin isotopes as they are for the superfluid properties of infinite 
neutron star matter (recall the discussion of Sec.~\ref{sec:NN_to_pairing}),
we study four different approximations to the shell-model
effective interaction, viz.,
\begin{enumerate}
  \item Our best approach to the effective interaction, 
   $V_{\mathrm{eff}}$, contains all one-body and two-body diagrams through 
       third order in the $G$-matrix, as discussed above, 
        see also  \cite{anne98}. 
  \item The effective interaction is given by the $G$-matrix only and inludes
        all partial waves up to $l=10$.
  \item We define an effective  interaction based on a $G$-matrix which now includes
        only the $^1S_0$ partial wave.
  \item Finally, we use an effective interaction based on a $G$-matrix which does
        not contain the  $^1S_0$ and $^3P_2$ partial waves, but all other waves
        up to $l=10$.  
\end{enumerate}
In all four cases the same NN interaction is used, viz., 
the CD-Bonn interaction described in  \cite{bonn-cd}.
Table \ref{tab:table_tincalc1} lists the results.  

We note from this table that the three first cases nearly produce a constant 
$2^+_1-0^+_1$ excitation energy, with our most optimal effective interaction
$V_{\mathrm{eff}}$ being closest to the experimental data. The bare $G$-matrix
interaction, with no folded diagrams 
as well, results in a slightly more compressed
spacing. This is mainly due to the omission of the core-polarization 
diagrams which typically render the $J=0$ matrix elements more attractive.
Such diagrams are included in $V_{\mathrm{eff}}$. 
Including only the $^1S_0$ partial wave in the construction of the  $G$-matrix
(case 3)
yields, in turn, a somewhat larger 
spacing. This can again be understood from the
fact that a $G$-matrix constructed with this partial wave  
only does not receive contributions from any entirely repulsive partial wave.
It should be noted that our optimal interaction, as demonstrated in 
 \cite{anne98}, shows a rather good reproduction of the 
experimental spectra for both even and odd nuclei. Although the approximations
made in cases 2 and 3 produce an almost constant $2^+_1-0^+_1$ excitation energy,
they reproduce poorly the properties of odd nuclei and other 
excited states in the even Sn isotopes. 

However, the fact that the first three  approximations result in a such a good
reproduction of the  $2^+_1-0^+_1$ spacing may hint to the fact that the 
$^1S_0$ partial wave is of paramount importance. 
If we now turn attention to case 4, i.e., we omit the
$^1S_0$ and $^3P_2$ partial waves in the construction of the $G$-matrix,
the results presented  in 
Table \ref{tab:table_tincalc1} exhibit  a spectroscopic 
catastrophe\footnote{Although we have singled out these
two partials waves, due to their connection to infinite matter, it is 
essentially the $^1S_0$ wave which is responsible for the behavior seen
in Table \ref{tab:table_tincalc1}.}. We also do not list eigenstates
with other quantum numbers. For $^{126}$Sn,
the ground state is no longer a $0^+$ state; rather it carries 
$J=4^+$ while for $^{124}$Sn the ground state 
has $6^+$. The first $0^+$ state for this nucleus is given at an excitation
energy of $0.1$ MeV with respect to the $6^+$ ground state.
The general picture for other eigenstates is that of 
an extremely poor agreement
with data.  
Since the agreement is so poor, even the qualitative reproduction of the 
$2^+_1-0^+_1$ spacing, we defer from performing time-consuming shell-model
calculations for $^{116,118,120,122}$Sn.

Since pairing is so prominent in such systems, 
we present a comparison of the SM with the generalized
seniority model \cite{talmi93}. The generalized seniority scheme is an 
extension of the seniority scheme, i.e., from involving only one single 
$j$--orbital, the model is generalized to involve a group of $j$--orbitals 
within a major shell. The generalized seniority scheme is a more simple 
model than the shell model since a rather limited number of configurations 
with a strictly defined structure are included, thus allowing a more
direct physical interpretation. States with seniority $v=0$ are by 
definition states where all particles are coupled in pairs. Seniority 
$v=2$ states have one pair broken, seniority $v=4$ states have two pairs 
broken, etc. The generalized seniority scheme is suitable for describing 
semi-magic nuclei where pairing plays an important role. The pairing picture 
and the generalized seniority scheme have been important for the description 
and understanding of the tin isotopes. A typical feature of the seniority 
scheme is that the spacing of energy levels is independent of the number of 
valence particles. For the tin isotopes, not only the spacing between the 
ground state and the $2^{+}_{1}$ state, but also the spacing beween the 
ground state and the $4^{+}_{1}$ and $6^{+}_{1}$ states is fairly constant 
throughout the whole sequence of isotopes. In fundamental works on generalized 
seniority by, for instance, Talmi \cite{talmi93}, the tin isotopes have been 
used as one of the major test cases. It is also worth mentioning the 
classical work on pairing by Kisslinger and Sorensen \cite{kiss60}, see also
the analyses of \cite{sandulescu97} and the review article of Bes
and Sorensen \cite{besreview}. 

If we, by closer investigation and comparison of the SM wave function and the
seniority states, find that the most important components are accounted for  
by the seniority scheme, we can benefit from this and reduce the SM basis. 
This would be particularly useful when we want to do calculations on systems 
with a large number of valence particles.

The operator for creating a generalized seniority $(v=0)$ pair is
\begin{equation}
    S^{\dagger}= \sum_{j} 
    \frac{1}{\sqrt{2j+1}}\alpha_{j}\sum_{m \geq 0} (-1)^{j-m} b^{\dagger}_{jm}
    b_{j-m}^{\dagger},
    \label{eq:s_dagger}
\end{equation}
where  $b^{\dagger}_{jm}$ is the creation operator for holes.
The generalized senitority $(v=2)$ operator for creating a broken pair 
is given by
\begin{equation}
     D_{J,M}^{\dagger}=
     \sum_{j \leq j'} (1+\delta_{j,j'})^{-1/2}
     \beta_{j,j'} \langle j m j' m'
     \left | JM \right \rangle b^{\dagger}_{jm}b^{\dagger}_{j'm'}.
\end{equation}
The coefficients $\alpha_{j}$ and $\beta_{jj'}$ are obtained from the $^{130}$Sn
ground state and the excited states, respectively.

We calculate the squared overlaps between the constructed generalized seniority
states and our shell-model states
\begin{equation}
\begin{array}{lccccl}
(v=0) & & & & &
|\langle ^{A}{\rm Sn(SM)} ;
0^{+}|(S^{\dagger})^{\frac{n}{2}}| \tilde{0} \rangle |^{2}, \\
(v=2) & & & & &
|\langle ^{A}{\rm Sn(SM)} ;
J_{i}|D^{\dagger}_{JM}(S^{\dagger})^{\frac{n}{2}-1}| \tilde{0} \rangle |^{2}. 
\end{array}
\end{equation}
The vacuum state $|\tilde{0} \rangle $ is the $^{132}$Sn--core and $n$ is
the number of valence particles. These quantities tell to what extent the 
shell-model states satisfy the pairing picture, or in other words, how well 
is generalized seniority conserved as a quantum number.

The squared overlaps are tabulated in Table~\ref{tab:seniority}, and vary
generally from 0.95 to 0.75. As the number of valence particles increases, the 
squared overlaps gradually decrease. The overlaps involving the $4^{+}$
states show a fragmentation. In $^{128}$Sn, the $4^{+}_{1}$ (SM)
state is mainly a seniority $v=2$ state. As approaching the middle of the 
shell, the next state, $4^{+}_{2}$, takes more and more over the 
structure of a seniority $v=2$ state. The fragmentation of seniority over these
two states can be understood from the fact that they are rather close in 
energy and therefore may have mixed structure.

In summary, these studies show clearly the prominence of pairing correlations
in nuclear systems with identical particles as effective degrees of freedom.
There is a clear link between superfluidity in infinite neutron star matter 
and spectra of finite nuclei such as the chain of tin isotopes.
This link is provided especially by the $^1S_0$ partial wave of the 
nucleon-nucleon interaction. Excluding this 
component from an effective interaction yields 
spectroscopy in poor agreement with experimental data.
Although the $^1S_0$ partial wave plays an important role, other many-body
effects arising e.g., from low-lying collective surface vibrations 
among nucleons can have important effects on properties of nuclei,
as demonstrated recently by Broglia {\em et al} 
\cite{broglia1999,broglia2002}. In order to interpret the results
of Table \ref{tab:table_tincalc1} one needs to analyze the core-polarizations
diagrams which are used to compute the effective interaction in terms
of the various partial waves. 

Generalized seniority provides an explicit measure of the degree of 
pairing correlations in the wave functions.  Furthermore, 
generalized seniority can serve as a useful starting point 
for large-scale shell-model calculations and is one among several ways 
of extracting information about pairing correlations. 
In the next subsections, we present further approaches.

\subsection{Isoscalar and isovector pair correlations}
\label{subsec:isoscalarvector}

Numerous phenomenological descriptions of nuclear collective motion describe
the nuclear ground state and its low-lying excitations in terms of bosons.
In one such model, the Interacting Boson Model
(IBM), $L=0$ (S) and $L=2$ (D) bosons are identified with
nucleon pairs having the same quantum numbers \cite{Arima},
and the ground state can be viewed as a condensate of such pairs.  
Shell-model studies of the pair
structure of the ground state and its variation with the number of valence
nucleons can therefore shed light on the validity and microscopic foundations
of these boson approaches. 

Generally speaking, nucleon-nucleon pairing may be
considered in several classes. A nucleon has a spin $j=1/2,j_z=\pm 1/2$ and an 
isospin $t=1/2,t_z=\pm 1/2$. Two protons (neutrons) thus are allowed to 
become paired to total $J,T=0,1$ and $T_z=-1$ ($T_z=1$). We shall 
call this isovector pairing. Isoscalar pairing delineates proton-neutron 
pairing for which $J,T = 1,0$ and $T_z=0$. 

While we concentrate here on shell model results, 
we do wish to point out to the reader several recent developments
in other model studies of nucleon-nucleon pairing.
Interesting studies of nucleon-nucleon pairing have been undertaken
in several models, including pseudo-$SU(4)$ symmetry studies for 
$pf$-shell nuclei \cite{isacker99}. These studies indicated that 
pseudo-$SU(4)$ is a reasonable starting point 
for the description of systems within the 
$pf$-shell larger than $^{56}$Ni. It is also the 
starting point for generating collective pairs within the framework 
of the Interacting Boson Model  that incorporates $T=0$ and 
$T=1$ bosons and a bosonic $SU(4)$ algebra \cite{elliott58}.
This symmetry dictates that pairing strengths are the 
same in the both the $T=0$ and $T=1$ channels. Extensive studies 
of pairing in the framework of Hartree-Fock-Bogoliubov theory have
also been undertaken (see, for example \cite{goodman00}). Recent 
work in this direction indicates that $T=0$ and $T=1$ pairing 
superfluids may develop near the mid-point of isotope chains (i.e., 
near $N=Z$ nuclei). 

In the framework of the shell model,
it appears sufficient for many purposes to study the BCS pair structure
in the ground state. 
The BCS pair operator for protons can be defined as
\begin{equation}
\hat{\Delta}^\dagger_p= \sum_{jm > 0} p^\dagger_{jm} p^\dagger_{j\bar m} \;,
\label{eqn_7}
\end{equation}
where the sum is over all orbitals with $m>0$ and
$p^{\dagger}_{j\bar{m}}=(-)^{j-m}p^{\dagger}_{jm}$ is the time-reversed
operator.
Thus, the observable
$\hat{\Delta}^\dagger \hat{\Delta}$
and its analog for neutrons are measures of the numbers of $J=0,T=1$ pairs in
the ground state. For an uncorrelated Fermi gas, we have simply
\begin{equation}
\langle \hat{\Delta}^\dagger \hat{\Delta} \rangle
= \sum_j {n_j^2 \over {2 (2j+1)}},
\label{eqn_8}
\end{equation}
where the $n_j= \langle p^\dagger_{jm} p_{jm} \rangle$
are the occupation numbers, so that any
excess of $\langle \hat \Delta^\dagger \hat \Delta \rangle$
in our SMMC calculations
over the Fermi-gas value indicates pairing correlations in
the ground state.

\begin{figure}
\includegraphics[scale=0.5,angle=0]{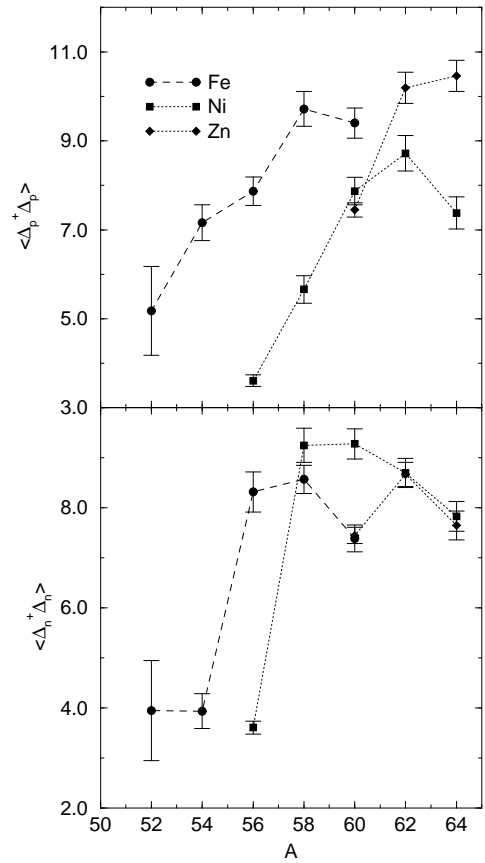}
\caption{SMMC expectation values of proton and neutron BCS-like pairs after
subtraction of the Fermi gas value. Taken from  \protect\cite{Langanke95}.}
\label{bcs_pairs}
\end{figure}

\begin{figure}
\includegraphics[scale=0.5,angle=0]{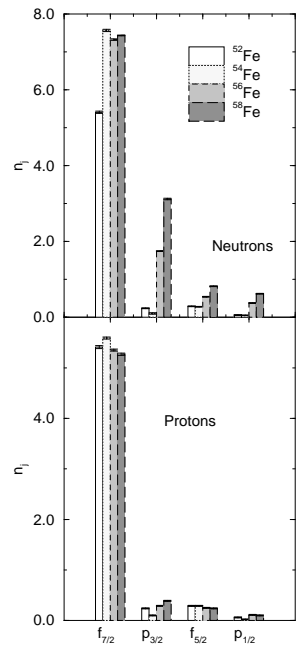}
\caption{Proton and neutron occupation numbers for various chains of isotopes.
Taken from  \protect\cite{Langanke95}.}
\label{fp_occs}
\end{figure}

In this analysis, we move from tin isotopes to nuclei which can
be described by the single-particle orbits of the $pf$-shell,
$1p_{3/2}$, $1p_{1/2}$, $0f_{5/2}$, and $0f_{7/2}$.
As effective interaction, we employ the phenomenological interaction of 
Brown and Richter \cite{Richter}.
Fig.~\ref{bcs_pairs} 
shows the SMMC expectation values of the proton and neutron
BCS-like pairs,
obtained after subtraction of the Fermi gas value (Eq.~\ref{eqn_8}), for three
chains of isotopes. As expected, these excess pair correlations are quite
strong and reflect the well-known coherence in the ground states
of even-even nuclei. Note that the proton BCS-like pairing fields are not
constant within an isotope chain, showing that there are important
proton-neutron correlations present in the ground state. The shell closure
at $N=28$ is manifest in the neutron BCS-like pairing. As is demonstrated
in Fig.~\ref{fp_occs}, the proton and neutron occupation numbers show a
much smoother behavior
with increasing $A$.

It should be noted that the BCS form (Eq.~\ref{eqn_7}) in which all
time-reversed pairs have equal amplitude is not necessarily the optimal one
and allows only the study of S-pair structure.
To explore the pair content of the ground state in 
a more general way \cite{gsoft96,Langanke95b},
we define proton pair
creation operators
\begin{equation}\hat{A}^\dagger_{J\mu}(j_aj_b)=
{1\over\sqrt{1+\delta_{ab}}}
[a^\dagger_{j_a}\times a^\dagger_{j_b}]_{J\mu}\;.
\end{equation}
These operators are boson-like in the sense that
\begin{equation}[\hat{A}^\dagger_{J\mu}
(j_aj_b),
\hat{A}_{J\mu} (j_aj_b)]=1
+{\cal O}(\hat n/2j+1)\;;
\end{equation}
i.e., they satisfy the expected commutation relations in the limit of an
empty shell. We may also construct from these operators a pair matrix
\begin{equation}
M^J_{\alpha\alpha'}=\frac{1}{\sqrt{2(1+\delta_{j_aj_b})}}\sum_M\langle
A^\dagger_{JM}(j_a,j_b)A_{JM}(j_c,j_d)\rangle
\label{pair_matrix}
\end{equation}

We construct bosons $\hat{B}^\dagger_{\alpha J\mu}$ as
\begin{equation}\hat{B}^\dagger_{\alpha J\mu}=
\sum_{j_aj_b}\psi_{\alpha\lambda}(j_aj_b)
\hat{A}^\dagger_{\lambda\mu}(j_aj_b)\;,
\end{equation}
where $\alpha=1,2,\ldots$ labels the particular boson and the ``wave
function'' $\psi$ satisfies
\begin{equation}\sum_{j_aj_b} \psi^\ast_{\alpha J}(j_aj_b)
\psi_{\beta J}(j_aj_b)=\delta_{\alpha\beta}\;.
\end{equation}
(Note that $\psi$ is independent of $\mu$ by rotational invariance.)

To find $\psi$ and \hbox{$n_{\alpha J}\equiv\sum_\mu\langle
\hat{B}^\dagger_{\alpha J\mu}
\hat{B}_{\alpha J\mu}\rangle$}, the number of bosons of type
$\alpha$, and multipolarity $J$, we compute the quantity
$\sum_\mu\langle \hat{A}^\dagger_{J\mu}(j_aj_b)\hat{A}_{J\mu} (j_cj_d)\rangle$,
which can be thought of as an hermitian matrix $M^J_{\alpha\alpha'}$
in the space of orbital pairs
$(j_aj_b)$; its non-negative eigenvalues define the $n_{\alpha J}$ (we
order them so that $n_{1 J}> n_{2 J}> \ldots$), while the
normalized eigenvectors are the $\psi_{\alpha J}(j_aj_b)$.
The index $\alpha$ distinguishes the various possible bosons.
For example, in the complete $pf$-shell the square matrix $M$ has
dimension $N_J=4$ for $J=0$, $N_J=10$ for $J=1$, $N_J=13$ for $J=2,3$.

\begin{figure}
\includegraphics[scale=0.5,angle=0]{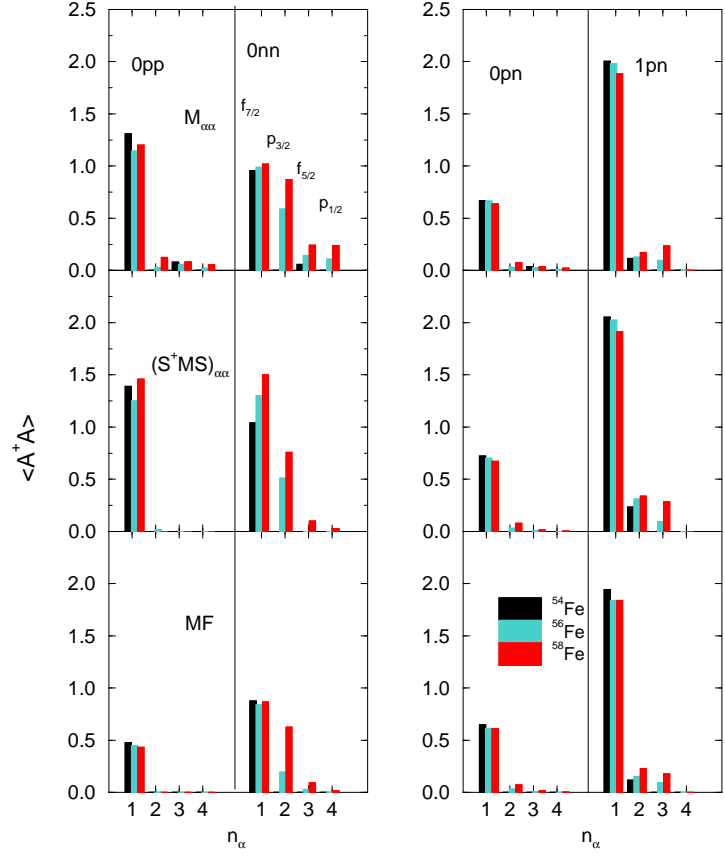}
\caption{Content of isovector $0^+$ pairs and isoscalar $1^+$ pairs
in the ground states of the isotopes
$^{54-58}$Fe. The upper panel shows the
diagonal matrix elements of the pair matrix $M_{\alpha\alpha}$.
The index $\alpha=1,..,4$ refers to $0^+$ pairs in the
$f_{7/2}$, $p_{3/2}$, $f_{5/2}$, and $p_{1/2}$ orbitals, respectively.
For the isoscalar pairs $\alpha=1,2,3$ refers to $(f_{7/2})^2$,
$(f_{7/2}f_{5/2})$, and $(f_{5/2})^2$ pairs, respectively.
The middle panel gives the eigenvalues of the pair matrix; for the
isoscalar pairs, only the 3 largest are shown. The lower panel
gives the eigenvalues of the pair matrix for the uncorrelated Fermi gas case
using Eq.~(\ref{equation_13}). Taken from  \protect\cite{Langanke95b}.\label{fig_88}}

\end{figure}

The presence of a pair condensate in a correlated ground state will be
signaled by the largest eigenvalue for a given $J$, $n_{1 J}$,
being much greater than any of the others; $\psi_{1J}$ will then be the
condensate wavefunction.
In Fig.~\ref{fig_88} we show the pair matrix eigenvalues $n_{\alpha J}$
for the three isovector $J=0^+$ pairing channels as calculated for the iron
isotopes $^{54-58}$Fe. We compare the SMMC results with those of an
uncorrelated Fermi gas, where
we can compute $\langle \hat{A}^\dagger \hat{A}\rangle$ using the
factorization
\begin{equation}\langle a^\dagger_\alpha a^\dagger_\beta a_\gamma
a_\delta\rangle=
n_\beta n_\alpha (\delta_{\beta\gamma} \delta_{\alpha\delta}-
\delta_{\beta\delta}\delta_{\alpha\gamma})\;,
\label{equation_13}
\end{equation}
where the $n_\beta=\langle a^\dagger_\beta a_\beta\rangle$ are the occupation
numbers. Additionally, Fig.~\ref{fig_88} 
shows the diagonal matrix elements of the
pair matrix $M_{\alpha\alpha}$. As expected, the protons occupy mainly
$f_{7/2}$ orbitals in these nuclei. Correspondingly,
the $\langle \hat{A}^\dagger \hat{A} \rangle$
expectation value is large for this orbital and small otherwise.
For neutrons, the pair matrix is also largest for the $f_{7/2}$ orbital.
The excess neutrons in $^{56,58}$Fe occupy the $p_{3/2}$ orbital,
signaled by a strong increase of the corresponding pair matrix element
$M_{22}$ in comparison to its value for $^{54}$Fe. Upon closer inspection,
we find that the proton pair matrix elements are not constant within the
isotope chain. This behavior is mainly caused by the isoscalar proton-neutron,
pairing. The dominant role is played by the isoscalar $1^+$ channel, which
couples protons and neutrons in the same orbitals and in spin-orbit partners.
As a consequence we find that, for $^{54,56}$Fe, the proton pair matrix in the
$f_{5/2}$ orbital, $M_{33}$, is larger than in the $p_{3/2}$ orbital,
although the latter is favored in energy. For $^{58}$Fe, this ordering
is inverted, as
the increasing number of neutrons in the $p_{3/2}$ orbital
increases the proton pairing in that orbital.

After diagonalization of $M$, the $J=0$
proton pairing strength is essentially found
in one large eigenvalue. Furthermore, we observe that this eigenvalue is
significantly larger than the largest eigenvalue on the mean-field level
(Fermi gas), supporting the existence of a proton pair condensate in the
ground state of these nuclei.
The situation is somewhat different for neutrons.
For $^{54}$Fe, only little additional coherence is found beyond the
mean-field value, reflecting the closed-subshell neutron structure.
For the two other isotopes, the neutron pairing exhibits
two large eigenvalues. Although the larger one exceeds the mean-field
value and signals noticeable additional coherence across the subshells,
the existence of a second coherent eigenvalue shows the shortcomings of the
BCS-like pairing picture.

\begin{figure}
\includegraphics[scale=0.5,angle=0]{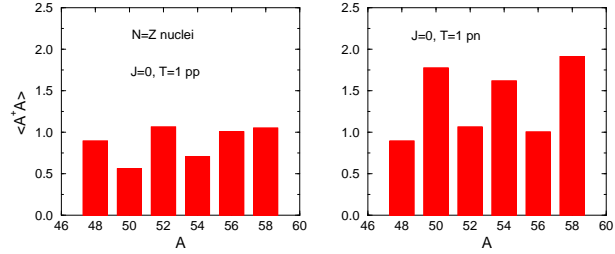}
\caption{Largest eigenvalue for the various isovector $0^+$ pairs
in the $N=Z$ nuclei in the mass region $A=48-56$. 
Taken from \cite{langanke97}.}
\label{fig_89}
\end{figure}

It has long been anticipated that $J=0^+$ proton-neutron correlations
play an important role in the ground states of $N=Z$ nuclei. To explore
these correlations, we have performed SMMC calculations of the $N=Z$ nuclei
in the mass region $A=48-56$ \cite{langanke97}. 
Note that for these nuclei the pair matrix
in all three isovector $0^+$ channels
essentially exhibits only one large eigenvalue, related to the
$f_{7/2}$ orbital. We will use this eigenvalue as a convenient measure of the
pairing strength. As the even-even $N=Z$ nuclei have isospin $T=0$,
the expectation values of ${\hat A}^\dagger {\hat A}$
are identical in all three isovector
$0^+$ pairing channels. This symmetry does not hold for the odd-odd $N=Z$
nuclei in this mass range, which have $T=1$ ground states, and
$\langle \hat{A}^\dagger \hat{A} \rangle$
can be different for proton-neutron pairs
than for like-nucleons pair (the expectation values for proton pairs
and neutron
pairs are identical). We find the proton-neutron pairing strength
significantly larger for odd-odd $N=Z$ nuclei than in even-even nuclei,
while the $0^+$ proton and neutron pairing shows the opposite behavior,
in both cases leading to an odd-even staggering, as displayed in
Fig.~\ref{fig_89}. This staggering is caused by a constructive interference
of the isotensor and isoscalar parts of $\hat{A}^\dagger \hat{A}$
in the odd-odd
$N=Z$ nuclei, while they interfere destructively in the even-even nuclei.
The isoscalar part is related to
the pairing energy, and is found to be about constant for the nuclei
studied here. Similar behavior was also demonstrated in a simplified $SO(5)$
seniority-like model \cite{engel96,engel1998}. This model is analytic, but shows the
same trends as the shell model results. Due to other correlations present
in the shell model, such as the inclusion of several orbits, isoscalar 
pairing, spin-orbit splitting, long-range correlations, deformation, 
etc., the shell model results are reduced in comparison to the simplified
model. 
Pairing correlations have also
been studied in heavier systems that require the
presence of the $0g_{9/2}$ orbital \cite{dean97,petrovici00}. 
In heavier odd-odd $N=Z$ nuclei the ground state becomes a $T=1$
(rather than $T=0$), as was found experimentally in $^{74}$Rb 
\cite{rudolph96}. The lowest $T=0$ and $T=1$ states in these systems
are very close in energy.
Recently, mean-field calculations that include both
$T=0$ and $T=1$ pairing correlations in odd-odd $N=Z$ nuclei 
\cite{satula01} showed that the interplay between quasiparticle 
excitations (relevant for the case of $T = 0$ states) and 
isorotations (relevant for the case of $T = 1$ states)
explains the near degeneracy of these states.

\subsection{Proton-neutron pairing and the Wigner energy}
\label{subsec:wigner}

So far, the strongest evidence for $np$ pairing comes from the
masses  of $N$=$Z$ nuclei. An additional binding (the so-called
Wigner energy) found in these nuclei  manifests itself as a
spike in the isobaric mass parabola as a function of
$T_z$=$\frac{1}{2}(N-Z)$ (see
the review \cite{Zel96} and references quoted therein).  
Gross estimates of the  magnitude of the Wigner energy
come from a large-scale fit to experimental binding energies
with the macroscopic-microscopic
approach \cite{Mye66,Kra79} and  from the analysis
of experimental masses \cite{Jen84}.
Several early attempts were made to incorporate the
effect of neutron-proton pair correlations in light 
nuclei in quasiparticle theory (for an early review see
\cite{goodman79}) with varying success. 
Satula {\em et al}, \cite{Sat97} presented  a
technique to extract the Wigner energy directly from the
experimental data and  gave empirical arguments that this
energy originates primarily from the $T$=0  part of the
effective interaction. To obtain deeper insight into the
structure of the Wigner term, they applied the
nuclear shell model to nuclei from the $sd$ and $fp$ shells.

The Wigner energy, 
$E_W$, is believed to 
represent the energy of collective
$np$-pairing correlations. It enters the
semi-empirical mass formula 
(see e.g. \cite{Kra79}) as an  
additional binding due to the 
$np$-pair correlations.
The Wigner energy can be decomposed into two parts:
\begin{equation}\label{EW1}
E_W = W(A)|N-Z| + d(A)\pi_{np}\delta_{NZ},
\end{equation}
The $|N-Z|$-dependence in Eq.~(\ref{EW1})
was first introduced by Wigner~\cite{Wig37} in
his analysis of the SU(4) spin-isospin symmetry of 
nuclear forces. In the supermultiplet approximation, there
appears
a term in the nuclear mass formula which is proportional to
$T_{\rm gs}(T_{\rm gs}+4)$, 
where $T_{\rm gs}$ denotes the isospin
of the ground state.
Empirically, $T_{\rm gs}$=$|T_z|$ for most nuclei except for
heavy odd-odd  $N$=$Z$ systems \cite{Jan65,Zel76}. 
Although the experimental
data indicate that the 
SU(4) symmetry is severely broken, 
and the masses behave according to the  
$T_{\rm gs}(T_{\rm gs}+1)$
dependence \cite{Jan65,Jen84}, 
the expression of Eq.~(\ref{EW1}) for the Wigner energy
is still very useful.
In particular, it accounts for a non-analytic behavior
of nuclear masses when an isobaric chain crosses the $N$=$Z$ 
line. An additional contribution to the Wigner term,
the $d$-term in  Eq.~(\ref{EW1}),
represents a 
correction for $N$=$Z$ odd-odd nuclei.
Theoretical justification of Eq.~(\ref{EW1}) has been given
in terms of basic properties of effective
shell-model interactions \cite{Zel96,Tal62}, and also by
using simple arguments based on the number of valence  
$np$-pairs \cite{Jen84,Mye77}. The estimates based on the 
number of $np$ pairs in identical spatial orbits suggest that
the ratio $d/W$ is constant and equal to one \cite{Mye77}.
A  different estimate has been given in 
 \cite{Jen84}:
$d/W$=0.56$\pm$0.27.

In the work of Satula \cite{Sat97},
the Wigner energy coefficient $W$ in an even-even nucleus
$Z$=$N$=$\frac{A}{2}$ was extracted by means of 
the indicator:
\begin{equation}\label{filtrw}
W(A) =   \delta V_{np}\left(\frac{A}{2},\frac{A}{2}\right)  -
{1\over 2} \left[ \delta V_{np}\left(\frac{A}{2},\frac{A}{2}-2
\right) +
\delta V_{np}\left(\frac{A}{2}+2,\frac{A}{2}\right) \right].
\end{equation}
The $d$-term  in
 an odd-odd  nucleus,
$Z$=$N$=$\frac{A}{2}$  [Eq.~(\ref{EW1})],
can be extracted using
another indicator:
\begin{equation}\label{filtrd}
d(A) =  
2\left[ \delta V_{np}\left(\frac{A}{2},\frac{A}{2}-2\right) +
\delta V_{np}\left(\frac{A}{2}+2,\frac{A}{2}\right) \right] -
4\delta V_{np}\left(\frac{A}{2}+1,\frac{A}{2}-1\right)\;,
\end{equation}
where the double-difference formula from \cite{Zha89} is
 \begin{eqnarray}\label{vnp}
 \delta V_{np} (N,Z) & = &
{1 \over 4} \left\{ B(N,Z) - B(N-2,Z) -B(N,Z-2)
 + B(N-2,Z-2)\right\}\nonumber \\
 & \approx & {\partial^2B\over{\partial N\partial Z}}.
 \end{eqnarray}
Although the recipe for
these third-order mass difference indicators
is not unique, the results appear to be very weakly dependent 
on the particular prescription used. 

\begin{figure}
\includegraphics[scale=0.45,angle=90,clip]{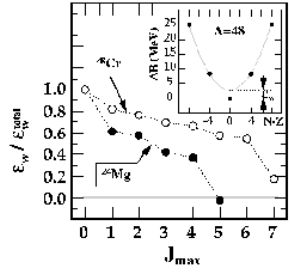}
\caption{Calculated displacement $\varepsilon_W$ of the binding energy of 
$^{24}$Mg and $^{48}$Cr
from the average parabolic $(N-Z)^2$
behavior along an isobaric chain.
 Shell-model calculations were performed in the 
$0\hbar\omega$ configuration  space.
The results of 
calculations for the binding energies
of even-even nuclei
along  the $A$=48 chain (normalized to  $^{48}$Cr)
are shown in the insert.
The values of  $\varepsilon_W$ 
were obtained using the shell-model Hamiltonian
with
the $J=1,2,..,J_{\rm max}, T=0$ matrix elements removed.
For instance, the result for $J_{\rm  max}$=3
corresponds to the variant of calculations in which
all the two-body  matrix elements between states
$|j_1j_2 JT$=$0\rangle$ with $J$=1,2,3, were put to zero.   
The results are normalized to  
the full shell-model value 
$\varepsilon_W^{\rm total}$ ($J_{\rm  max}$=0). 
The Coulomb contribution to the binding energy has been 
disregarded. Taken from  \protect\cite{Sat97}.
\label{satula_pn}}
\end{figure}

To visualize the influence of
the $T$=0 part of the effective nuclear
interaction on the Wigner term,
we have performed a set of shell-model
calculations while {\em switching off}
 sequentially the $J=1,
2,..., J_{\rm max}$,
$T$=0 two-body matrix elements $\langle 
j_1j_2JT|\hat{H}|j_1'j_2'JT\rangle$
of the shell-model Hamiltonian $\hat{H}$
for different values of $J_{\rm max}$.
Figure~\ref{satula_pn} shows a  ratio
$\varepsilon_W/\varepsilon_W^{\rm total}$,
where $\varepsilon_W^{\rm total}$ denotes the 
result of full shell-model
calculations versus  $J_{\rm max}$. The calculations were
performed for 
two representative examples, namely, the $fp$-shell
nucleus $^{48}$Cr and the $sd$-shell nucleus $^{24}$Mg.
The  largest contribution to the Wigner energy comes from
the part of the $T$=0 interaction between 
deuteron-like ($J$=1) and
`stretched' [$J$=5 ($sd$) and 7 ($pf$)] pairs.
The importance of these matrix elements
is well known; it is precisely for  $J$=1 and 
stretched  pair-states
that experimentally determined effective 
$np$  $T$=0 
interactions are strongest \cite{Ana71,Sch71,Mol75}. 
Note also that
the deuteron-like correlations contribute more strongly
to $\varepsilon_W$
in $sd$-nuclei than in $fp$-nuclei, and that matrix
elements corresponding to  intermediate values of $J$
give non-negligible contributions.
This reveals the complex structure of the Wigner energy and
suggests that models which ignore high-$J$ components of
the $np$ interaction (e.g., by considering only
$J$=0, $T$=1 and $J$=1, $T$=0 $np$ pairs \cite{Eva81})
are not too useful for discussing the
actual  $np$ pair
correlations.

%\vskip.5truecm
\noindent
\subsection{Thermal properties of $pf$-shell nuclei}
%\vskip.5truecm

The properties of nuclei at finite temperatures are of considerable
experimental (for reviews, see \cite{Suraud,Snover})
and theoretical interest \cite{Alhassid1,Egido}.
How thermal excitations influence the pairing properties will be
the main focus of this section. We address this topic in 
Sec.~\ref{sec:leveldensities_sec3} as well but with an emphasis on
an analysis based on experimental data on level densities.

Theoretical studies of nuclei at finite temperature have been based mainly
on mean-field approaches and thus only consider the temperature dependence
of the most probable configuration in a given system. These approaches
have been criticized due to their neglect of quantum and statistical
fluctuations \cite{Dukelsky}. The SMMC method does not suffer this defect and
allows the consideration of model spaces large enough to account
for the relevant nucleon-nucleon correlations at low and moderate
temperatures.

SMMC calculations were performed to study the 
thermal properties of several even-even  and odd-A nuclei in the mass 
region $A=50-60$ \cite{Dean95,Langanke95b} in an $fp$-shell model 
space using realistic interactions.
More recently, Alhassid {\em et al.}, carried out thermal 
calculations in a larger model-space which included the $0g_{9/2}$ shell.
As a typical example, we discuss
in the following our SMMC results for the nucleus $^{54}$Fe, which
is very abundant in the presupernova core of a massive star.

Our calculations include the complete set of $1p_{3/2,1/2}0f_{7/2,5/2}$
states interacting through the realistic Brown-Richter Hamiltonian
\cite{Richter}.
(SMMC calculations using the modified KB3 interaction \cite{kb3a,kb3b} 
give essentially
the same results.) 
Some
$5\times10^9$ configurations of the 8 valence neutrons and 6 valence protons
moving in these 20 orbitals are involved in the canonical ensemble. 
The
results presented below have been obtained with a
time step of $\Delta\beta=1/32~{\rm MeV}^{-1}$ using 5000--9000 independent
Monte Carlo samples.

\begin{figure}
\includegraphics[scale=0.5,angle=0]{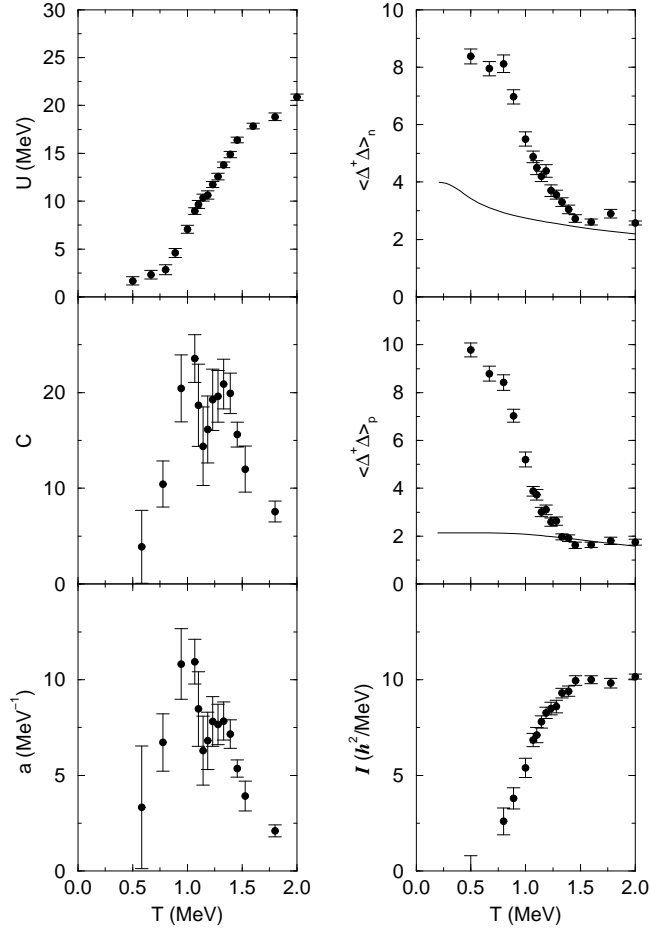}
\caption{Temperature dependence of various observables in ${}^{54}$Fe. Monte
Carlo points with statistical errors are shown at each temperature $T$. In
the left-hand column, the internal energy, $U$,
is calculated as $\langle {\hat H}
\rangle -E_0$, where ${\hat H}$ is the many-body Hamiltonian and $E_0$ the
ground state energy. The heat capacity $C$ is calculated by a finite-difference
approximation to $dU/dT$, after $U(T)$ has been subjected to a three-point
smoothing, and the level density parameter is $a\equiv C/2T$.
In the right-hand column, we show the expectation values of the
squares of the proton and neutron BCS pairing fields.
For comparison, the pairing fields
calculated in an uncorrelated Fermi gas are shown by the solid curve. The
moment of inertia is obtained from the expectation values of the square of
the total angular momentum by $I=\beta \langle {\hat J}^2 \rangle/3$.
Taken from  \protect\cite{Dean95}. \label{fig_810}}
\end{figure}

The calculated temperature dependence of various observables is shown in
Fig.~\ref{fig_810}. 
In accord with general thermodynamic principles, the internal energy
$U$ steadily increases with increasing temperature \cite{Dean95}. It shows an
inflection point around $T \approx 1.1$~MeV, leading to a peak in the heat
capacity, $C\equiv dU/dT$, whose physical origin we will discuss below. The
decrease in $C$ for $T \gtrsim 1.4$~MeV is due to our finite model space (the
Schottky effect \cite{Schottky});
we estimate that limitation of the model space to only
the $pf$-shell renders our calculations of ${}^{54}$Fe quantitatively
unreliable for temperatures above this value (internal energies $U\gtrsim
15$~MeV). The same behavior is apparent in the level density parameter,
$a\equiv C/2T$. The empirical value for $a$ is $A/8~{\rm MeV} =6.8~{\rm
MeV}^{-1}$, which is in good agreement with our results for $T \approx
1.1$--1.5~MeV.

\begin{figure}
\includegraphics[scale=0.5,angle=0]{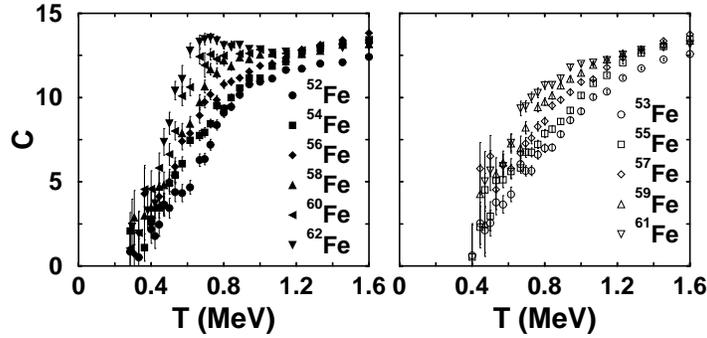}
\caption{The heat capacity of even-even (left panel) and
odd-even (right panel) iron isotopes. Taken from  \protect\cite{liu01}.
\label{fig:figyoram}}
\end{figure}

More recent calculations confirm these basic findings. Liu and Alhassid
calculated \cite{liu01} the heat capacity for iron isotopes in a complete $0f1p-g_{9/2}$
model space. They used a phenomenological pairing-plus-quadrupole model 
for the two-body interaction and found that the pairing transition
in the heat capacity is correlated with the suppression of the number
of spin-zero neutron pairs as the temperature increases. The results were
obtained using a novel method to calculate the heat capacity that decrease
the statistical error bars in the calculation. We show results of this
calculation for Fe isotopes in Fig.~\ref{fig:figyoram}. 
While the 
original calculations of \cite{Dean95} indicate a possible phase
transition (along with a pairing collapse in the measured 
$\langle \Delta^\dagger\Delta \rangle$ pairing expectation), this effect
appears to be delayed to more neutron-rich nuclei in the calculations of
\cite{liu01}. Several factors likely contribute to this difference. 
First, the  interactions are obviously different. Second, the 
extrapolation techniques used for realistic interactions likely 
over-estimate the influence of pairing in the region between $0.5$ 
and 1.0~MeV.  Finally, the model space is smaller in the early calculation,
although the Schottky peak is seen to appear at about 1.4 MeV. This 
makes the interpretation of the low temperature peak more difficult 
in \cite{Dean95}.
Nevertheless, it should be clear that both the original calculations 
with realistic interactions and the more recent work in \cite{liu01}
both indicate interesting physics related to pairing 
phenomena in the $T=1.0$~MeV region.

We also show in Fig.~\ref{fig_810} the expectation values of the BCS-like
proton-proton and neutron-neutron pairing fields,
$\langle \hat \Delta^\dagger
\hat \Delta\rangle$.
At low temperatures, the pairing fields are significantly
larger than those calculated for a non-interacting Fermi gas, indicating a
strong coherence in the ground state. With increasing temperature, the
pairing fields decrease, and both approach the Fermi gas values for $T\approx
1.5$~MeV and follow it closely for even higher temperatures. Associated with
the breaking of pairs is a dramatic increase in the moment of inertia,
$I\equiv \langle J^2\rangle/3T$,
for $T=1.0$--1.5~MeV; this is analogous to the rapid increase in magnetic
susceptibility in a superconductor. At temperatures above 1.5~MeV, $I$ is in
agreement with the rigid rotor value, $10.7\hbar^2$/MeV; at even higher
temperatures, it decreases linearly due to our finite model space.

\begin{figure}
\includegraphics[scale=0.5,angle=270]{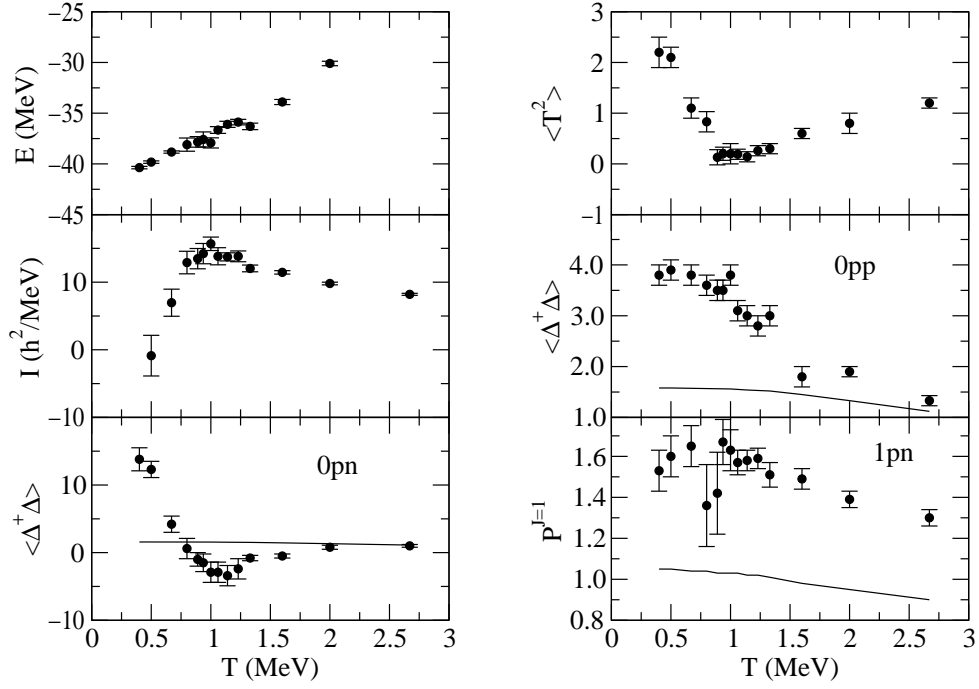}
\caption{Temperature dependence of various observables in $^{50}$Mn.
The left panels show (from top to bottom) the total energy,
the moment of inertia, and the proton-neutron BCS pairing fields,
while the right panels exhibit
the expectation values of the isospin operator
$\langle \hat T^2 \rangle$, the isovector $J=0$ proton-proton 
BCS pairing fields, and the isoscalar $J=1$ pairing
strength, as defined in the text. For comparison,
the solid lines indicate the
Fermi gas values of the BCS pairing fields and $J=1$ pairing strength.
}
\label{fig_812}
\end{figure}

Although the results discussed above are typical for even-even nuclei
in this mass region
(including the $N=Z$ nucleus $^{52}$Fe), they are not for odd-odd $N=Z$
nuclei. This is illustrated in Fig.~\ref{fig_812} which shows the thermal
behavior of several observables for $^{50}$Mn ($N=Z=25$), calculated
in a SMMC study within the complete $pf$-shell using the KB3 interaction
\cite{kb3a,kb3b}.
A closer inspection of the isovector
$J=0$ and isoscalar $J=1$ pairing correlations holds the key to the
understanding of these differences. The $J=0$ isovector correlations
are studied using the BCS pair operators, Eq.~(\ref{eqn_7}) 
with a similar definition
for proton-neutron pairing. For the isoscalar $J=1$ correlations, we have
interpreted the trace of the pair matrix $M^{J=1}$ 
(defined in Eq.~ (\ref{pair_matrix})) as an overall
measure for the pairing strength,
\begin{equation}
P_{sm}^J = \sum_{\beta} \lambda_{\beta}^J = \sum_\alpha M^J_{\alpha\alpha}.
\label{pair_cor_mat}
\end{equation}

Note that at the level of the non-interacting Fermi gas, proton-proton,
neutron-neutron, and proton-neutron $J=0$ correlations are identical
for $N=Z$ nuclei. However, the residual interaction breaks the symmetry
between like-pair correlations and proton-neutron correlations
in odd-odd $N=Z$ nuclei. As is obvious from Fig.~\ref{fig_812}, 
at low temperatures
proton-neutron pairing dominates in $^{50}$Mn, while pairing among
like nucleons shows only a small excess over the Fermi gas values, in
strong contrast to even-even nuclei.

A striking feature of Fig.~\ref{fig_812} is that the isovector
proton-neutron correlations decrease strongly with temperature
and have essentially vanished
at $T=1$ MeV, while the isoscalar pairing strength
remains about constant in this temperature region
(as it does in even-even nuclei) and greatly exceeds the Fermi gas values.
We also note that the pairing between like
nucleons is roughly constant at $T<1$ MeV.
The change of importance between isovector and isoscalar proton-neutron
correlations with temperature is nicely reflected in the isospin expectation
value, which decreases from $<\hat T^2>=2$ at temperatures around 0.5 MeV,
corresponding to the dominance
of isovector correlations, to $<\hat T^2>=0$ at temperature $T=1$ MeV, when
isoscalar proton-neutron correlations are most important.

The temperature dependence of the excitation energy $E=\langle H\rangle$
in the odd-odd nucleus $^{50}$Mn is significantly different than that in 
even-even nuclei. The difference is due to the uniqueness of isospin
properties in odd-odd $N=Z$ nuclei. It is only in these nuclei that one 
finds states of different isospin, $T=1$ and $T=0$, that are close to
each other at low excitation energies. The $^{50}$Mn ground state
is $T=1$, $T_z=0$, and $J^\pi=0^+$. In that state $pn$ pair correlations 
dominate, and the like-particle correlations are reduced \cite{langanke97}.
However, at relatively low excitation energy these nuclei exhibit a 
multiplet of $T=0$ states with nonvanishing angular momenta. These 
states contribute efficiently to corresponding thermal averages. 
On the other hand, it follows from isospin symmetry that in the $T=0$
states all three pairing strengths (in $T_Z$) must be equal. Thus, at
temperatures where the $T=0$ states dominate the thermal average, the $pn$
pair correlations are substantially reduced when compared to ground
state values.  This argument appears to be a generic feature
of odd-odd $N=Z$ nuclei beyond $^{40}$Ca. For a further discussion  
see \cite{zheng97}.  For a different point of view from the 
perspective of mean-field calculations, see, for example \cite{ropke00}.

%\vskip.5truecm
\noindent
\subsection{Pair correlations and thermal response}
%\vskip.5truecm

All SMMC calculations of even-even nuclei
in the mass region $A=50-60$ show
that the BCS-like pairs
break at temperatures around 1 MeV.
Three observables exhibit a particularly interesting behavior
at this phase transition: a) the moment of inertia rises sharply; b) the
M1 strength shows a sharp rise, but unquenches only partially; and c)
the Gamow-Teller strength remains roughly constant (and strongly quenched).
Note that
the $B(M1)$ and $B(GT_+)$ strengths unquench at temperatures larger than
$\approx 2.6$ MeV and
in the high-temperature limit
approach the appropriate values for our adopted model space.

\cite{Langanke95b}
has studied the pair correlations in the four nuclei
$^{54-58}$Fe and $^{56}$Cr for the various
isovector and isoscalar pairs up to $J=4$,
tentatively interpreting
the sum
of the eigenvalues of the matrix $M^J$ \ref{pair_matrix}
as an overall measure for the pairing strength.
Note that
the pairing strength, as defined in \ref{pair_cor_mat},
is non-zero at the
mean-field level. The physically relevant pair correlations
$P_{\rm corr}^J$ are then defined as the difference of the SMMC
and mean-field pairing strengths.

\begin{figure}
\includegraphics[scale=0.5,angle=0]{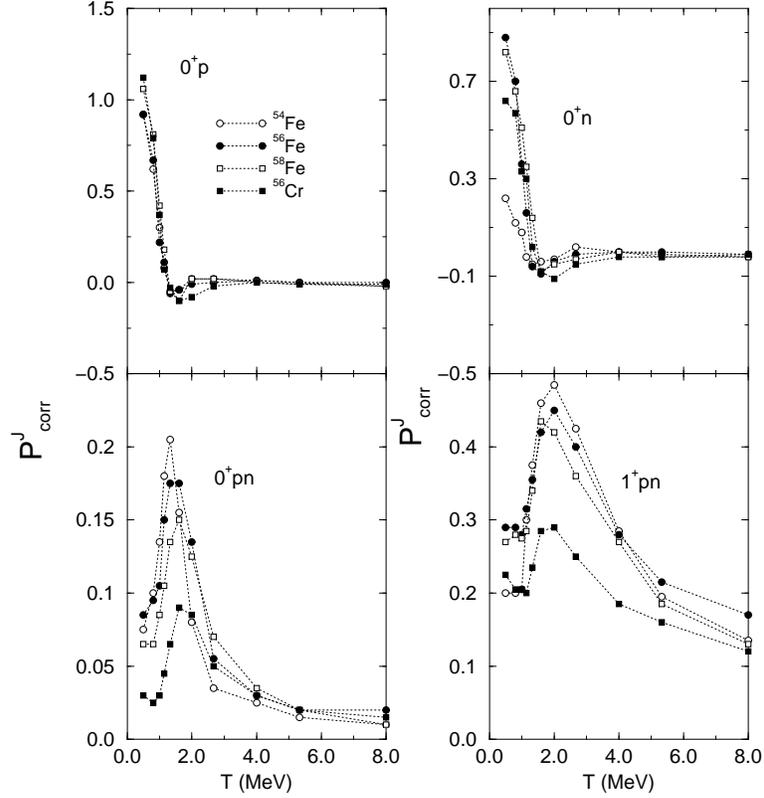}
\caption{Pair correlations, as defined in Eq. \ref{pair_cor_mat},
for isovector $0^+$ and isoscalar $1^+$ pairs for $^{54-58}$Fe
and $^{56}$Cr, as functions of temperature. Taken from  \protect\cite{Langanke95b}.
}
\label{fig_813}
\end{figure}

Detailed calculations of the pair correlations have been
performed for selected temperatures in the region between $T=0.5$ MeV and 8
MeV.
Fig.~\ref{fig_813} shows the
temperature dependence of those pair correlations
that
play an important role in
understanding the thermal behavior of the moment of inertia and the
total $M1$ and Gamow-Teller strengths.

The most interesting behavior is found in the $J=0$ proton and neutron pairs.
There is a large excess of this pairing at low
temperatures, reflecting the ground state coherence of even-even nuclei.
With increasing temperature, this excess diminishes and vanishes at
around $T=1.2$ MeV.
We observe further from Fig.~\ref{fig_813} that the temperature dependence
of the $J=0$ proton-pair correlations are remarkably independent
of the nucleus, while the neutron pair correlations show interesting
differences. At first, the correlation excess is smaller in the semimagic
nucleus $^{54}$Fe than in the others. When comparing the iron isotopes, the
vanishing of the neutron $J=0$ correlations occurs at higher temperatures
with increasing neutron number.

The vanishing of the $J=0$ proton and neutron pair correlations is
accompanied by an  increase in the correlations of the other pairs.
For example, the isovector $J=0$ proton-neutron correlations
increase by about a factor of 3 after the $J=0$
proton and neutron pairs have vanished. The correlation peak is reached
at higher temperatures with increasing neutron number,
while the
peak height decreases with neutron excess.

The isoscalar proton-neutron $J=1$ pairs show an interesting temperature
dependence. At low temperatures, when the nucleus is still dominated
by the $J=0$ proton and neutron pairs, the isoscalar proton-neutron
correlations show a
noticeable excess but, more interestingly, they are roughly constant and
do not directly reflect  the vanishing of the $J=0$ proton and neutron pairs.
However, at $T>1$ MeV, where the $J=0$ proton- and neutron-pairs have broken,
the isoscalar
$J=1$ pair correlations significantly increase and have their maximum
at around 2 MeV, with peak values of about twice the correlation excess
in the ground state.
In contrast to the isovector $J=0$ proton-neutron pairs, the correlation peaks
occur at lower temperatures with increasing neutron excess. We also
observe that these correlations fade rather slowly with
increasing temperature.

A further discussion on thermal properties through recent experimental
information on level densities will be given in 
Sec.~\ref{sec:leveldensities_sec3}.

%%%  section 4
% october 2002 mhj included comments from Hubert

\section{RANDOM INTERACTIONS AND PAIRING}

\label{sec:randoms}

We have seen that all even-even
nuclei have a $J^{\pi}=0^+$ ground state. Pairing in
even-even systems was also shown in previous sections 
to be a major contributor to the ground-state correlations. 
Furthermore, a property like the $0^+$ nature of all even nuclei
can be explained within the simple seniority model, based itself 
on the short-range nature of the effective interaction. It is therefore 
interesting to see whether such a general property is specific of this
hamiltonian or whether it could also 
emerge from 
a random ensemble of rotationally and isospin invariant random two-body
interactions. This question was first posed in 
\cite{jbd98}, where the low-lying spectral properties
of random interactions were first studied from the shell-model
perspective. Several interesting results were obtained including
highly likely $0^+$ ground states emerging from the random ensembles,
an enhanced phonon collectivity, strongly correlated pairing 
phenomena \cite{jbdt00}, odd-even staggering \cite{papen02},
and a likelihood of generating rotational and vibrational spectra. 

Similar results
were also obtained in the interacting
boson model (IBM) \cite{bf00}. 
The IBM Hamiltonian used in these calculations is given by 
\begin{eqnarray}
H &=& \epsilon\hat{n}_d - \kappa\hat{Q}(\chi)\times \hat{Q}(\chi) \; \\
\hat{Q}_\mu(\chi)&=&\left(s^\dagger\tilde{d}+d^\dagger s\right)_\mu^{(2)}
+\chi \left(d^\dagger \tilde{d}\right)^{(2)}_\mu \;, 
\end{eqnarray}
where only spin 0 ($s$, monopole) and spin 2 ($d$, quadrupole) 
bosons are considered. This interaction is randomized by introducing
a scaling parameter $\eta = \epsilon/\left[\epsilon + 4\kappa(N-1)\right]$
and $\bar{\chi}=2\chi/\sqrt{7}$, and choosing $\bar{\chi}$ and $\eta$ 
randomly on the intervals $[-1,1]$ and $[0,1]$, respectively. 
The IBM calculations also 
gave a predominance of $0^+$ ground states as well as 
strong evidence for the occurence of both vibrational and
rotational band structures. Within the shell model, these structures
appear within a continuum of rotational bands, but the nature of the
IBM model restricts the structures to be of these two particular
forms. In this section we will briefly review the present 
status of research into this interesting phenomena. 

For fermions, we define the two-body matrix elements $V_{JT}(ab,cd)$ through
an ensemble of two-particle Hamiltonians and require that the
ensemble be invariant under changes in the basis of two-particle
states.  This is achieved by taking the matrix elements to be
Gaussian distributed about zero with the widths possibly depending
on $J$ and $T$ such that:
\begin{equation}
\matrix{
\displaystyle
  \langle V_{\alpha,\alpha'}^2\rangle
  = c_{J_\alpha,T_\alpha} ( 1+ \delta_{\alpha,\alpha'} )  \bar{v}^2,\hfill\cr
\noalign{\vskip 1.25em}
\displaystyle
  \langle V_{\alpha,\alpha'}V_{\beta,\beta'}\rangle =
  0, \,\,\,(\alpha,\alpha')\ne(\beta,\beta').\hfill\cr
}
\end{equation}
Here $\bar{v}$ is an overall energy scale that we generally ignore (except
for scaling single-particle energies for the RQE-SPE defined below).
The coefficients $c_{J}$ then define the ensemble.
We emphasize that $J,T$ refer to quantum
numbers of  {\it two-body} states and not of the final
many-body states (typically 4-10 valence particles).

Several basic ensembles may be defined by the choice of
the form of the $c_{J,T}$ coefficients
and the single-particle Hamiltonian, if present. One 
ensemble is called the
Random Quasiparticle Ensemble (RQE).
In this case \smash{$c_{J,T} =
\left[(2T+1)(2J+1)\right]^{-1}$}.
This relation between the $c_{J,T}$, which was discussed in \cite{jbd98},
came from imposing on the ensemble the constraint that
it should remain the same for the particle-particle
interaction as for the particle-hole interaction. A different
ensemble in this class is called the 
two-body random ensemble (TBRE) for which \smash{$c_{JT}=$} constant.
Historically, this was the
first two-particle random ensemble to be employed in studying statistical
properties of many-particle
spectra \cite{fr70}. These two ensembles assume degenerate
single-particle energies.
Realistic interactions do have nondegenerate single-particle
energies that will, in principle, affect various
spectral properties.  For calculations in the {\it sd} shell one
uses single-particle energies from the Wildenthal interaction
\cite{wildenthal}, scaling
$\bar{v}=3.84$ MeV
to best match the widths of the two-particle matrix elements.
The resulting interactions with the single-particle splitting included
are called the RQE-SPE and TBRE-SPE.

\begin{figure}
\begin{center}
\includegraphics[scale=0.55,angle=270]{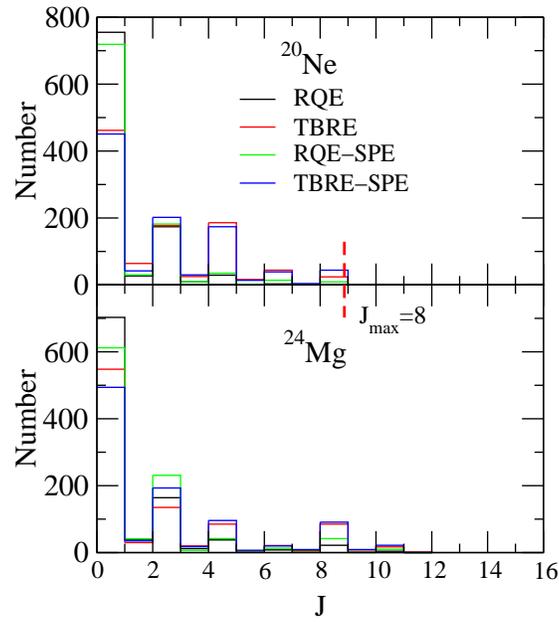}
\end{center}
\caption{The distribution of ground state spins in the various
random ensembles.}
\label{fig:j0s}
\end{figure}

The first, and perhaps most striking, feature of all of these 
random interactions is the preponderance of $J^\pi=0+$ ground states. 
In Fig.~\ref{fig:j0s} we show the distribution of ground-state spins
in the various ensembles for the two systems $^{20}$Ne and $^{24}$Mg. 
We generated 1000 random interactions from each ensemble. These results
are typical and consistent with calculations with only one type
of particle (for example, neutrons only), or fermions in which the 
$\vec{l}\cdot\vec{s}$ force is not present \cite{kaplan01}.

Fig.~\ref{fig:j0s} also shows that the even spins are preferred. 
In some cases, higher even spin states are preferred over medium
spin states. For example, in $^{24}$Mg, the $8^+$ state is preferred
over the $6^+$. The single-particle splitting tends to lower slightly
the number of $0^+$ ground states. The RQE clearly obtains the highest
number of $0^+$ ground states in each case. 

Various research efforts have been undertaken to 
understand the preponderance of the $0^+$ ground-state. 
Ensembles of interactions derived from a Gaussian unitary ensemble (GUE) 
distribution are not
time-reversal invariant, but both the Gaussian orthogonal ensemble (GOE) 
and GUE random ensembles 
yield $0^+$ dominance \cite{bijker99}. This apparent paradox was recently
resolved \cite{vz02} by noting that the $J^2$ operator commutes with 
the ${\cal T}$ time-reversal operator for either ensemble. 
For bosons, Kusnezov \cite{kus00} was able to map the $U(4)$ vibron
model onto random polynomials on the unit interval. 
Kusnezov was then able to show analytically the origin of $0^+$ ground
states. While the $U(4)$ model is extremely simplified and only describes
bosons (rather than fermions),  it points to the 
interesting link between random polynomials and the two-body interaction. 
The $0^+$ predominance in the fermion case was recently studied by 
Mulhall {\em et al} \cite{mulhall00}. These authors used  
a single $j$-shell to show that statistical 
correlations of fermions in a finite system with random interactions 
drive the ground state spin to its minimum or 
maximum value. The effect is universal and related to the 
geometric chaoticity, or the assumption of pseudorandom coupling 
of angular momentum \cite{volodya1996}, 
of the spin coupling of 
individual fermions. While a rigorous derivation of these findings for 
general orbital schemes is not yet available, the research is pointing towards
an understanding of why an ensemble of 
random interactions posesses predominantly $0^+$ ground states. 

A second feature concerns the likelihood of finding rotational or vibrational
spectra from the ensembles. The relevant measure for these states is the
ratio of the first $4^+$ excitation energy to the first $2^+$ energy. This
ratio, $\rho=[E(4^+)-E(0^+)]/[E(2^+)-E(0^+)]$ 
is 2 for a vibrational spectrum and $10/3$ for
a rotational one. The results from the RQE show a broad distribution
of various kinds of spectra peaked towards vibrational spectra.  
Using random interactions in the IBM,
virtually all random interactions yield a vibrational or rotational spectrum
in nearly equal proportions.  
The difference is due to the restricted nature of the random IBM interaction 
in which only $s$ and $d$ boson pairs and couplings are included in the 
Hamiltonian.  Kusnezov {\em et al.} \cite{kzc00} confronted the 
results of the IBM model with known experimental data 
and found two interesting results. They found
that both the interaction and the number of relevant valence nucleons
were key to understanding the distribution of $\rho_{\rm IBM}$. They also
found that experimental data favor rotational spectra over vibrational
spectra and were able to place limits on the choice of random variables
that would allow for a reproduction of experimental data. 

Of the three general properties of random interactions we discuss here, 
the enhancement of the $B(E2)$ strength is not spontaneously produced by
our choices of random interactions \cite{horoi01}. This is particularly
true for strongly deformed nuclei. 
The problem can be cured \cite{vz02} by
choosing a constant displacement of all the matrix elements which is 
essentially the same as adding some coherence to the choice of the random
Hamiltonian. Velazques and Zuker \cite{vz02} were able to do this and 
showed how one may obtain good rotational spectra from the displaced-TBRE. 

A third feature of the random shell-model interactions involves the pair 
content of the $0^+$ ground-state wave functions. The pairing content of
the wave function was measured by calculating the pair-transfer operator. 
Interestingly, for a given interaction, the same coherent pair connected
several even-even nuclei in a given isotopic chain. This feature appears
to be robust.  On the other hand, studies \cite{mulhall00} 
made in a single $j$ shell relate the origin of regularities in 
the spectra to incoherent interactions
rather than to coherent pairing. The origin of order in the spectra,
attributed to geometric chaocity,
tends to prove that the role of pairing is minimal.
In order to better understand these two seemingly conflicting
observations, a further analysis of the pairing properties of 
the system has been performed by Bennaceur {\em et al.} \cite{ben02}
who compared shell-model results to those obtained 
from Hartree-Fock-Bogoliubov (HFB) calculations using the same set of
random interactions. 
The aim of this study was to determine whether
the interactions support static pairing or whether the effect is 
more dynamical.  Because the HFB solution generally breaks all the 
symmetries required by the many-body Hamiltonian, it is not a 
physical state, but an indicator of the intrinsic structure in 
the many-body system.  

In the HFB approach
one does not explore the full Hilbert space; the trial function
is constrained to be a superposition of pairs of single particles.
Moreover, the Hamiltonian of \cite{ben02} omits
the terms which represent the residual interaction between quasiparticles, 
and proton-neutron pairing is not taken into account.
Finally, the HFB approach is an unprojected
variational method, so $J$ is not a good quantum number
and neither are the particle numbers $N$ and $Z$.
In the shell model the particle numbers are well defined, while in HFB
approximation, only the average particle numbers are constrained.

The HFB approximation describes
only {\it static} pairing. On the contrary, in the SM model picture, the
wave functions contain all the possible correlations inside a given model
space. For that reason, a shell-model ground-state 
wave function can show some strong pairing
properties while the corresponding HFB solution can be totally unpaired.
In that case the pair structure can be due to {\it dynamical} pairing
which cannot be described in the HFB method but require going beyond
the mean field.

The three 
systems $^{24}$Mg, $^{22}$Ne, and $^{20}$O were considered. 
Only those interactions generated from the RQE-SPE ensemble
that lead to a SM ground state with $J^\pi=0^+$ were used,
and 
the pairing properties of the ground state
wave functions were investigated. The simultaneous use of the shell model 
and the HFB approximation gives a better understanding of the pairing
induced by the random interactions.

In order to investigate the pairing properties of the shell-model
solutions, one introduces the pair transfer coefficient
\begin{equation}
<P^+>\,\equiv\,<A|P^+_{00}|A-2>
=\sum_{\alpha}<A|\big[a_\alpha^+a_{\bar\alpha}^+\big]^0_0|A-2>\;,
\end{equation}
where $|A>$ and $|A-2>$ represent the ground states of the $A$ and $A-2$
particle systems obtained from the shell model (isospin $T=1$ is understood).
This quantity is compared
to the mean pairing strength in the HFB approximation $<\kappa>$ defined
by
\begin{equation}
<\!\kappa\!>={\rm Tr}\;\kappa\;.
\end{equation}
The three systems considered correspond
in the $(sd)$ shell to $N-Z=0$, $N-Z=2$, and
$N-Z=4$. 

If the dominance of $J^\pi=0^+$ ground state is due to the pairing
properties of the system, then we expect to obtain a significant value
for $<\!\kappa\!>$ in most of the cases. Moreover if the pairing plays
an important role for the structure of the ground state, then it must
be related with the pair transfer coefficient, and in that case one can
also expect a clear correlation between $<\!\kappa\!>$ and $<\!P^+\!>$.

\begin{figure}
\begin{center}
\includegraphics[scale=0.7,angle=0]{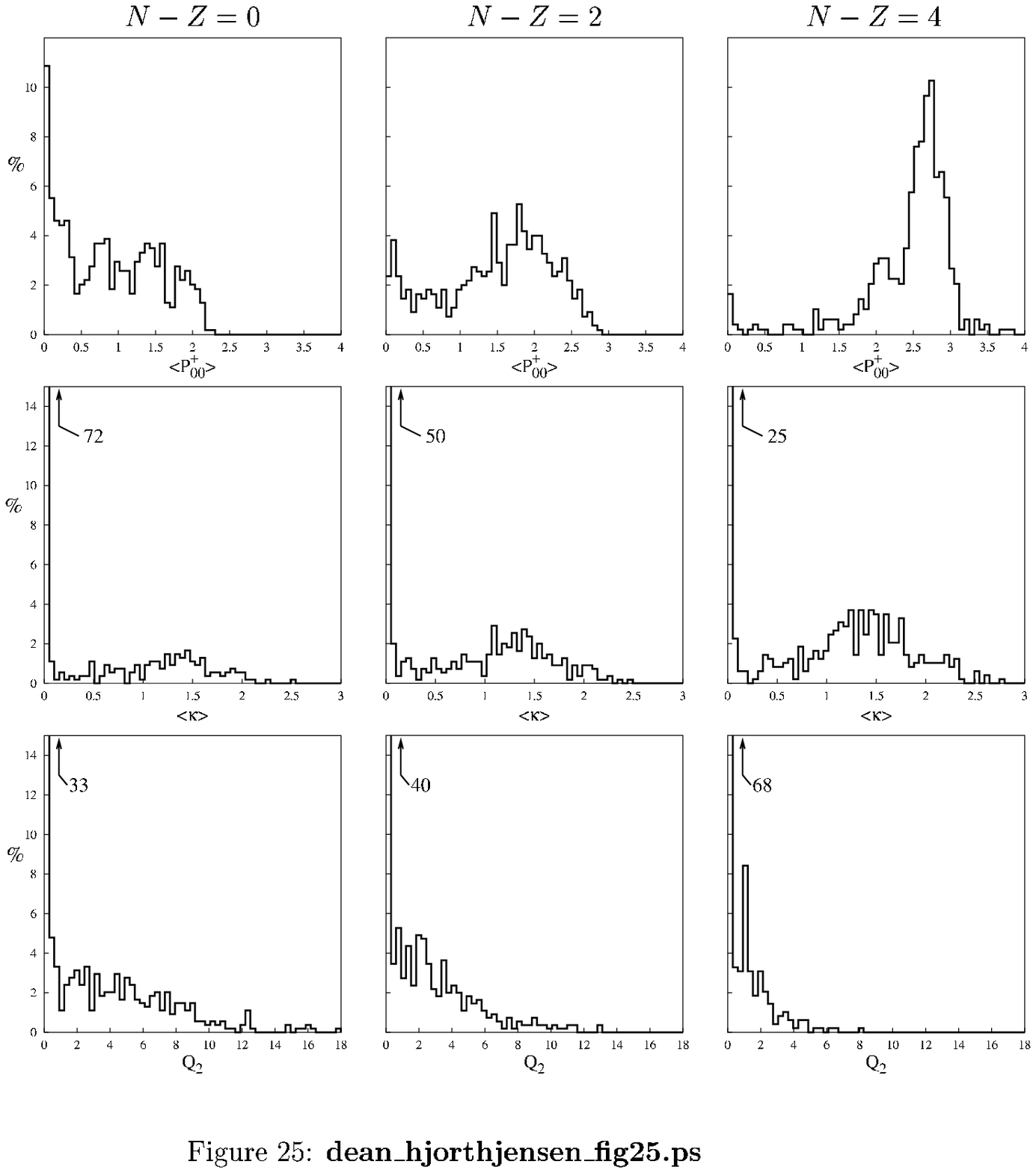}
\end{center}
%\begin{center}
%$\matrix{
%  \ \ \ \ \ \ N-Z=0 & \!N-Z=2 & N-Z=4\ \cr
%  \!\includegraphics*[scale=0.57,angle=0,clip,bb=24 46 298 296]{dean_hjorthjensen_figp1b.ps} &
%  \includegraphics*[scale=0.57,angle=0,clip,bb=72 46 298 296]
%  {dean_hjorthjensen_figp2b.ps} &
%  \includegraphics*[scale=0.57,angle=0,clip,bb=72 46 298 296]
%  {dean_hjorthjensen_figp3b.ps} \cr
%  \!\includegraphics*[scale=0.57,angle=0,clip,bb=24 46 298 296]
%  {dean_hjorthjensen_figk1b.ps} &
%  \includegraphics*[scale=0.57,angle=0,clip,bb=72 46 298 296]
%  {dean_hjorthjensen_figk2b.ps} &
%  \includegraphics*[scale=0.57,angle=0,clip,bb=72 46 298 296]
%  {dean_hjorthjensen_figk3b.ps} \cr
%  \!\includegraphics*[scale=0.57,angle=0,clip,bb=24 46 298 296]
%  {dean_hjorthjensen_figq1b.ps} &
%  \includegraphics*[scale=0.57,angle=0,clip,bb=72 46 298 296]
%  {dean_hjorthjensen_figq2b.ps} &
%  \includegraphics*[scale=0.57,angle=0,clip,bb=72 46 298 296]
%  {dean_hjorthjensen_figq3b.ps} \cr
%}$
%\end{center}
\caption{Distribution of the pair transfer coefficient $<\!\!P^+_{00}\!\!>$,
pairing strength $<\!\!\kappa\!\!>$, and deformation $Q_2$ for the random 
interactions
leading to a $J^\pi=0^+$ ground state in the shell-model description.
From left to right we report the results for $N-Z=4$, 2, and 0. The arrows
indicate the number of results in a bin when it is out of scale.}
\label{kfig4}
\end{figure}

In Fig.~\ref{kfig4}, we represent the distribution of the number
of interactions according to the results obtained for $<\!P^+\!>$,
$<\!\kappa\!>$,
and $Q_2$ \cite{ben02}. 
In the case with only one kind of active particle in the model
space ($N-Z=4$), almost all the interactions lead to a strong
pair transfer coefficient. This property can be related to the fact
that when we consider only one kind of particle, the deformation effects
are (almost) zero, and pairing can develop more easily. Netherless, for the
three sets, we see once again that $<\!P^+\!>$ has a significant value
in most cases.  The distribution of the number of 
interactions according to the
static pairing strength, measured via $<\!\kappa\!>$, follow the same
evolution ({\it i.e.} the number of interactions that give a significant
value of $<\!\kappa\!>$ increases with $N-Z$), but the number of interactions
for which $<\!\kappa\!>$ is small is always important. It is then very
unlikely to relate the origin of the spin 0 of the ground states with the
static pairing created by the interactions.

It is also instructive to consider the evolution of the plots when
one changes the asymmetry of the system from $N-Z=4$ to $N-Z=0$.
In the mean-field description (center and lower
part of Fig.~\ref{kfig4}), the pairing strength is concentrated
into regions $<\!\kappa\!>\,\sim 0$ and
$<\!\kappa\!>\,\sim 0.5$ to 2. This property
does not strongly change as a function of $N-Z$. Nevertheless, we notice
that $<\!\kappa\!>$ is more often close to zero and the non-zero values are
less scattered when $N=Z$. This effect can probably be attributed to
the deformations which play a more important role when $N\sim Z$.
When the pairing is weak, the deformation effects prevail and so
{\it decrease} the pairing strength in the region between 0 and 1.5.

In the shell-model description (upper
part of Fig.~\ref{kfig4}), we notice a clear evolution of the pair transfer
coefficient with the asymmetry of the system. This property seems to be
mainly due to the deformation effects. Indeed for the system with $N-Z=4$,
the coefficient $<\!P^+\!>$ is peaked at around 2.5, and no interactions
give a value close to zero. The cases with $N-Z=2$ and $N-Z=0$ are similar
and indicate that the $T=0$ part of the pairing
interaction does not play a crucial role in these systems.
This last remark tends to show that the $T=0$ part of the pairing interaction,
like the $T=1$ part, 
does not play a significant role for the relative abundance
of $0^+$ ground states. This conclusion is in agreement with the statement
made by Mulhall {\it et al.} \cite{mulhall00}, in which the origin of the
abundance of the $0^+$ states in the even fermion systems is related with
the {\it geometric chaocity} rather than with the pairing properties of the
system.

%%%%%   section 5

%\input{levden}

% latest upgrade 2/april/02 mhj
% 4-4-02 djd. Included some smmc stuff on Dy162. typo fixing. 
% 4/7/02 mhj more typo fixing plus removing od eq 55 on back-shifted eq
% june 2002 djd typo correcting
% july-aug  mhj ref correction + further typos
% october 2002 mhj included comments from Hubert
\section{THERMODYNAMIC PROPERTIES AND PAIRING CORRELATIONS IN NUCLEI}
\label{sec:leveldensities_sec3}

\subsection{Introduction}

One of the 
most interesting problems in the context of phase transitions in small systems
is the question of the 
existence and classification of a possible phase transition from a hadronic 
phase to a quark-gluon plasma in high-energy physics. The answer to this 
question has far-reaching consequences into many other fields of research like,
e.g., cosmology, since it has been argued that hadronization of the quark-gluon
plasma should be a first-order phase transition in order to allow for possible 
super cooling and consequently the emergence of large-scale inhomogeneities in 
the cosmos within the inflationary big-bang model. 

In nuclear physics, different phase transitions have been discussed in the 
literature. A first-order phase transition has been reported in the 
multifragmentation of nuclei \cite{agostino00}, thought to be the analogous 
phenomenon in a finite system to a liquid-gas phase transition in the 
thermodynamical limit. A pivotal role in these studies is played by the 
presence of a convex intruder in the microcanonical entropy curve 
\cite{gross97,gross99}. This leads to a 
negative branch of the microcanonical heat 
capacity which is used as an indicator of a first-order phase transition in 
small systems. Negative heat capacities have indeed been observed in the 
multifragmentation of atomic nuclei, though the heat capacity curve has not 
been derived directly from the caloric curve, but by means of energy 
fluctuations \cite{agostino00,CD00}. 
Another finding of a negative branch of the heat
capacity curve has been in sodium clusters of 147 atoms \cite{schmidt01}, indicating
a possible first-order phase transition. On the other hand, it is not clear 
whether the observed negative heat capacities are simply not due to the changing 
volume of the system under study that is progressively evaporating particles 
\cite{moretto01}. In general, great care should be 
taken in the proper extraction of
temperatures and other thermodynamical quantities of a multifragmenting system.

Another transition discussed for atomic nuclei has been anticipated 
for the transition from a phase with strong pairing correlations to a phase 
with weak pairing correlations \cite{SY63}. Early schematic calculations have 
shown that pairing correlations can be quenched by temperature as well as by 
the Coriolis force in rapidly 
rotating nuclei \cite{Go81a,Go81b,TS80,TS82,shimizu89}. 
This makes the quenching of pairing correlations in atomic nuclei very similar 
to the breakdown of superfluidity in $^3$He (due to rapid rotation and/or 
temperature) or of superconductivity (due to external magnetic fields and/or 
temperature). Recently, structures in the heat capacity curve related to the 
quenching of pairing correlations have been obtained within the relativistic 
mean-field theory \cite{AT00,AT01}, the finite-temperature random phase 
approximation \cite{Ng90}, the finite-temperature Hartree-Fock-Bogoliubov 
theory \cite{ER00}, and the shell-model Monte Carlo (SMMC) approach 
\cite{DK95a,NA97,RH98,WK00,liu01}. An S-shaped structure in the heat capacity 
curve could also be observed experimentally \cite{schiller2001} and has been 
interpreted as a fingerprint of a second-order phase 
transition in the thermodynamic limit from a phase 
with strong pairing correlations to one with no pairing correlations. For finite systems
there will be a gradual transition from strong pairing correlations to weak pairing 
correlations, implying a finite order parameter, here the pairing gap $\Delta$.
Indeed, the analogy of the quenching of pairing correlations in atomic nuclei 
with the breakdown of superfluidity in $^3$He and the breakdown of 
superconductivity suggests a second-order phase transition and a schematic 
calculation might support this assumption (see the discussion below). 
Interestingly, similar 
structures of the heat capacity curve as observed for atomic nuclei in 
\cite{schiller2001} have been seen in 
small metallic grains undergoing a second-order 
phase transition from a superconductive to a normal conductive phase 
\cite{tinkham95,tinkham96,tinkham98,delft2000,LA93}, thereby supporting the analogous findings for atomic nuclei. On 
the other hand, breaking of nucleon pairs has been experimentally shown to 
cause a series of convex intruders in the microcanonical entropy curve of rare 
earth nuclei \cite{oslo3,MG01}, leading to several negative branches of the 
microcanonical heat capacity. 
This finding might, in analogy to the discussion 
of nuclear multifragmentation, be taken as an indicator of several first-order
transitions.
Other possible phase transition-like behaviors are, e.g., shape transitions
from a collective to an oblate aligned-particle structure 
at higher temperatures (see for example the recent work of Ma {\em et al.}
\cite{tenglek2000}).

For a finite isolated many-body system such as a nucleus, the correct
thermodynamical ensemble is the microcanonical one. In this ensemble,
the nuclear level density, the density of eigenstates of a nucleus
at a given excitation energy, is the important quantity that should be
used to describe thermodynamic properties of nuclei, such as the 
nuclear entropy, specific heat, and temperature. 
Bethe first described the level density using a 
non-interacting fermi gas model
for the nucleons \cite{bethe1}. Modifications to this picture, such as the 
back-shifted fermi gas which includes pair and
shell effects \cite{back_shift}
not present in Bethe's original formulation, are in 
wide use.  We note that several approaches based on mean-field 
theory have recently been developed to 
investigate nuclear level densities including recent work 
that incorporates BCS pairing into the mean-field picture 
to derive level densities for nuclei across the periodic table 
\cite{demetriou01}. Other mean-field applications based on the
Gogny effective nucleon-nucleon interaction (which includes 
pairing due to the finite range of the interaction) have also been 
developed recently \cite{hilaire01}. 
The level density\footnote{Hereafter we use $\rho$ for the level 
density in the microcanonical ensemble.} 
$\rho$ defines the partition function for 
the microcanonical ensemble and the entropy through the well-known
relation $S(E)=k_Bln(\rho(E))$.
Here $k_B$ is Boltzmann's constant and $E$ is the energy.  In 
the microcanonical ensemble, we could then extract expectation values 
for thermodynamical quantities like temperature $T$, or the 
heat capacity $C$. The temperature in the microcanonical ensemble 
is defined as 
\begin{equation}
      T=\left(\frac{dS(E)}{dE}\right)^{-1}.
      \label{eq:temp}
\end{equation}
It is a function of the excitation energy, which is 
the relevant variable of interest in the microcanonical ensemble. 
However, since the extracted level density is given only at discrete 
energies, the calculation of expectation values 
like $T$, involving derivatives of the partition function, is not
reliable unless a strong smoothing over energies is 
performed. This case is discussed in detail in  \cite{oslo3} 
and below. Another possibility
is to perform a transformation to the canonical ensemble.
The partition function for the 
canonical ensemble is related to that of the 
microcanonical ensemble through a Laplace transform
\begin{equation}
     Z(\beta)=\int_0^{\infty}dE\rho(E)\exp{(-\beta E)}.
     \label{eq:zcan}
\end{equation}
Here we have defined $\beta=1/k_BT$. Since we will 
deal with discrete energies, the 
Laplace transform of Eq.\ (\ref{eq:zcan}) takes the form
\begin{equation}
         Z(\beta)=\sum_E \Delta E\rho(E)\exp{(-\beta E)},
         \label{eq:zactual}
\end{equation}
where $\Delta E$ is the energy bin used.
With $Z$ we can evaluate the 
entropy in the canonical ensemble using the definition of the 
free energy 
\begin{equation}
     F(T)= -k_B T \ln Z(T)=\langle E(T)\rangle - TS(T).
\end{equation}
Note that the temperature $T$ is now the variable of 
interest and the energy $E$ is given by the expectation 
value $\langle E\rangle$ as a function of $T$. Similarly, 
the entropy $S$ is also a function of $T$.
For finite systems, fluctuations in various 
expectation values can be large.
In nuclear and solid state physics, thermal properties have 
mainly been studied in the canonical and grand-canonical ensemble. 
In order to obtain the level density, the inverse transformation 
\begin{equation}
      \rho(E) =\frac{1}{2\pi i}\int_{-i\infty}^{i\infty}
 d\beta Z(\beta) \exp{(\beta E)},
      \label{eq:zbigcan}
\end{equation}
is normally used. Compared with Eq.\ (\ref{eq:zcan}), this 
transformation is rather difficult to perform since 
the integrand $\exp{\left(\beta E+ \ln Z(\beta)\right)}$ is a 
rapidly varying function of the integration parameter. In order to obtain 
the density of states, approximations like the saddle-point 
method, viz., an expansion of the exponent in the integrand 
to second order around the equilibrium point and subsequent integration, 
have been used widely,
see for example,  \cite{kdl97,yoram,WK00}. 
For the ideal Fermi gas (FG), this gives the following density of states
\begin{equation}
      \rho_{\rm FG}(E)=\frac{\exp{(2\sqrt{aE})}}{E\sqrt{48}},
      \label{eq:omegaideal}
\end{equation}
where $a$ in nuclear physics is a factor 
typically of the order $a=A/8$ with dimension 
MeV$^{-1}$, $A$ being the mass number of a given nucleus.

To obtain an experimental level density is a rather hard task.
Ideally, we would like an experiment to provide us with the level 
density as a function of excitation energy and thereby 
the `full' partition function for the microcanonical ensemble. 
It is only rather recently that 
experimentalists have been able to develop methods 
\cite{oslo1,oslo2} for extracting level densities at low spin from
measured $\gamma$-spectra. 
These measurements were performed at the Oslo 
Cyclotron Laboratory. Detection of gamma-rays are obtained with the
CACTUS multidetector array \cite{oslo3} using the $(^3$He,$\alpha \gamma$)
and $(^3$He,$^3$He')
reactions on several rare-earth nuclei.
Assuming that the
experimental analysis is correct,   
the resulting microcanonical 
level density reveals structures between 1-5 MeV of excitation energy
and were interpreted as indications of pair breaking in these systems. 

The Oslo experimental results lead us to ask whether we can simultaneously
understand the thermodynamic and pairing properties of a nuclear many-body
system. We are also led to questions concerning the nature of phase 
transitions in a finite many-body system.  
After these introductory words, we briefly delineate the aim of this
section. 
In the next subsection we present experimental level densities for several
rare-earth nuclei together with a thermodynamical analysis and possible
interpretations. 
In Subsec.~\ref{subsec:statprops} the simple pairing model
of Eq.~(\ref{eq:pairHamiltonian1_sec2}) is used in a similar analysis
in order to see whether such a simplified pairing Hamiltonian can mimick
some of the features seen from the experimental level densities.
Since this is a simplified model, we also present results from
shell-model Monte Carlo calculations of level densities 
in the rare-earth region with pairing-plus-quadrupole effective interactions 
in realistic model spaces.

\subsection{Level densities from experiment and thermal properties}

The Oslo cyclotron group has developed a method to 
extract nuclear level densities at low spin from 
measured $\gamma$-ray spectra 
\cite{oslo2,oslo3,schiller2001,oslo1,andreas2000}. 
The main advantage of utilizing $\gamma$-rays as a probe for 
level density is that the nuclear system is likely thermalized prior 
to the $\gamma$-emission. In addition, the method 
allows for the simultaneous extraction of level density 
and $\gamma$-strength function over a wide energy region. 

The experiments were carried out with 45~MeV $^3$He-projectiles 
accelerated by the MC-35 cyclotron at the University of 
Oslo. The experimental data were recorded with the CACTUS multidetector 
array using the $(^3$He,$\alpha \gamma$)
reaction on several rare-earth nuclei such as 
$^{149}$Sm, 
$^{162}$Dy, $^{163}$Dy, $^{167}$Er, 
$^{172}$Yb, and $^{173}$Yb, yielding as result the nuclei
 $^{148}$Sm,  
$^{161}$Dy, $^{162}$Dy, $^{166}$Er, , $^{171}$Yb, and $^{172}$Yb.
The $(^3$He,$^3$He')
reaction was used to obtain the nuclei $^{149}$Sm and $^{167}$Er. 
For a critical discussion of the latter reaction see 
\cite{andreasheherect}.
The charged ejectiles were detected with eight particle telescopes placed 
at an angle of 45$^{\circ}$ relative to the beam direction. 
Each telescope comprises one Si~$\Delta E$ front and one Si(Li) $E$ 
back detector with thicknesses of 140 and 3000~$\mu$m, respectively.

From the reaction kinematics, the measured 
$\alpha$-particle energy can be transformed to 
excitation energy $E$. Thus, each coincident $\gamma$-ray can 
be assigned to a $\gamma$-cascade originating from a 
specific excitation energy. The data are sorted into a 
matrix of $(E,E_{\gamma})$ energy pairs. 
The resulting matrix $P(E,E_{\gamma})$, which describes 
the primary $\gamma$-spectra obtained at initial excitation energy 
$E$, is factorized according to the Brink-Axel 
hypothesis \cite{brink_thesis,axel62} by 
\begin{equation}
P(E,E_{\gamma}) \propto \rho (E -E_{\gamma})\sigma (E_{\gamma}), 
\label{eq:firstgamma1}
\end{equation}
where the level density $\rho$ and the $\gamma$-energy-dependent
function $\sigma$ are unknown. The iterative procedure for obtaining
these two functions and 
the assumptions behind the factorization of this expression are 
described in more detail in \cite{oslo1,andreas2000}.
The experimental level density $\rho(E)$ at excitation energy 
$E$ is proportional to the number of levels accessible in 
$\gamma$-decay. For the present reactions, the spin 
distribution is centered around $\langle J\rangle \sim$ 4.4 $\hbar$ 
with a standard deviation of $\sigma_J \sim$ 2.4 $\hbar$. 
Hence, the entropy\footnote{The experiment reveals 
the level density and not the state density. Thus, also the 
observed entropy reveals the number of levels. The state density can 
be estimated by $\rho_{\rm state} 
\sim (2J+1)\rho_{\rm level} \sim$ 9.8 $\rho_{\rm level}$.} can 
be deduced within the microcanonical ensemble, using 
\begin{equation}
S(E) = k_B \ln N(E) = k_B \ln \frac{\rho(E)}{\rho_0},
\label{eq:s_experimental}
\end{equation}
where $N$ is the number of levels in the 
energy bin at energy $E$. The normalization factor 
$\rho_0$ can be determined from the ground state band 
in the even-even nuclei, where one  has $N(E) \sim$ 1 within a 
typical experimental energy bin of $\sim$ 0.1 MeV. 

\begin{figure}
\includegraphics[totalheight=12cm,angle=0,bb=0 20 350 730]{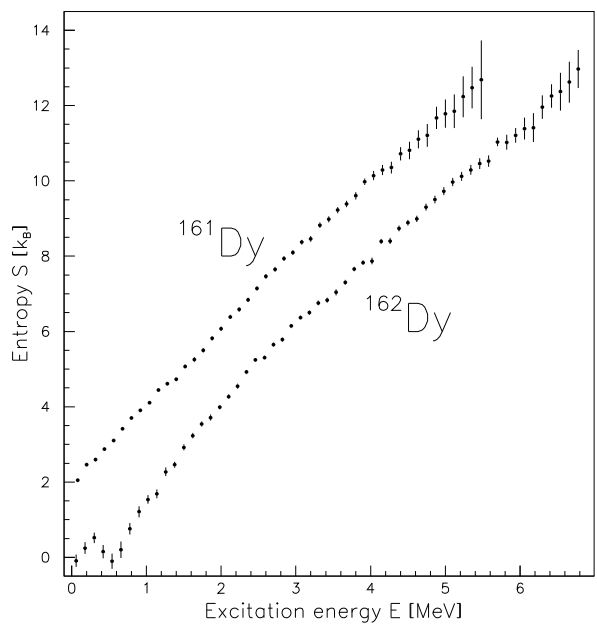}
\caption{Observed entropy for $^{161,162}$Dy as 
function of excitation energy $E$. Taken from \cite{entropy2000}.}
\label{fig:fig7_sec3}
\end{figure}

\begin{figure}
\includegraphics[totalheight=12cm,angle=0,bb=0 20 350 730]{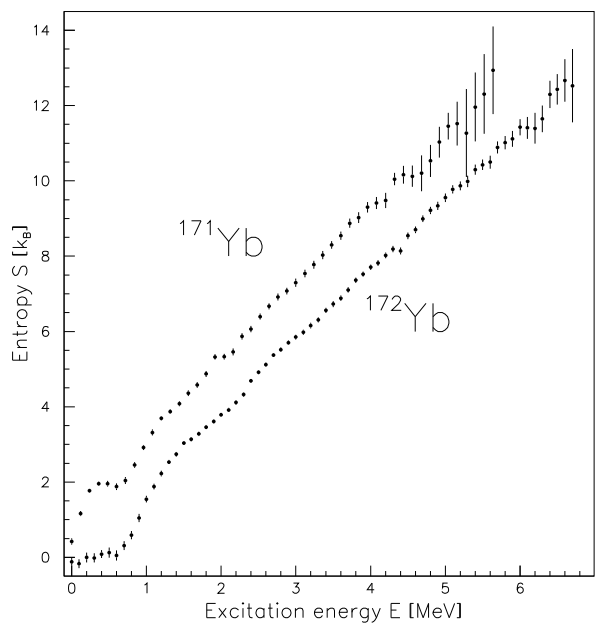}
\caption{Observed entropy for $^{171,172}$Yb as function 
of excitation energy $E$. Taken from \cite{entropy2000}.}
\label{fig:fig8_sec3}
\end{figure}

The extracted entropies for the $^{161,162}$Dy and $^{171,172}$Yb 
nuclei are shown in Figs.\ \ref{fig:fig7_sec3} and \ref{fig:fig8_sec3}. 
In the transformation from level density to entropy, $\rho_0$ was set to
$\rho_0 \sim$ 3 MeV$^{-1}$ in
Eq.~(\ref{eq:s_experimental}). The entropy curves 
are rather linear, but with small oscillations or bumps 
superimposed. The curves terminate around 1 MeV below 
their respective neutron binding energies due to the 
experimental cut excluding $\gamma$-rays with $E_{\gamma} <$ 1 MeV. 
All four curves reach $S \sim$ 13 $k_B$, which by extrapolation correspond 
to $S \sim$ 15 $k_B$ at the neutron binding energy $B_n$. 

\begin{figure}
\includegraphics[totalheight=12cm,angle=0,bb=0 80 350 730]{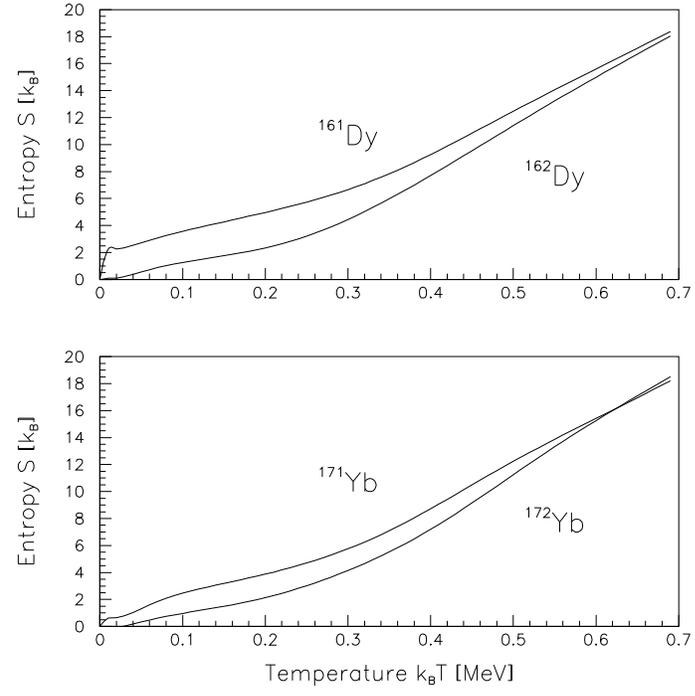}
\caption{Semi-experimental entropy 
$S$ for $^{161,162}$Dy and $^{171,172}$Yb calculated in 
the canonical ensemble as a 
function of temperature $k_BT$. Taken from \cite{entropy2000}.}
\label{fig:fig10_sec3}
\end{figure}

Note that there is an entropy excess 
for the odd systems, since many low-lying
states can be reached without needing to break a pair.
The experimental level density can be used to determine the canonical 
partition function $Z(T)$. However, in the evaluation of 
Eq.\ (\ref{eq:zactual}), one needs  to extrapolate the 
experimental $\rho$ curve to $\sim$ 40 MeV. The 
back-shifted level density formula of \cite{back_shift,esb88} 
was employed (for further details see \cite{andreas2000}).
From this semi-experimental partition function, the 
entropy can be determined from Eq.\ (\ref{eq:canentropy}). The results 
are shown in Fig.\ \ref{fig:fig10_sec3}. The entropy curves show 
a splitting at temperatures below  $k_BT = 0.5 - 0.6$ MeV 
which reflects the experimental splitting shown in the microcanonical 
plots of Figs.~\ref{fig:fig7_sec3}  and \ref{fig:fig8_sec3}.

\begin{figure}\centering
\includegraphics[totalheight=16cm]{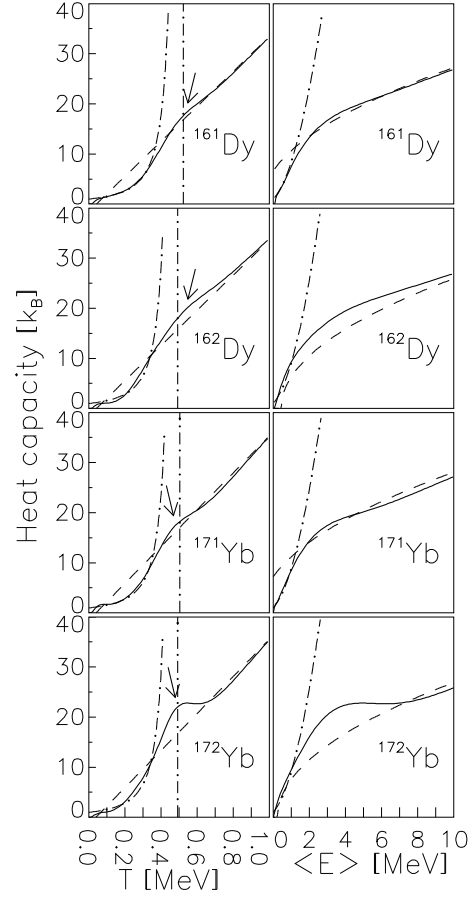}
\caption{Semi-experimental heat capacity as a function of temperature (left 
panels) and energy $\langle E\rangle$ (right panels) in the canonical ensemble 
for $^{161,162}$Dy and $^{171,172}$Yb. The dashed lines describe the 
approximate Fermi gas heat capacity. The arrows indicate the first local 
maxima of the experimental curve relative to the Fermi gas estimates. The 
dashed-dotted lines describe extrapolated estimates of 
the critical temperature $T_c$. $T_c$ is indicated
by the vertical lines. 
Taken from \cite{schiller2001}.}
\label{fig:heatcapacity} 
\end{figure}

The merging together of the entropy curves at roughly
$k_BT = 0.5 - 0.6$ MeV can also be seen in the analysis of the 
heat capacity in the canonical ensemble.  
The extraction of the microcanonical heat capacity $C_V(E)$ gives large 
fluctuations which are difficult to interpret \cite{oslo3}. Therefore, the heat 
capacity $C_V(T)$ is calculated within the canonical ensemble as function of
temperature $T$. The heat 
capacity is then given by 
\begin{equation}
C_V(T)=\frac{\partial\langle E\rangle}{\partial T}.
\end{equation}
The deduced heat capacities for the $^{161,162}$Dy and $^{171,172}$Yb nuclei 
are shown in Fig.~\ref{fig:heatcapacity}. All four nuclei exhibit similarly 
S-shaped $C_V(T)$-curves with a local maximum relative to the Fermi gas 
estimate at $T_c\approx 0.5$~MeV. The S-shaped curve is interpreted as a 
fingerprint of a phase transition-like behavior in a 
finite system from a phase with strong 
pairing correlations to a phase without such correlations. Due to the strong 
smoothing introduced by the transformation to the canonical ensemble, we do not
expect to see discrete transitions between the various quasiparticle regimes, 
but only the transition where all pairing correlations are quenched as a whole.
In the right panels of Fig.~\ref{fig:heatcapacity}, we see that 
$C_V(\langle E\rangle)$ has an excess in the heat capacity distributed over 
a broad region of excitation energy and is not giving a clear signal for 
quenching of pairing correlations at a certain energy \cite{oslo3}.

In passing, we note that the results displayed in Fig.~\ref{fig:heatcapacity}
are similar to those of Liu and Alhassid \cite{liu01} 
shown in Fig.~\ref{fig:figyoram}.

\subsection{Thermodynamics of a simple pairing model}
\label{subsec:statprops}

In this section, we will try to analyze the results from the previous
subsection in terms of the simple pairing model presented in 
Eq.~(\ref{eq:pairHamiltonian1_sec2}). 
As stated in Sec.~\ref{sec:NN_to_pairing}, seniority is
a good quantum number, which 
means that we can subdivide the full eigenvalue problem 
into minor blocks with given seniority and diagonalize these 
separately. 
If we consider an even  system of $N=12$ 
particles which are distributed over $L=12$ two-fold 
degenerate levels, we obtain a total of $2.704.156$.
Of this total, for seniority
$\cal{S}$ $=0$, i.e. no broken pairs, we have $924$
states. Since the Hamiltonian does not connect states with 
different seniority $\cal{S}$, we can diagonalize a 
$924\times 924$ matrix and obtain all eigenvalues with 
$\cal{S}$ $=0$. Similarly, we can subdivide the Hamiltonian 
matrix into $\cal{S}$ $=2$, $\cal{S}$ $=4$,... and 
$\cal{S}$ $=12$ (all pairs broken) blocks and obtain 
{\em all} $2.704.156$ eigenvalues for a system with $L=12$ 
levels and $N=12$ particles. As such, we have the 
{\em exact density of levels} and can compute observables like the 
entropy, heat capacity, etc. This numerically solvable model 
enables us to compute exactly the entropy in the microcanonical 
and the canonical ensembles for systems with odd and even numbers 
of particles.  In addition, varying the relation $\delta=d/G$  between the 
level spacing $d$ and the pairing 
strength $G$ may reveal features of the entropy that are 
similar to those of the experimentally extracted entropy discussed
in the previous subsection. Recall that the experimental level densities  
represent both even-even and even-odd nucleon systems.

Here we study two systems in order to extract differences 
between odd and even systems, namely by fixing the number of 
doubly degenerated single-particle levels to $L=12$, whereas the 
number of particles is set to $N=11$ and $N=12$.  

These two systems result in a total of 
$\sim 3 \times 10^6$ eigenstates. In the calculations, we 
choose a single-particle level spacing of $d=0.1$ MeV, which is close 
to what is expected for rare-earth nuclei. 
We select three values of the pairing strength, 
namely $G=1$, $0.2$, and $0.01$, 
($\delta=d/G=0.1$, $\delta=d/G=0.5$, and $\delta=d/G=10$), 
respectively. The first case represents a strong 
pairing case, with almost degenerate single-particle levels. The 
second is an intermediate case where the level spacing is of 
the order of the pairing strength, while the last case results in 
a weak pairing case. As shown below, the results for the latter 
resemble to a certain extent those for an ideal gas.

\subsubsection{Entropy}

The numerical procedure is rather straightforward. 
First we diagonalize the large Hamiltonian matrix (which is 
subdivided into seniority blocks) and obtain all 
eigenvalues $E$ for the odd and even particle case. This defines also 
the density of levels $\rho(E)$, the partition function, and 
the entropy in the microcanonical ensemble. Thereafter, we can 
obtain the partition function $Z(T)$ in the canonical ensemble through 
Eq.\ (\ref{eq:zactual}). The partition function 
$Z(T)$ enables us, in turn, to compute the entropy $S(T)$ by
\begin{equation}
    S(T)=k_B \ln Z(T)+\langle E(T)\rangle/T.
    \label{eq:canentropy}
\end{equation}
Since this is a model with a finite number of levels and 
particles, unless a certain smoothing is done, 
the microcanonical entropy may vary strongly from energy to energy
(see for example the discussion in \cite{entropy2000}).
Thus, rather than performing a 
certain smoothing of the results in the microcanonical ensemble, 
we will choose to present further results 
for the entropy in the canonical ensemble, using the 
Laplace transform of Eq.\ (\ref{eq:zactual}). 

\begin{figure}
\includegraphics[scale=0.5,angle=0]{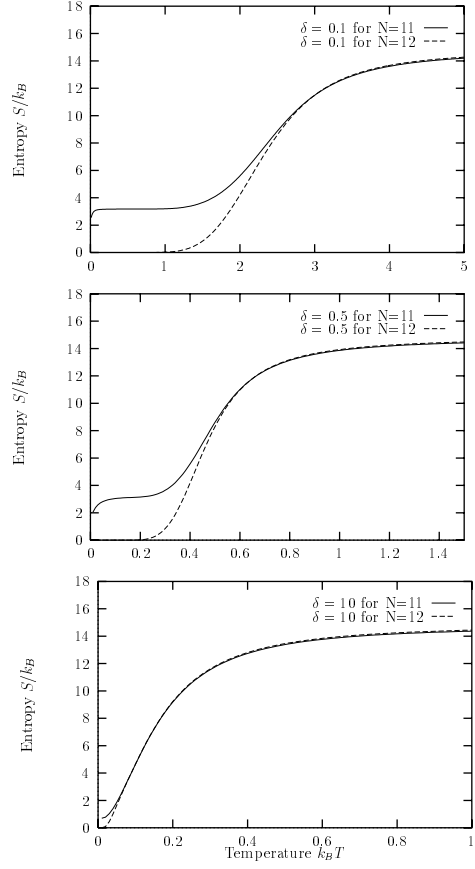}
\caption{Entropy in the canonical ensemble as a function 
of temperature $k_BT$ for odd and even systems for $\delta=0.1$ 
(upper panel), $\delta=0.5$ (central panel),
and $\delta=10$ (lower panel).
If we wish to make contact with experiment, one can assign units 
of MeV to $k_BT$.  The entropy $S/k_B$ is dimensionless. Taken from \cite{entropy2000}.
} 
\label{fig:fig5_sec3}
\end{figure}

The results for the entropy in the canonical ensemble 
as functions of $T$ for the above three sets of 
$\delta=d/G$ are shown in Fig.\ \ref{fig:fig5_sec3}. For the two 
cases with strong pairing, we see a clear difference in entropy 
between the odd and the even system. The difference in 
entropy between the odd and even systems can be easily 
understood from the fact that the lowest-lying states in 
the odd system involve simply the excitation of one single-particle 
to the first unoccupied single-particle state and is 
interpreted as a single-quasiparticle state. These states are 
rather close in energy to the ground state and explain 
why the entropy for the odd system has a finite value 
already at low temperatures. Higher-lying excited 
states include also breaking of pairs and can be 
described as three-, five-, and more-quasiparticle states. 
For $\delta=10$, the odd and even systems 
merge together already at low temperatures, indicating that 
pairing correlations play a negligible role. For a small single-particle 
spacing, also the difference in energy between the first 
excited state and the ground state for the odd system is rather small.

For $\delta=0.5$, we note that at a 
temperature of $k_BT \sim 0.5-0.6$, the even and odd systems
approach each other\footnote{If we wish to make contact with experiment, 
we could assign units of MeV to $k_BT$ and $E$.}. The 
temperature where this occurs corresponds to an excitation energy 
$\langle E\rangle$ in the canonical ensemble of 
$\langle E\rangle \sim 4.7-5.0$. This corresponds to excitation 
energies where we have $4-6$ quasiparticles, 
seniority ${\cal S}$ $=4-6$, in the even system and 
$5-7$ quasiparticles, seniority ${\cal S}$ $=5-7$, in the odd 
system (see for example the discussion in \cite{entropy2000}).
For the two cases 
with strong pairing ($\delta=0.1$ and $\delta=0.5$), 
Fig.\ \ref{fig:fig5_sec3} tells us that at temperatures where we 
have  $4-6$ quasiparticles in the even system and $5-7$  quasiparticles 
in the odd system, the odd and even system tend to merge 
together. This reflects the fact that pairing correlations tend to be less 
important as we approach the non-interacting case.  
In a simple 
model with just pairing interactions, it is thus easy to see where, 
at given temperatures and excitation energies, certain degrees 
of freedom prevail. For the experimental results, 
this may not be the case since the interaction between nucleons is 
much more complicated. The hope, however, is that pairing may dominate 
at low excitation energies and that the physics behind the 
features seen 
in Fig.\ \ref{fig:fig5_sec3} is qualitatively similar to 
the experimental information conveyed in  Fig.\ \ref{fig:fig10_sec3}.

\subsubsection{The free energy}

We may also investigate the free energy of the system.
Using the density of states, we can define
the  free energy $F(E)$ in the microcanonical ensemble 
at a fixed temperature $T$ (actually an expectation value in this ensemble), 
\begin{equation}
    F(E)=-T\ln\left[\Omega_N(E)e^{-\beta E}\right]\;.
    \label{eq:freenergy}
\end{equation}
Note that here we include only
configurations at a particular $E$.
 
The free energy was used by
Lee and Kosterlitz \cite{lk90,lk91},
based on the histogram approach for studying
phase transitions developed by Ferrenberg and Swendsen \cite{fs88,fs89}
in their studies of phase transitions
of classical spin systems.
If a phase transition is present, a plot of $F(E)$ versus $E$ will show
two local minima which correspond to configurations that are
characteristic of the high and low temperature phases.
At the transition temperature $T_C$, the value of $F(E)$ at the
two minima equal, while at temperatures below $T_C$, the low-energy
minimum is the absolute minimum. At temperatures above $T_C$, the high-energy
minimum is the largest. If there is no
phase transition, the system develops only one minimum for all temperatures.
Since we are dealing with finite systems, we can study the development
of the two minima as a function of the size of the system and thereby
extract information about the nature of the phase transition. If we are dealing
with a second-order phase transition, the behavior of $F(E)$ does not change
dramatically as the size of the system increases. However, if the transition
is first order, the difference in free energy, i.e.,
the distance between the maximum and minimum values,
will increase with increasing size of the system.

We calculate exactly the  
free energy $F(E)$ of Eq.~(\ref{eq:freenergy})
through diagonalization of the pairing Hamiltonian of 
Eq.~(\ref{eq:pairHamiltonian1_sec2})
for systems with up to 16 particles in $16$ doubly degenerate
levels, yielding a total of $\sim 4\times 10^8$ configurations. The
density of states hence defines the microcanonical partition function.

\begin{figure}
%\begin{center}
%\setlength{\unitlength}{1mm}
%   \begin{picture}(140,220)
%   \put(0,0){\epsfxsize=18cm \epsfbox{dean_hjorthjensen_figfreeenergy.ps}}
%   \end{picture}
\includegraphics[scale=0.5,angle=0]{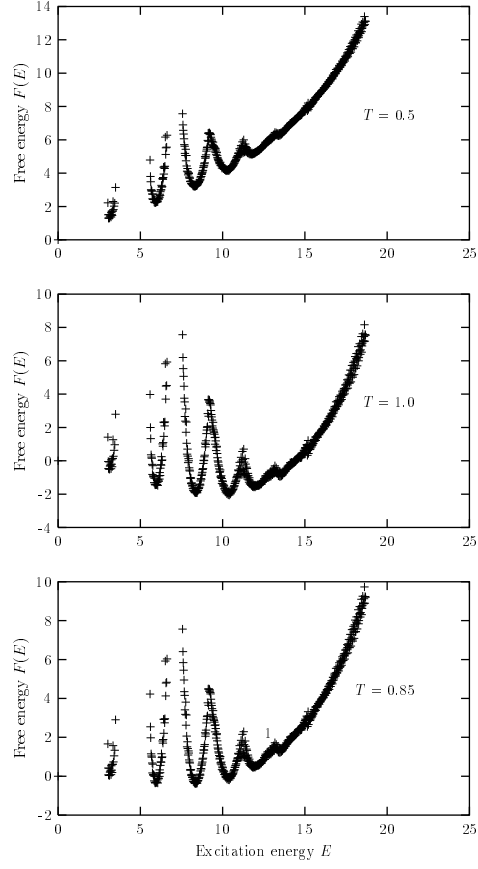}
%\end{center} 
\caption{Free energy from Eq.~(\ref{eq:freenergy}) at $T=0.5$, $0.85$ and
         $T=1.0$ MeV  with 
         $d/G=0.5$ with 16 particles in 16 doubly degenerate
         levels. All energies are in units of MeV and 
         an energy bin of $10^{-3}$ MeV has been chosen.}
\label{fig:free_energy16}
\end{figure}

For $d/G=0.5$ and 16 single-particle levels, the calculations yield
two clear  minima for the free energy.
This is seen in
Fig.~\ref{fig:free_energy16} where we show the free energy as a function of 
excitation energy
using Eq.~(\ref{eq:freenergy}) at temperatures $T=0.5$, $T=0.85$, and $T=1.0$.
The first minimum corresponds to the case where we break one pair.
The second and third minima  correspond
to cases where two and three pairs are broken, respectively. 
When two pairs are broken, corresponding to seniority ${\cal S}=4$, 
the free energy minimum is made up of contributions
from states with ${\cal S}=0,2,4$. It is this contribution from states 
with lower seniority which contributes to the lowering of the free
energy of the second minimum. Similarly, with three pairs
broken, we have a free energy minimum which receives contributions
from ${\cal S}=0,2,4,6$, yielding a new minimum. 
At higher excitation energies, population
inversion takes place, and our model is no longer realistic. 

We note that for $T=0.5$, the minima at lower excitation
energies are favored. 
At $T=1.0$, the higher energy
phase (more broken pairs) is favored.
We see also that at $T=0.85$ for our system with 
16 single-particle states and $d/G=0.5$,
the free-energy minima where we break two and three pairs 
equal. 
Where two minima coexist, we may have an
indication  of a phase transition. Note, however, that this is not a 
phase transition in the ordinary thermodynamical sense.
There is no abrupt transition from a purely paired phase to a 
nonpaired phase.  
Instead, our system develops several such intermediate steps
where different numbers of broken pairs can coexist. 
At e.g., $T=0.95$, we find again two equal minima. For this case,
seniority ${\cal S}=6$ and ${\cal S}=8$ yield two equal minima.
This picture repeats itself for higher seniority and higher temperatures.

If we then focus on the second and third minima, i.e., where we break
two and three pairs, respectively, the difference $\Delta F$ between the 
minimum and the maximum of the free energy can aid us in distinguishing
between a first-order and a second-order phase transition. If $\Delta F/N$,
with $N$ being the number of particles present, remains constant as $N$
increases, we have a second-order transition. An increasing $\Delta F/N$
is, in turn, an indication of a first-order phase transition. 
It is worth noting that the features
seen in Fig.~\ref{fig:free_energy16} apply to the cases with $N=10$, 12, 
and 14 as well, with $T=0.85$ being the temperature where the second and
third minima equal. This means that the temperature where the transition
is meant to take place remains stable as a function of number of single-particle
levels and particles. This is in agreement with the simulations of 
Lee and Kosterlitz \cite{lk90,lk91}. We find a similar result for the minima
developed at $T=0.95$, where both ${\cal S}=6$ and ${\cal S}=8$.
However, due to population inversion, these minima are only seen clearly
for $N=12$, $14$, and $16$ particles.
In Table \ref{tab:free_energy10_16} we display $\Delta F/N$ for 
$N=10$, 12, 14, and 16 at $T=0.85$ MeV. 

Table \ref{tab:free_energy10_16} reveals that $\Delta F/N$ is nearly
constant, with  $\Delta F/N\approx 0.5$~MeV, indicating a 
transition of second order. This result is in 
agreement with what is expected for an infinite system. 

Before proceeding to the next method for classifying a phase transition
in a finite system, we note the important result that for $d/G > 1.5$, 
our free energy, for $N\le 16$, develops
only one minimum for all temperatures. That is, for larger single-particle
spacings, there is no sign of a phase transition. This means that there
is a critical relation between $d$ and $G$ for the appearance of a phase 
transition-like behavior, being a  reminiscence of the thermodynamical limit.
This agrees also with e.g., the results for ultrasmall metallic grains
\cite{delft2000}. 

\subsubsection{Distribution of zeros of the partition function}

\label{subsec:doz}

Another way to classify the thermal behavior of finite systems 
requires the analytic continuation of the partition function to the
complex plane. Grossmann {\it et al.} \cite{gr67,gr69,gl69} first 
introduced this technique for infinite systems. In these early works,
the authors were able to indicate the nature of phase transitions 
by studying the density of zeros (DOZ) of the partition function. 
Borrmann {\it et al.} recently extended this idea to finite many-body
systems \cite{bmh00}. We implement the method by extending 
the inverse temperature to the complex plane $\beta \rightarrow {\cal B}=
\beta +i\tau$. The partition function is then given by
\begin{equation}
Z({\cal B})= \int dE \rho(E) \exp(-{\cal B}E).
\end{equation}
Since the partition function is an integral function, the zeros 
${\cal B}_k = {\cal B}_{-k}^*=\beta_k + i\tau_k \;\; (k =1,\cdots, {\cal N})$
are complex conjugated. 

Different phases are represented by regions of holomorphy that are
separated by zeros of the partition function. These zeros typically lie
on lines in the complex temperature plane. For a finite system, the zeros
do not fall exactly on lines (they can be quite distinguishable depending
on the size of the system), and therefore the separation between two 
phases is more blurred than in an infinite system. The distribution of zeros
contains the complete thermodynamic information about the system, and 
all thermodynamic properties are derivable from it. For example, in the
complex plane, we define the specific heat as
\begin{equation}
C_v({\cal B})=\frac{\partial^2 \ln Z({\cal B})}{\partial {\cal B}^2}\;.
\end{equation}
Hence, the zeros of the partition function become poles of $C_v({\cal B})$. 
A pole approaching the real axis  infinitely closely causes a divergence
at a real critical temperature $T_C$. The contribution of a zero ${\cal B}_k$
to the specific heat decreases with increasing imaginary part $\tau_k$, so 
that thermodynamic properties of a system are governed by the zeros of 
$Z$ lying close to the real axis. 

The distribution of zeros close to the real axis is approximately described
by three parameters. Two of these parameters reflect the order of the
phase transition, while the third indicates the size of the system. 
Let us assume that the zeros lie on a line. We label the zeros according
to their closeness to the real axis. Thus $\tau_1$ reflects the discreteness
of the system. The density of zeros for a given $\tau_k$ is given by
\begin{equation}
\phi\left(\tau_k\right)=\frac{1}{2}
\left(\frac{1}{\mid {\cal B}_k -{\cal B}_{k-1}\mid} + 
\frac{1}{\mid {\cal B}_{k+1}-{\cal B}_k \mid}\right)\;,
\end{equation}
with $k=2,3,4,\cdots$. A simple power law describes the density of
zeros for small $\tau$, namely $\phi(\tau)\sim \tau^\alpha$. If we use
only the first three zeros, then $\alpha$ is given by
\begin{equation}
\alpha=\frac{\ln \phi(\tau_3) - \ln\phi(\tau_2)}{\ln\tau_3 -\ln\tau_2}\;.
\end{equation}
The final parameter that describes the distribution of zeros is given
by $\gamma=\tan\nu\sim (\beta_2-\beta_1)/(\tau_2-\tau_1)$. 

In the thermodynamic limit $\tau_1\rightarrow 0$ in which case the parameters
$\alpha$ and $\gamma$ coincide with the infinite system limits discussed 
by Grossman {\it et al.} \cite{gr67,gr69,gl69}. For the infinite system,
$\alpha=0$ and $\gamma=0$ yield a first-order phase transition, while
for $0<\alpha<1$ and $\gamma=0$ or $\gamma\ne 0$ indicates a second-order
transition. For arbitrary $\gamma$ third-order transitions occur when 
$1 \le \alpha < 2$. For systems approaching infinite particle number, 
$\alpha$ cannot be smaller than zero since this causes a divergence of the
internal energy. In small systems with finite $\tau_1$, $\alpha<0$ is 
possible. 

Continuation of the partition function to the complex plane is best
interpreted by invoking a quantum-mechanical interpretation, namely
\begin{equation}
Z(\beta+i\tau)=\hat{Tr}_A\left[\exp\left(-i\tau\hat{H}\right)
\exp\left(-\beta\hat{H}\right)\right]\;,
\end{equation}
where the quantum-mechanical trace of an operator, projected on a
specified particle number is given by
\begin{equation}
\hat{Tr}_A\hat{\xi}=\sum_\alpha 
\langle\alpha\mid\hat{P}_A\hat{\xi}\mid\alpha\rangle\;,
\end{equation}
$\hat{P}_A$ is the number projection operator, and $\alpha$ runs
over all many-body states. Since $\beta$ represents the inverse
temperature, the thermally averaged many-body state is
a linear combination of many-body states weighted by a Boltzmann factor,
$\mid\Psi(\beta,t=0)\rangle=\exp(-\beta E_\alpha)\mid\alpha\rangle$, so 
that the partition function may be compactly written as
\begin{equation}
Z(\beta+i\tau)=\langle\Psi(\beta,t=0)\mid\Psi(\beta,t=\tau)\rangle\;.
\end{equation}
Thus the zeros represent those times for which the overlap of the 
initial canonical state with the time-evolved state vanishes. In the
$\tau$ direction, the zeros represent a memory boundary for the system.

\begin{figure}
%\begin{center}
%   \setlength{\unitlength}{1mm}
%   \begin{picture}(140,220)
%   \put(30,0){\epsfxsize=9cm \epsfbox{dean_hjorthjensen_figcontourplot.ps}}
\includegraphics[scale=1.0,angle=0]{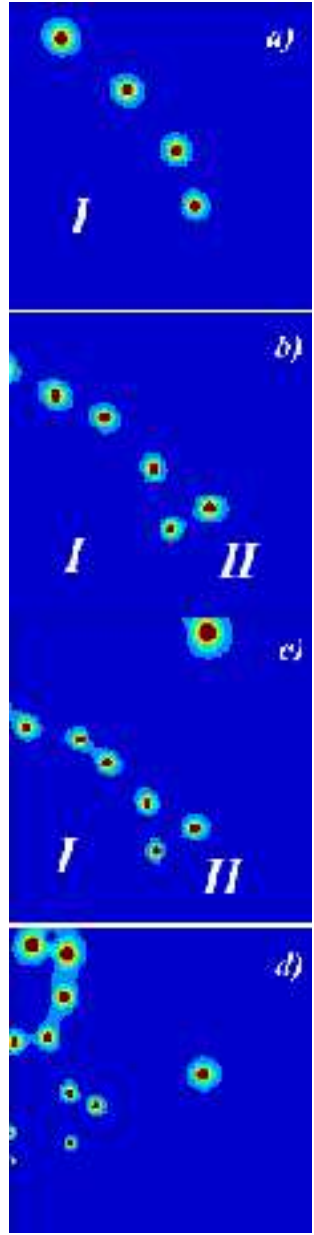}
%   \end{picture}
%\end{center} 
\caption{Contour plots of the specific heat in the complex temperature plane
for a) $N=11$, b) $N=14$, and c) $N=16$ particles. Panel d) 
shows the $N=14$ case with weak pairing.  
The spots indicate the locations of the 
zeros of the canonical partition function.} 
\label{fig:contourplot}
\end{figure}

In Fig.~\ref{fig:contourplot} we show contour 
plots of the specific heat $\mid C_v({\cal B})\mid $
in the complex temperature plane
for $N=11$ (a), $14$ (b), and $16$ (c) particles at normal 
pairing $d/G=0.5$ and
the $N=14$ (d) in the strong pairing limit, $d/G=1.5$. 
The poles are at the center of the dark contour regions.  We see evidence
of two phases in these systems. The first phase, labeled $I$ in 
Fig.~\ref{fig:contourplot}, is a mixed seniority phase, while the 
second phase, $II$, is a paired phase with zero seniority and exists
only in even-$N$ systems.  No paired phase exists in 
the $N=11$ system, and no clear boundaries are evident 
in the strong pairing case. We find that for (b) and (c) the 
DOZ are apparently distributed along two lines where the 
intersection occurs at $\tau_1$, which is the point closest to 
the real axis. As the pairing branch (for $\beta >\beta_1$) 
only encompasses two points, we are unable to precisely 
determine $\alpha$ along this branch while $\gamma>0$. Based on
our free energy results discussed above, we believe $\alpha$ 
along this branch will be positive. In the mixed phase branch
(for $\beta<\beta_1$) we find $\gamma<0$, and $\alpha < 0$ in
all normal-pairing cases. 

\subsection{Level densities from shell-model Monte Carlo calculations}

We also applied shell-model Monte Carlo (SMMC) 
techniques to survey rare-earth nuclei
in the Dy region \cite{WK00}. The goal of this extensive study 
was to examine how the phenomenologically motivated ``pairing-plus-quadrupole''
interaction compares in exact shell-model solutions with other methods.
We also examined how the shell-model solutions compare with
experimental data.  

We discuss here one particular aspect of that work,
namely level density calculations. Details may be found in
\cite{WK00}.
We used the Kumar-Baranger Hamiltonian with parameters appropriate
for this region. Our single-particle space included the 50-82 subshell
for the protons and the 82-126 shell for the neutrons.
While several interesting aspects of these systems were studied in
SMMC, we limit our discussion here to the level densities obtained for
$^{162}$Dy.

SMMC is an excellent way to calculate level densities.  $E(\beta)$ is
calculated for many values of $\beta$ which determine the partition
function, $Z$, as
\begin{equation}
\ln[Z(\beta)/Z(0)]=-\int_0^\beta d\beta'E(\beta')\;.
\end{equation}
$Z(0)$ is the total number of available states in the space.  The
level density is then computed as an inverse
Laplace transform of $Z$.  Here, the last step is performed with a
saddle point approximation with $\beta^{-2}C\equiv -dE/d\beta$:
\begin{eqnarray}
S(E)& = & \beta E + \ln Z(\beta)\;, \\
\rho(E)&=&(2\pi\beta^{-2}C)^{-1/2}\exp(S)\;.
\label{eq:rho}
\end{eqnarray}

\begin{figure}
{\hskip 0.5in \psfig{file=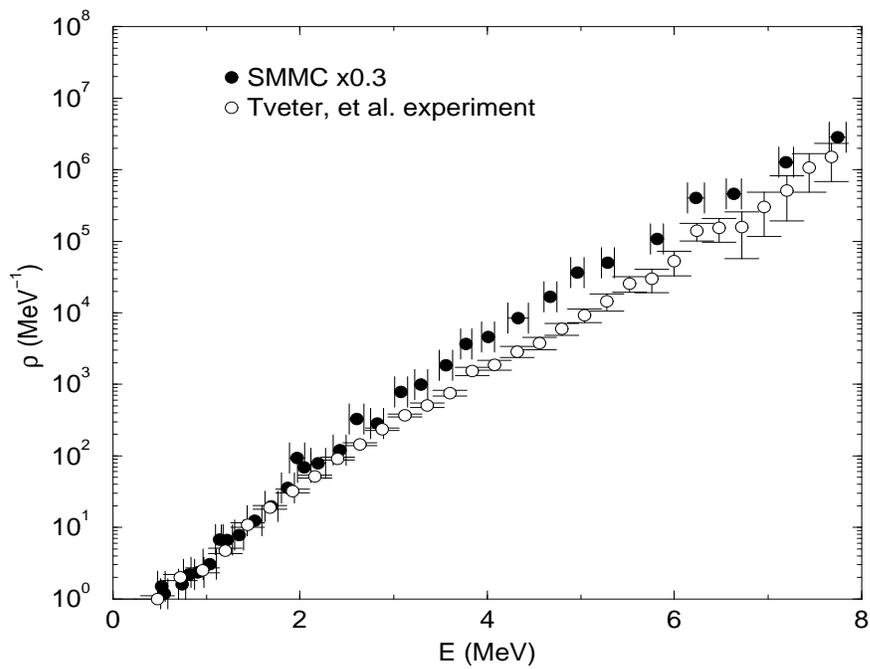,height=3.5in,width=4.5in}}
\caption{SMMC density vs. experimental
data in $^{162}$Dy.
\label{fig:dy162}}
\end{figure}

The comparison of SMMC density in $^{162}$Dy with the Tveter 
{\em et al.}~\cite{oslo2} data is displayed in Fig.~\ref{fig:dy162}.
The experimental
method can reveal fine structure, but does not determine the absolute
density magnitude.  The SMMC calculation is scaled by a factor to
facilitate comparison.  In this case, the factor has been chosen to
make the curves agree at lower excitation energies.  From 1-3 MeV,
the agreement is very good.  From 3-5 MeV, the SMMC density
increases more rapidly than the
data.  This deviation from the data cannot be accounted for by
statistical errors in either the calculation or measurement.  Near 6
MeV, the measured density briefly flattens before increasing and this
also appears in the calculation, but the measurement errors are larger
at that point.

The measured density includes all states
from the theoretical calculation plus some others, so that one
would expect the measured density to be greater than or equal to the
calculated density and never smaller.  We may have instead chosen our
constant to match the densities for moderate excitations and let the
measured density be higher than the SMMC density for lower energies
(1-3 MeV). Comparing structure between SMMC and data is difficult
for the lowest energies due to statistical errors in the calculation
and comparison at the upper range of the SMMC calculation, i.e.,
$E\approx15$ MeV is unfortunately impossible since the data only
extend to about $8$ MeV excitation energy.

%%%  sect 6 conclusion

%\input{conclusion}

% first iteration 4/7/02 mhj  
% typos + added mf stuff 4/10/02 mhj
% october 2002 mhj included comments from Hubert
\section{CONCLUSIONS AND OUTLOOK}
\label{sec:conclusion}

Pairing is an essential feature of nuclear systems, with 
several interesting and unsettled theoretical and experimental   
consequences, such  as superfluidity and neutrino
emission in neutron stars or pairing transitions in finite nuclei.

This review is by no means an exhaustive overview; rather, our focus has been
on the link between the nuclear many-body problem, and the underlying 
features of the nuclear force, and selected experimental interpretations
and manifestations of pairing in nuclear systems. 
Our preferred many-body tools have been  the   
nuclear shell model with its  effective interactions 
and various many-body approaches to infinite matter. The common starting 
point for all these
many-body approaches is, however, the free nucleon-nucleon interaction.

Within this setting, we have 
tried to present and expose several features of pairing 
correlations in nuclear systems. 
In particular, we have shown that in neutron star matter 
(Sec.~\ref{sec:NN_to_pairing}), pairing and superfluidity is synonymous with
singlet $^1S_0$ and triplet $^3P_2$ pairing up to densities 2-3 times nuclear
matter saturation density. 
For singlet pairing it is the central part of the nucleon-nucleon 
force which matters, which within a meson-exchange picture  
can be described in terms of multi-pion exchanges. 
For triplet pairing, it is the two-body spin-orbit force which provides
the attraction necessary for creating a positive pairing gap.
Hyperon pairing, especially $\Sigma^-$ pairing, is also
very likely. However, the actual size of these 
nucleon and/or hyperon pairing gaps
in infinite neutron star matter is an unsettled 
problem and awaits further theoretical studies.
A proper treatment of both short-range and long-range correlations 
is central to this problem.
It will have significant consequences on the emissivity of 
neutrinos in a neutron star. Color superconductivity in the interior of such 
compact objects is also an entirely open topic. A similarly unsettled issue 
is the size of 
the triplet $^3S_1$ gap in symmetric matter or asymmetric nuclear matter.

The above partial waves are also important for our understanding of pairing 
properties in finite nuclei. In Sec.~\ref{sec:pairing_correlations} 
we showed, for example, that the near-constancy 
of the excitation energy between the ground state 
with $J=0$ and the first excited state with $J=2$ 
for the tin isotopes from $^{102}$Sn to $^{130}$Sn is essentially due to the 
same partial waves which yield a finite pairing gap in neutron star matter.
Moreover, a seniority analysis of the pairing content of the wave functions 
for these states shows that we can very well approximate the ground state 
with a seniority ${\cal S} = 0$ state (no broken pairs) and the first 
excited state in terms of a 
seniority   ${\cal S} = 2$ state (one broken pair).

Furthermore, we have used results from large-scale shell-model Monte Carlo 
and diagonalization
calculations to extract information about isoscalar and isovector pairing
and thermal response 
for $fp$-shell nuclei. 
One key result here was the decrease of $T=1$ pairing correlations 
as a function of increasing temperature (up to about 1~MeV) 
and a commensurate buildup of structure in the specific heat 
curves at the same temperature. 
Information about proton-neutron pairing and the Wigner 
energy has also been presented in Sec.~\ref{sec:pairing_correlations}.
The important result here was that all $J$ channels of the 
interaction contribute to the
Wigner energy, and that the $J=1$ and $J=J_{\rm max}$ channels contributed
most (see also \cite{poves98}).
Proton-neutron pairing is, however, a much more elusive  aspect 
of the nuclear pairing problem. 
Its actual size, as is also the case in infinite 
matter, needs further analysis. 
Our results from Sec.~\ref{sec:randoms} may
indicate that the $T=0$ part of the pairing
interaction does not play a crucial role. For more information, see
\cite{volodya1996,volya_2}. 

Finally, we have tried to analyze recent experimental data on nuclear level
densities in terms of pairing correlations. This was done in 
Sec.~\ref{sec:leveldensities_sec3}. These data reveal structures in the level
density of rare-earth nuclei that can be interpreted as a gradual breaking
of pairs. The experimental level densities can also be used to compute the 
thermal properties such as the entropy or the specific heat.
The even systems exhibit a clear bump in the heat capacity. The temperature
where this bump appears can be interpreted as a critical temperature 
for the quenching of pairing correlations. Similar features were also noted in
Sec.~\ref{sec:pairing_correlations} (see especially Figs.~\ref{fig_810} and 
\ref{fig:figyoram}).
More information was also obtained by studying the 
experimental entropy for even and odd nuclei with those
extracted from a simple pairing model with a given number of particles 
and number of doubly-degenerate particle levels. We showed in 
Sec.~\ref{sec:leveldensities_sec3} that the entropy of the odd and even system
merge at a temperature which corresponds to the observed bumps in the 
heat capacity. This temperature occurs typically where we have 2-3 
broken pairs.

Within the framework of this simple pairing model, we 
showed also that for a finite system there is no sudden and abrupt transition
to another phase, as we have in an infinite system. Rather, there is a
gradual breaking of pairs as temperature increases. However, studying
systems with different numbers of particles and levels, we 
presented also two 
possible methods for classifying the order of the transition. 
In order to perform these studies, we needed all eigenvalues
from the simple pairing model in order to compute thermodynamical
properties. 
Albeit there have been several interesting theoretical developments
of the solution of the model Hamiltonian of Eq.~(\ref{eq:pairH2_sec})
or related models,
see for example 
\cite{richardson2002,dukelsky2002,volya_1,volya_3}, we would like to 
stress that the investigation of thermodynamic properties requires a
knowledge of all eigenvalues. 

Furthermore, an obvious deficiency of this simple model in nuclear physics
is the lack of long-range correlations, which could, e.g., 
be expressed via quadrupole terms. A pairing-plus-quadrupole model,
as discussed in  Sec.~\ref{sec:pairing_correlations}, would however, spoil
the simple block-diagonalization feature in terms of seniority as a good 
quantum number. Such a model is however necessary, since the nuclear force
is particular in the sense that the ranges of its short-range and long-range
parts are rather similar. This means that short-range contributions arising
from e.g., strongly paired particle-particle terms and 
long-range terms from particle-hole excitations  
are central for a correct many-body
description of nuclear systems, from nuclear matter to finite nuclei. 
The difficulty connected with these aspects of the nuclear force means
that further 
analysis of the thermodynamics of e.g.,
rare-earth nuclei can presently only be done  
in terms of large-scale shell-model Monte Carlo methods.

It should be fairly obvious that we have only been able to cover
a very limited aspect of pairing in nuclear systems.
We have limited the attention to stable systems. However, 
pairing correlations are expected to play a special role
in drip-line nuclei \cite{dob_96}. There is currently a considerable
experimental effort in nuclear physics, especially due to the 
advances from radioactive-beam and heavy-ion 
facilities, which  have stimulated an exploration of 
nuclei far from stability. Many of these nuclei are weakly bound systems.
Hence, due to strong surface effects, the properties of such nuclei
are perfect laboratories for studies of the density dependence of 
pairing interactions. An experimental observable that may probe
pairing correlations is the pair transfer factor, which is directly related 
to the pairing density (see \cite{jacekwitek1998} for more details). 
The difference in the asymptotic behavior of 
the single-particle density and the pair density in a weakly bound system can 
be probed by comparing the energy dependence 
of one-particle and two-particle transfer cross sections.
Such measurements, when performed on both stable and neutron rich nuclei,
can hopefully shed some light on the asymptotic behavior of pairing. 
An interesting system here is the chain of tin isotopes beyond $^{132}$Sn.
Various mean-field calculations \cite{dob_96} indicate that there is a
considerable increase in the pair transfer form factors for nuclei 
between  $^{150}$Sn and  $^{172}$Sn \cite{dob_96}. At the moment of writing,
$\beta$-decay properties of nuclei like  $^{136}$Sn have just been studied
\cite{shergur2002}. 
 
From a many-body point of view, a 
correct treatment of these weakly bound systems
entails an approach which encompasses a proper description of bound states and eventually features from the continuum. Such calculations have recently been mounted within the framework of mean-field and Hartree-Fock-Bogoliubov (HFB) 
models (see \cite{sandulescu2001,sandulescu2002}). A 
finite-range pairing 
interaction was included explicitely in the calculations. 
We mention here that the pairing terms in such mean-field calculations
can be parameterized from microscopic many-body calculations, as demonstrated
by Smerzi {\em et al.} \cite{smerzi1997}.
However, 
to include
the continuum in a many-body description such as the shell-model with an
appropriate effective interaction is highly non-trivial. Even the determination
of the effective two-body interaction is an open problem. The low-density
studies of singlet $^1S_0$ pairing in dilute Fermi systems \cite{henning2000}
clearly demonstrate that polarization terms cannot be neglected. 

We conclude by pointing to the strong similarities between pairing in 
the nuclear 
many-body problem and pairing in systems of trapped fermions (see
\cite{henning_ben,bruun2000} for recent examples). 
There are also very strong couplings
to the experimental and theoretical developments of our understanding of
ultra-small superconducting grains 
\cite{tinkham95,tinkham96,tinkham98,delft2000,balian1999,mastellone98,sierra99}.

\section*{Acknowledgments}
This work would have been impossible without the 
invaluable and fruitful
collaborations and discussions over many years with several colleagues: 
Yoram Alhassid, Marcello Baldo, George F.~Bertsch, Alexandar
Beli\'c, Alex Brown, John Clark, 
Magne Guttormsen, \O ystein Elgar\o y, Jonathan Engel, Torgeir Engeland, 
Lars Engvik, Henning Heiselberg, 
Anne Holt, Calvin Johnson, Steven Koonin, 
Karlheinz Langanke, Umberto Lombardo, Ben Mottelson, Witek Nazarewicz, 
Erich Ormand, 
Eivind Osnes, Thomas Papenbrock, Chris Pethick, 
Alfredo Poves, P.B. Radha, Nicolae Sandulescu, Andreas Schiller, 
Peter Schuck, Hans-Joseph Schulze, Jody White, and Andres Zuker.
Research at Oak Ridge National Laboratory was sponsored by the Division
of Nuclear Physics, U.S. Department of Energy under contract DE-AC05-00OR22725 
with UT-Battelle, LLC. 

\bibliography{dean_hjorthjensen}
\bibliographystyle{apsrmp}

\newpage

\begin{table}[hbt]
\begin{center}
\caption{ $2^+_1-0^+_1$ excitation energy for the 
even tin isotopes $^{130-116}$Sn for various approaches
to the effective interaction. See text for further details. 
Energies are given in MeV. \label{tab:table_tincalc1}}
\begin{tabular}{lcccccccc}\hline
 & {$^{116}$Sn} & {$^{118}$Sn} & {$^{120}$Sn} &{$^{122}$Sn} & {$^{124}$Sn} & {$^{126}$Sn} & {$^{128}$Sn} & {$^{130}$Sn} \\ \hline
Expt & 1.29 & 1.23 & 1.17 & 1.14 & 1.13 & 1.14 & 1.17 & 1.23 \\
$V_{\mathrm{eff}}$ & 1.17 & 1.15 & 1.14 & 1.15 & 1.14 & 1.21 & 1.28 & 1.46 \\
$G$-matrix &1.14 & 1.12& 1.07 & 0.99 & 0.99 & 0.98 & 0.98 & 0.97  \\
$^1S_0$ $G$-matrix &1.38 &1.36 &1.34 &1.30 & 1.25& 1.21 &1.19 &1.18 \\
No $^1S_0$ \& $^3P_2$ in $G$ &     &     &     &      &0.15 &-0.32 &0.02 &-0.21  \\\hline
\end{tabular}
\end{center}
\end{table}

\begin{table}[htbp]
\caption{ Seniority $v=0$ overlap (first row)  
         $|\langle ^{A}Sn;0^{+}|(S^{\dagger})^{\frac{n}{2}}| 
         \tilde{0} \rangle |^{2}$ and the seniority $v=2$ 
         overlaps (remaining rows) $|\langle ^{A}Sn ;J_{f}|
         D^{\dagger}_{JM}(S^{\dagger})^{\frac{n}{2} - 1}| 
         \tilde{0} \rangle |^{2}$ for the lowest--lying eigenstates 
         of $^{128-120}$Sn.\label{tab:seniority}}
\begin{center}
\begin{tabular}{cccccc}
\hline
 & A=128 & A=126 & A=124 & A=122 & A=120 \\ 
\hline
$0^{+}_{1}$ & 0.96 & 0.92 & 0.87 & 0.83 & 0.79 \\ 
$2^{+}_{1}$ & 0.92 & 0.89 & 0.84 & 0.79 & 0.74 \\ 
$4^{+}_{1}$ & 0.73 & 0.66 & 0.44 & 0.13 & 0.00 \\ 
$4^{+}_{2}$ & 0.13 & 0.18 & 0.39 & 0.66 & 0.74 \\
$6^{+}_{1}$ & 0.81 & 0.85 & 0.83 & 0.79 & 0.64 \\
\hline
\end{tabular}
\end{center}
\end{table}

\begin{table}[b]
\caption{ $\Delta F/N$ for $T=0.85$ MeV. 
See text for further details. \label{tab:free_energy10_16}} 
\begin{tabular}{ccccc}\hline
$N$ & 10 & 12 & 14& 16 \\ \hline
$\Delta F/N$ [MeV]   &0.531 & 0.505 & 0.501 & 0.495 \\\hline
\end{tabular} 

\end{table}

%%%   figure captions, no embedded figures

\clearpage


\end{document}